\documentclass[reprint,superscriptaddress,frontmatterverbose,noshowpacs,preprintnumbers,longbibliography,amsmath,amssymb,aps,pre]{revtex4-1}

\usepackage{graphicx}
\usepackage{times}
\usepackage{hyperref}
\hypersetup{
    unicode=false,                                                                 
    pdftoolbar=false,                                                              
    pdfmenubar=false,                                                              
    pdffitwindow=false,                                                            
    pdfstartview={FitH},                                                           
    pdftitle={Deep learning of contagion dynamics on complex networks},            
    pdfauthor={Charles Murphy, Edward Laurence and Antoine Allard},                
    pdfsubject={},                                                                 
    pdfkeywords={machine learning} {complex networks} {stochastic processes},      
    pdfnewwindow=true,                                                             
    colorlinks=true,                                                               
    linkcolor=seaborn_deep_blue,                                                   
    citecolor=seaborn_deep_blue,                                                   
    filecolor=seaborn_deep_blue,                                                   
    urlcolor=seaborn_deep_blue                                                     
}

\renewcommand{\vec}[1]{\boldsymbol{#1}}
\newcommand{\avg}[1]{\langle #1\rangle}
\newcommand{\param}{\vec{\Theta}}

\newcommand{\set}[1]{\left\{#1\right\}}

\usepackage{xcolor}
\definecolor{seaborn_deep_blue}{RGB}{76, 114, 176}
\definecolor{seaborn_deep_orange}{RGB}{221, 132, 82}
\definecolor{seaborn_deep_green}{RGB}{85, 168, 104}
\definecolor{seaborn_deep_red}{RGB}{196, 78, 82}

\usepackage{pdfpages}
\makeatletter
\AtBeginDocument{\let\LS@rot\@undefined}
\makeatother

\begin{document}
\title{Deep learning of contagion dynamics on complex networks}
\author{Charles Murphy}%
\affiliation{D\'epartement de Physique, de G\'enie Physique, et d'Optique, Universit\'e Laval, Qu\'ebec (Qu\'ebec), Canada G1V 0A6}%
\affiliation{Centre interdisciplinaire en mod\'elisation math\'ematique, Universit\'e Laval, Qu\'ebec (Qu\'ebec), Canada G1V 0A6}%
\author{Edward Laurence}%
\affiliation{D\'epartement de Physique, de G\'enie Physique, et d'Optique, Universit\'e Laval, Qu\'ebec (Qu\'ebec), Canada G1V 0A6}%
\affiliation{Centre interdisciplinaire en mod\'elisation math\'ematique, Universit\'e Laval, Qu\'ebec (Qu\'ebec), Canada G1V 0A6}%
\author{Antoine Allard}%
\email[]{antoine.allard@phy.ulaval.ca}%
\affiliation{D\'epartement de Physique, de G\'enie Physique, et d'Optique, Universit\'e Laval, Qu\'ebec (Qu\'ebec), Canada G1V 0A6}%
\affiliation{Centre interdisciplinaire en mod\'elisation math\'ematique, Universit\'e Laval, Qu\'ebec (Qu\'ebec), Canada G1V 0A6}%
\date{\today}
\begin{abstract}
  Forecasting the evolution of contagion dynamics is still an open problem to which mechanistic models only offer a partial answer.  To remain mathematically or computationally tractable, these models must rely on simplifying assumptions, thereby limiting the quantitative accuracy of their predictions and the complexity of the dynamics they can model.  Here, we propose a complementary approach based on deep learning where the effective local mechanisms governing a dynamic on a network are learned from time series data.  Our graph neural network architecture makes very few assumptions about the dynamics, and we demonstrate its accuracy using different contagion dynamics of increasing complexity. By allowing simulations on arbitrary network structures, our approach makes it possible to explore the properties of the learned dynamics beyond the training data. Finally, we illustrate the applicability of our approach using real data of the COVID-19 outbreak in Spain.  Our results demonstrate how deep learning offers a new and complementary perspective to build effective models of contagion dynamics on networks.
\end{abstract}
\maketitle
\section*{Introduction}
Our capacity to prevent or contain outbreaks of infectious diseases is directly linked to our ability to accurately model contagion dynamics.  Since the seminal work of Kermack and McKendrick almost a century ago~\cite{kermack1927contribution}, a variety of models incorporating ever more sophisticated contagion mechanisms has been proposed~\cite{hethcote2000mathematics, siettos2013mathematical, Kiss2017, brauer2019mathematical}.  These mechanistic models have provided invaluable insights about how infectious diseases spread, and have thereby contributed to the design of better public health policies.  However, several challenges remain unresolved, which call for contributions from new modeling approaches~\cite{grassly2008mathematical, pastoreypiontti2019charting, viboud2019future}.

For instance, many complex contagion processes involve the nontrivial interaction of several pathogens~\cite{morens2008predominant, Sanz2014,nickbakhsh2019virus,hebert-dufresne2020macroscopic}, and some social contagion phenomena, like the spread of misinformation, require to go beyond pairwise interactions between individuals~\cite{centola2010spread, lehmann2018complex, iacopini2019simplicial}.  Also, while qualitatively informative, the forecasts of most mechanistic models lack quantitative accuracy~\cite{biggerstaff2018results}.  Indeed, most models are constructed from a handful of mechanisms which can hardly reproduce the intricacies of real complex contagion dynamics.  One approach to these challenges is to complexify the models by adding more detailed and sophisticated mechanisms.  However, mechanistic models become rapidly intractable as new mechanisms are added.  Moreover, models with higher complexity require the specification of a large number of parameters whose values can be difficult to infer from limited data.

There has been a recent gain of interest towards using machine learning to address the issue of the often-limiting complexity of mechanistic models~\cite{Brunton2016,Kutz2017,Pathak2017,Lusch2018,Pathak2018,desilva2020discovery,Chen2020a,hebert-dufresne2020macroscopic}. This new kind of approach aims at training predictive models directly from observational time series data. These data-driven models are then used for various tasks such as making accurate predictions~\cite{Pathak2017,Pathak2018}, gaining useful intuitions about complex phenomena~\cite{hebert-dufresne2020macroscopic} and discovering new patterns from which better mechanisms can be designed~\cite{Brunton2016,Kutz2017}. Although these approaches were originally designed for regularly structured data, this new paradigm is now being applied to epidemics spreading on networked systems~\cite{Dutta2018,Shah2020}, and more generally to dynamical systems~\cite{Rodrigues2019,Salova2019,Laurence2020}. Meanwhile, the machine learning community has dedicated a considerable amount of attention on deep learning on networks, structure learning and graph neural networks (GNN)~\cite{Zhang2018i, Zhou2018a, Xu2018}.  Recent works showed great promise for GNN in the context of community detection~\cite{perozzi2014deepwalk}, link prediction~\cite{Hamilton2017}, network inference~\cite{zhang2019general}, as well as for the discovery of new materials and drugs~\cite{Fout2017,Zitnik2018}. Yet, others have pointed out the inherent limitations of a majority of GNN architectures in distinguishing certain network structures~\cite{Xu2018}, in turn limiting their learning abilities. Hence, while recent advances and results~\cite{Kapoor2020,skarding2020foundations,Fritz2021,Gao2021} suggest that GNNs could be prime candidates for building effective data-driven dynamical models on networks, it remains to be shown if, how and when GNNs can be applied to dynamics learning problems.

In this paper, we show how GNN, usually used for structure learning, can also be used to model contagion dynamics on complex networks.  Our contribution is threefold.  First, we design a training procedure and an appropriate GNN architecture capable of representing a wide range of dynamics, with very few assumptions.  Second, we demonstrate the validity of our approach using various contagion dynamics on networks of different natures and of increasing complexity, as well as on real epidemiological data.  Finally, we show how our approach can provide predictions for previously unseen network structures, therefore allowing the exploration of the properties of the learned dynamics beyond the training data.
Our work generalizes the idea of constructing dynamical models from regularly structured data to arbitrary network structures, and suggests that our approach could be accurately extended to many other classes of dynamical processes.

\section*{Results}
In our approach, we assume that an unknown dynamical process, denoted $\mathcal{M}$, takes place on a known network structure---or ensemble of networks---, denoted $G=(\mathcal{V},\mathcal{E}; \vec{\Phi}, \vec{\Omega})$, where $\mathcal{V}=\set{v_1,\cdots, v_{N}}$ is the node set and $\mathcal{E}=\set{e_{ij}|\text{$v_j$ is connected to $v_i$}\wedge (v_i,v_j)\in \mathcal{V}^2}$ is the edge set. We also assume that the network(s) can have some metadata, taking the form of node and edge attributes denoted $\Phi_i = (\phi_1(v_i), \cdots,\phi_Q(v_i))$ for node $v_i$ and $\Omega_{ij} = (\omega_1(e_{ij}), \cdots,\omega_P(e_{ij}))$ for edge $e_{ij}$, respectively, where $\phi_q:\mathcal{V}\to \mathbb{R}$ and $\omega_p:\mathcal{E}\to \mathbb{R}$. These metadata can take various forms like node characteristics or edge weights. We also denote the node and edge attribute matrices $\vec{\Phi} = (\Phi_i|v_i\in\mathcal{V})$ and $\vec{\Omega} = (\Omega_{ij}|e_{ij}\in\mathcal{E})$, respectively.

Next, we assume that the dynamics $\mathcal{M}$ has generated a time series $D$ on the network $G$. This time series takes the form of a pair of consecutive snapshots $D = (\vec{X},\vec{Y})$ with $\vec{X} = \big(X_1, \cdots, X_T\big)$ and $\vec{Y} = \big(Y_1, \cdots, Y_T\big)$, where $X_t\in \mathcal{S}^{|\mathcal{V}|}$ is the state of the nodes at time $t$, $Y_t\in\mathcal{R}^{|\mathcal{V}|}$ is the outcome of $\mathcal{M}$ defined as
\begin{equation}
    Y_t = \mathcal{M}\big(X_t, G\big)\,,
\end{equation}
$\mathcal{S}$ is the set of possible node states, and $\mathcal{R}$ is the set of possible node outcomes. This way of defining $D$ allows us to formally concatenate multiple realizations of the dynamics in a single dataset. Additionally, the elements $x_i(t) \equiv (X_t)_i $ and $y_i(t) \equiv (Y_t)_i$ correspond to the state of node $v_i$ at time $t$ and its outcome, respectively. Typically, we consider that the outcome $y_i(t)$ is simply the state of node $v_i$ after transitioning from state $x_i(t)$. In this case, we have $\mathcal{S} = \mathcal{R}$ and $x_i(t+\Delta t) = y_i(t)$ where $\Delta t$ is the length of the time steps. However, if $\mathcal{S}$ is a discrete set---i.e. finite and countable---$y_i(t)$ is a transition probability vector conditioned on $x_i(t)$ from which the following state, $x_i(t+\Delta t)$, will be sampled. The element $\big(y_i(t)\big)_m$ corresponds to the probability that node $v_i$ evolves to state $m\in\mathcal{S}$ given that it was previously in state $x_i(t)$---i.e. $\mathcal{R} = [0, 1]^{|\mathcal{S}|}$. When $\mathcal{M}$ is a stochastic dynamics, we do not typically have access to the transition probabilities $y_i(t)$ directly,  but rather to the observed outcome state---e.g. $x_i(t+\Delta t)$ in the event where $\vec{X}$ is temporally ordered---, we therefore define the observed outcome $\tilde{y}_i(t)$ as
\begin{equation}
    (\tilde{y}_i(t))_m = \delta\big(x_i(t+\Delta t), m\big)\,,\,\,\forall m\in\mathcal{S}
\end{equation}
where $\delta(x,y)$ is the Kronecker delta.  Finally, we assume that $\mathcal{M}$ acts on $X_t$ locally and identically at all times, according to the structure of $G$. In other words, knowing the state $x_i$ as well as the states of all the neighbors of $v_i$, the outcome $y_i$ is computed using a time independent function $f$ identical for all nodes
\begin{equation}
    \label{eq:locality}
    y_i \equiv f\big(x_i, \Phi_i, x_{\mathcal{N}_i}, \Phi_{\mathcal{N}_i}, \Omega_{i{\mathcal{N}_i}}\big)\,,
\end{equation}
where $x_{\mathcal{N}_i} = \big(x_{j}|v_j\in\mathcal{N}_i\big)$ denotes the states of the neighbors, $\mathcal{N}_i := \set{v_j|e_{ij}\in\mathcal{E}}$ is the set of the neighbors, $\Phi_{\mathcal{N}_i} := \set{\Phi_{j}|v_{j}\in\mathcal{N}_i}$ and $\Omega_{i{\mathcal{N}_i}} := \set{\Omega_{ij}|v_{j}\in\mathcal{N}_i}$.  As a result, we impose a notion of locality where the underlying dynamics is time invariant and invariant under the permutation of the node labels in $G$, under the assumption that the node and edge attributes are left invariant.

Our objective is to build a model $\hat{\mathcal{M}}$, parametrized by a GNN with a set of tunable parameters $\param$ and trained on the observed dataset $D$ to mimic $\mathcal{M}$ given $G$, such that
\begin{equation} \label{eq:objective}
  \hat{\mathcal{M}}(X_t', G';\param) \approx \mathcal{M}(X_t', G')\,,
\end{equation}
for all states $X_t'$ and all networks $G'$. The architecture of $\hat{\mathcal{M}}$, detailed in Sec.~\ref{sub:gnn}, is designed to act locally similarly to $\mathcal{M}$. In this case, the locality is imposed by a modified attention mechanism inspired by Ref.~\cite{Velickovic2017}. The advantage of imposing locality allows our architecture to be \emph{inductive}: If the GNN is trained on a wide range of local structures---i.e. nodes with different neighborhood sizes (or degrees) and states---it can then be used on any other networks within that range. This suggests that the topology of $G$ will have a strong impact on the quality of the trained models, an intuition that is confirmed below. Simiarly to Eq.~\eqref{eq:locality}, we can write each individual node outcome computed by the GNN using a function $\hat{f}$ such that
\begin{equation} \label{eq:GNN_outcome}
    \hat{y}_i \equiv \hat{f}\big(x_i, \Phi_i, x_{\mathcal{N}_i}, \Phi_{\mathcal{N}_i}, \Omega_{i{\mathcal{N}_i}};\param\big)\,
\end{equation}
where $\hat{y}_i$ is the outcome of node $v_i$ predicted by $\hat{\mathcal{M}}$.

\begin{figure}[t]
  \centering
  \includegraphics[width=0.45\textwidth]{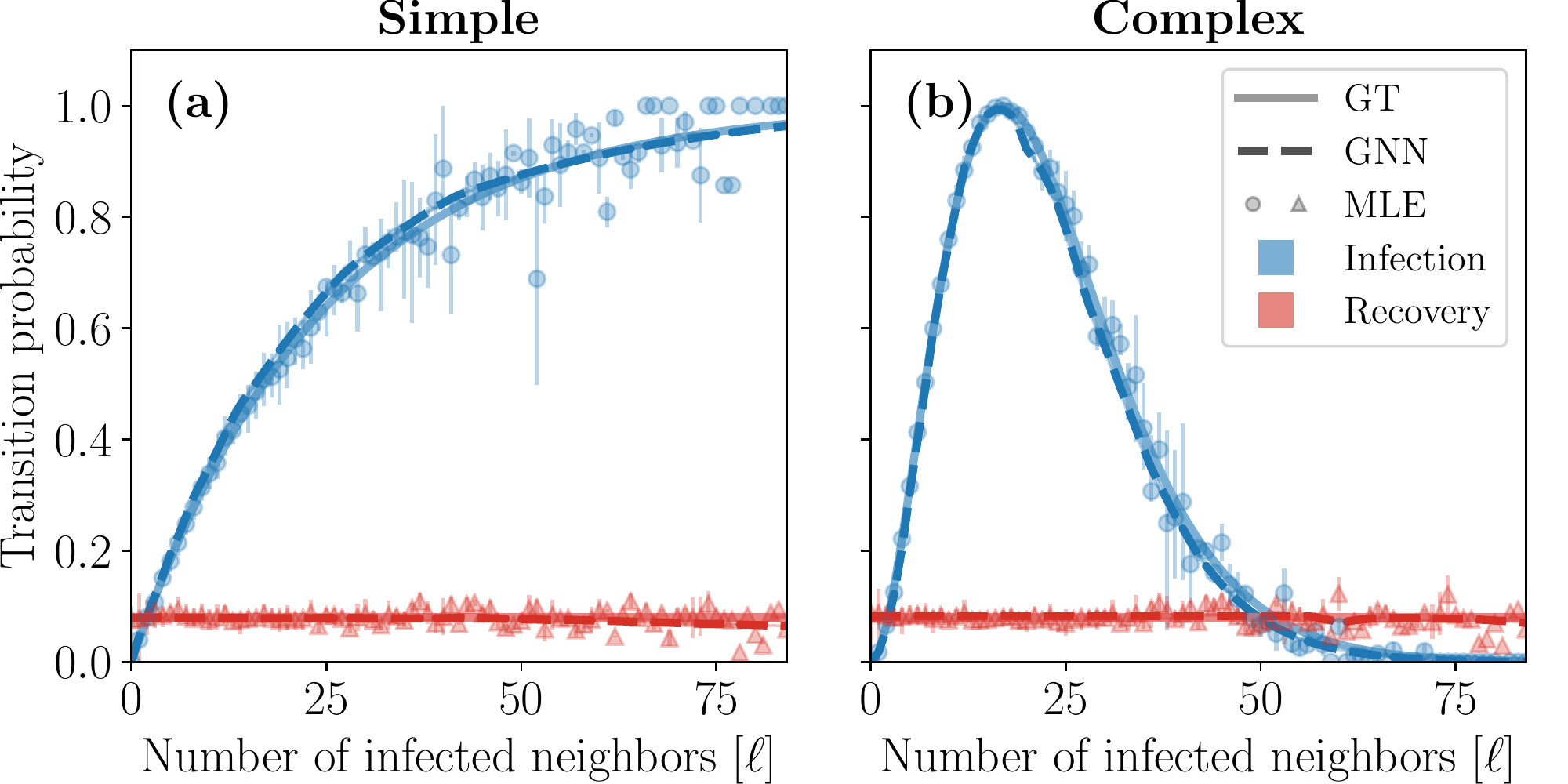}
  \caption{\textbf{Predictions of GNN trained on a Barab\'asi-Albert random network (BA)~\cite{Barabasi2016} for the (a) simple and (a) complex contagion dynamics.}
  The solid and dashed lines correspond to the transition probabilities of the dynamics used to generate the training data (labeled GT for ``ground truth''), and predicted by the GNN, respectively.  Symbols correspond to the maximum likelihood estimation (MLE) of the transition probabilities computed from the dataset $D$.
  The colors indicate the type of transition: infection ($S\to I$) in blue and recovery ($S\to I$) in red.
  The standard deviations, as a result of averaging the outcomes given $\ell$, are shown using a colored area around the lines (typically narrower than the width of the lines) and using vertical bars for the symbols.
  }
  \label{fig:ltp}
\end{figure}

\begin{figure*}[t]
  \centering
  \includegraphics[width=1.\linewidth]{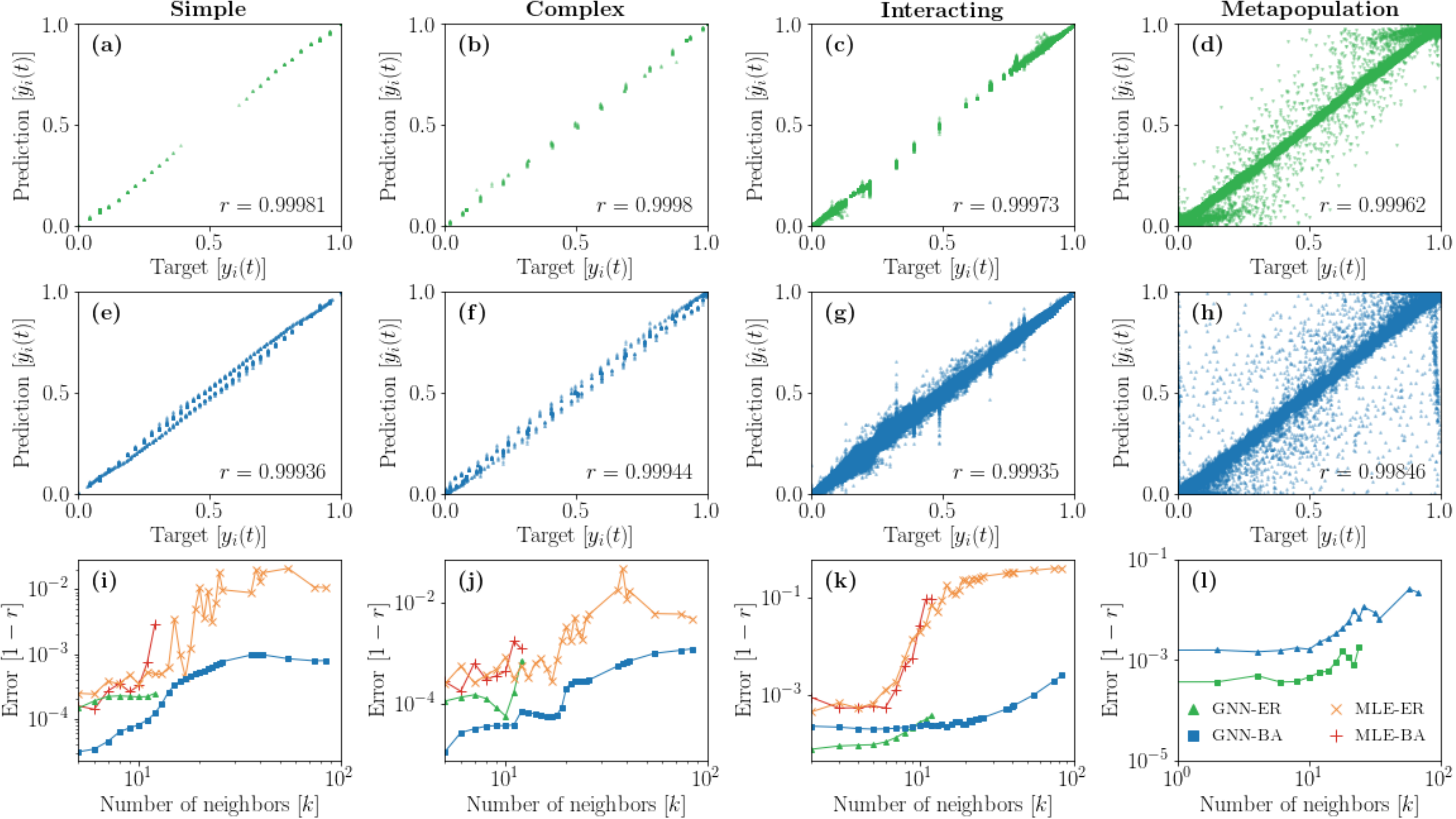}
  \caption{\textbf{Comparison between the targets and the predictions of GNN trained on Erd\H{o}s-R\'enyi networks (ER, top row) and on Barab\'asi-Albert networks~\cite{Barabasi2016} (BA, middle row) for the (a, e, i) simple, (b, f, j) complex, (c, g, k) interacting and (d, h, l) metapopulation dynamics.}
  Each point shown on the panels (a--h) corresponds to a different pair $(y_i(t), \hat{y}_i(t))$ in the complete dataset $D$.  We also indicate the Pearson coefficient $r$ on each panel to measure the correlation between the predictions and the targets and use it as a global performance measure.
  The panels (i--l) show the errors ($1 - r$) as a function of the number of neighbors for GNN trained on ER and BA networks, and those of the corresponding MLE. These errors are obtained from the Pearson coefficients computed from subsets of the prediction-target pairs where all nodes have degree $k$.
  }
  \label{fig:generalization}
\end{figure*}

The objective described by Eq.~\eqref{eq:objective} must be encoded into a global loss function, denoted $\mathcal{L}(\param)$. Like the outcome functions, $\mathcal{L}(\param)$ can be decomposed locally, where the local losses of each node $L(y_i, \hat{y}_i)$ are arithmetically averaged over all possible node inputs $(x_i, \Phi_i, x_{\mathcal{N}_i}, \Phi_{\mathcal{N}_i}, \Omega_{i{\mathcal{N}_i}})$, where $y_i$ and $\hat{y}_i$ are given by Eqs.~\eqref{eq:locality}~and~\eqref{eq:GNN_outcome}, respectively. By using an arithmetic mean to the evaluation of $\mathcal{L}(\param)$, we assume that the node inputs are distributed uniformly. Consequently, the model should be trained equally well on all of them. This is important because in practice we only have access to a finite number of inputs in $D$ and $G$, for which the node input distribution is typically far from being uniform. Hence, in order to train effective models, we recalibrate the inputs  using the following global loss
\begin{equation}
    \label{eq:loss}
  \mathcal{L}(\param) = \sum_{t\in\mathcal{T}'}\,\sum_{v_i\in\mathcal{V}'(t)} \frac{w_i(t)}{Z'}L\big(y_i(t), \hat{y}_i(t)\big) \,
\end{equation}
where $w_i(t)$ is a weight assigned to node $v_i$ at time $t$, and $Z' = \sum_{t\in\mathcal{T}'} \sum_{v_i\in\mathcal{V}'(t)}w_i(t)$ is a normalization factor. Here, the training node set $\mathcal{V}'(t)\subseteq \mathcal{V}$ and the training time set $\mathcal{T}'\subseteq [1, T]$ allow us to partition the training dataset for validation and testing when required.

The choice of weights needs to reflect the importance of each node at each time.  Because we wish to lower the influence of overrepresented inputs and increase that of rare inputs, a sound choice of weights is
\begin{equation}
    \label{eq:ideal_weights}
    w_i(t) \propto \rho\Big(k_i, x_i, \Phi_i, x_{\mathcal{N}_i}, \Phi_{\mathcal{N}_i}, \Omega_{i{\mathcal{N}_i}}\Big)^{-\lambda}\,
\end{equation}
where $k_i$ is the degree of node $v_i$ in $G$, and $0\leq\lambda\leq 1$ is an hyperparameter.  Equation~\eqref{eq:ideal_weights} is an ideal choice, because it corresponds to a principled importance sampling approximation of Eq.~\eqref{eq:loss}~\cite{Rubinstein2016}, which is relaxed via the exponent $\lambda$. We obtain a pure importance sampling scheme when $\lambda=1$. Note that the weights can rarely be exactly computed using Eq.~\eqref{eq:ideal_weights}, because the distribution $\rho$ is typically computationally intensive to obtain from data, especially for continuous $\mathcal{S}$ with metadata. We illustrate various ways to evaluate the weights in Sec.~\ref{sub:weights} and in Supplementary Information.

\begin{figure*}
  \centering
  \includegraphics[scale=0.4]{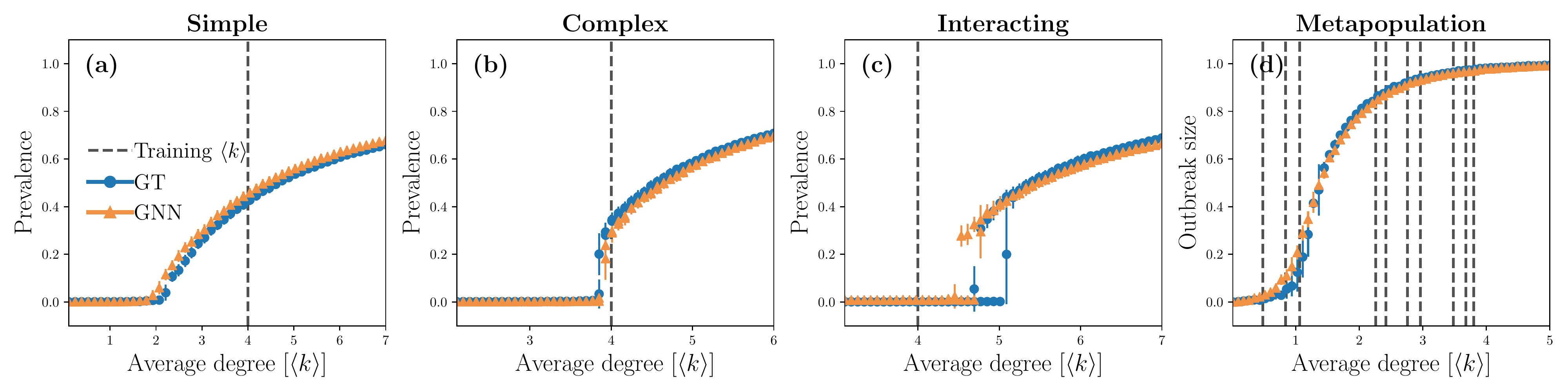}
  \caption{\textbf{Bifurcation diagrams of the (a) simple, (b) complex, (c) interacting and (d) metapopulation dynamics on Poisson networks~\cite{Barabasi2016} composed of $|\mathcal{V}| = 2000$ nodes with different average degrees $\langle k \rangle$.}
  The prevalence is defined as the average fraction of nodes that are asymptotically infected by at least one disease and the outbreak size corresponds to the average fraction of nodes that have recovered.  These quantities are obtained from numerical simulations using the ``ground truth`` (GT) dynamics (blue circles) and the GNN trained on Barab\'asi-Albert networks (orange triangles).  The error bars correspond to the standard deviations of these numerical simulations.  The trained GNN used are the same ones as those used for Fig.~\ref{fig:generalization}.  As a reference, we also indicate with dashed lines the value(s) of average degree $\avg{k}$ corresponding to the network(s) on which the GNN were trained. On panel (d), more than one value of $\avg{k}$ appear as multiple networks with different average degrees were used to train the GNN.
  }
  \label{fig:phasetransition}
\end{figure*}

We now illustrate the accuracy of our approach by applying it to four types of synthetic dynamics of various natures (see Sec.~\ref{sub:dynamics} for details on the dynamics). We first consider a \emph{simple contagion} dynamics: The discrete-time susceptible-infected-susceptible (SIS) dynamics. In this dynamics, nodes are either susceptible ($S$) or infected ($I$) by some disease, i.e. $\mathcal{S}=\set{S, I}=\set{0,1}$, and transition between each state stochastically according to an infection probability function $\alpha(\ell)$, where $\ell$ is the number of infected neighbors of a node, and a constant recovery probability $\beta$. A notable feature of simple contagion dynamics is that susceptible nodes get infected by the disease through their infected neighbors independently. This reflects the assumption that disease transmission behaves identically whether a person has a large number of infected neighbors or not.

Second, we relax this assumption by considering a \emph{complex contagion} dynamics with a nonmonotonic infection function $\alpha(\ell)$ where the aforementioned transmission events are no longer independent~\cite{lehmann2018complex}.  This contagion dynamics has an interesting interpretation in the context of the propagation of a social behavior, where the local popularity of a behavior (large $\ell$) hinders its adoption.
%
The independent transmission assumption can also be lifted when multiple diseases are interacting~\cite{Sanz2014}. Thus, we also consider an asymmetric \emph{interacting contagion} dynamics with two diseases. In this case, $\mathcal{S} = \set{S_1S_2, I_1S_2, S_1I_2, I_1I_2} = \set{0,1,2,3}$ where $U_1V_2$ corresponds to a state where a node is in state $U$ with respect to the first disease and in state $V$ with respect to the second disease. The interaction between the diseases happens via a coupling that is active only when a node is infected by at least one disease, otherwise it behaves identically to the simple contagion dynamics. This coupling may increase or decrease the virulence of the other disease.

Whereas the previously presented dynamics capture various features of contagion phenomena, real datasets containing this level of detail about the interactions among individuals are rare~\cite{Isella2011,Mastrandrea2015,Genois2018}. A class of dynamics for which dataset are easier to find is that of mass-action \emph{metapopulation} dynamics~\cite{Colizza2007,Balcan2010,Ajelli2018,soriano-panos2018spreading}, where the status of the individuals are gathered by geographical regions. These dynamics typically evolve on the weighted networks of the individuals' mobility between regions and the state of a region consists in the number of people that are in each individual health state.  As a fourth case study, we consider a type of deterministic metapopulation dynamics where the population size is constant and where people can either be susceptible ($S$), infected ($I$) or recovered from the disease ($R$). As a result, we define the state of the node as three-dimensional vectors specifying the fraction of people in each state---i.e. $\mathcal{S}=\mathcal{R}=[0,1]^3$.

\begin{figure*}
    \centering
    \includegraphics[width=1.\textwidth]{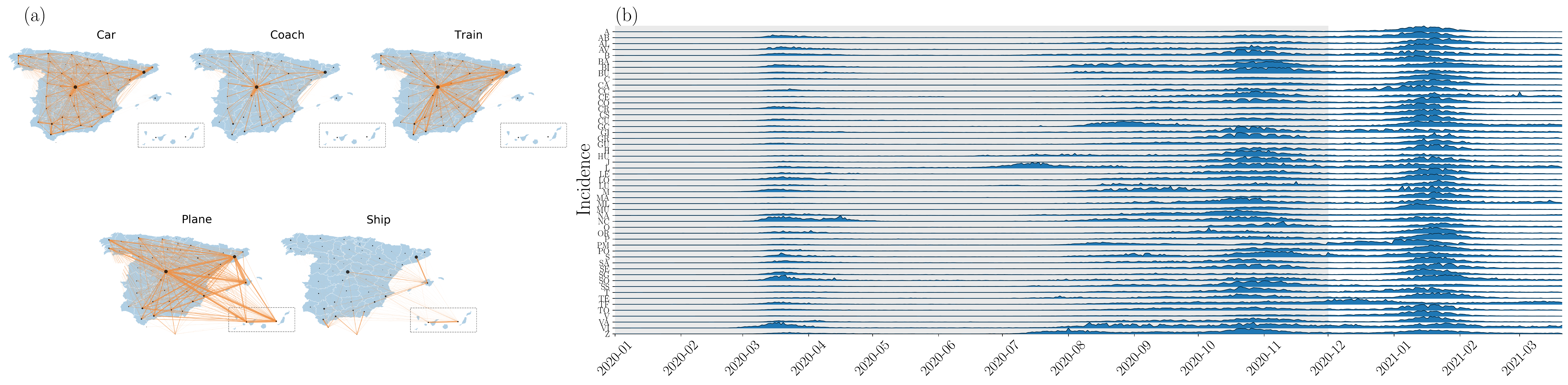}
    \caption{\textbf{Spain COVID-19 dataset.}
    (a) Spain mobility multiplex network~\cite{SpainMobility}.  The thickness of the edges is proportional to the average number of people transitioning between all connected node pairs.  The size of the nodes is proportional to the population $N_i$ living in the province.
    (b) Time series of the incidence for the 52 provinces of Spain between January 2020 and March 2021~\cite{SpainCOVID}.  Each province is identified by its corresponding ISO code.  Each incidence time series has been rescaled by its maximum value for the purpose of visualization.  The shaded area indicates the training and validation datasets (in-sample) from January 1\textsuperscript{st} 2020 to December 1\textsuperscript{st} 2021.  The remaining of the dataset is used for testing.
    }
    \label{fig:covid-data}
\end{figure*}

Figure~\ref{fig:ltp} shows the GNN predictions for the infection and recovery probabilities of the simple and complex contagion dynamics as a function of the number of infected neighbors $\ell$. We then compare them with their ground truths, i.e. Eq.~\eqref{eq:outcome_sis} using Eqs.~\eqref{eq:infection_sis}--\eqref{eq:infection_planck} for the infection functions. We also show the maximum likelihood estimators (MLE) of the transition probabilities computed from the fraction of nodes in state $x$ and with $\ell$ infected neighbors that transitioned to state $y$ in the complete dataset $D$. The MLE, which are typically used in this kind of inference problem~\cite{Eichelsbacher2002}, stands as a reference to benchmark the performance of our approach.

We find that the GNN learns remarkably well the transition probabilities of the simple and complex contagion dynamics.  In fact, the predictions of the GNN seem to be systematically smoother than the ones provided by the MLE.  This is because the MLE is computed for each individual pair $(x,\ell)$ from disjoint subsets of the training dataset.  This implies that a large number of samples of each pair $(x, \ell)$ is needed for the MLE to be accurate; a condition rarely met in realistic settings, especially for high degree nodes.  This also means that the MLE cannot be used directly to interpolate beyond the pairs $(x, \ell)$ present in the training dataset, in sharp contrast with the GNN which, by definition, can interpolate within the dataset $D$. Furthermore, all of its parameters are hierarchically involved during training, meaning that the GNN benefits from any sample to improve all of its predictions, which are then smoother and more consistent.
Further still, we found that not all GNN architectures can reproduce the level of accuracy obtained in the Fig.~\ref{fig:ltp} (see Sec.~III F of the Supplementary Information). In fact, we showed that many standard GNN aggregation mechanisms are ineffective at learning the simple and complex contagion dynamics, most likely because they were specifically designed with structure learning in mind rather than dynamics learning.

It is worth mentioning that the GNN is not specifically designed nor trained to compute the transition probabilities as a function of a single variable, namely the number of infected $\ell$. In reality, the GNN computes its outcome from the complete multivariate state of the neighbors of a node. The interacting contagion and the metapopulation dynamics, unlike the simple and complex contagions, are examples of such multivariate cases. Their outcome is thus harder to visualize in a representation similar to Fig.~\ref{fig:ltp}.
%
Figures~\ref{fig:generalization}(a--h) address this issue by comparing each of the GNN predictions $\hat{y}_i(t)$ with its corresponding target $y_i(t)$ in the dataset $D$. We quantify the global performance of the models in different scenarios, for the different dynamics and underlying network structures, using the Pearson correlation coefficient $r$ between the predictions and targets (see Sec.~\ref{sub:gnn}). We also compute the error, defined from the Pearson coefficient as $1 - r$ for each degree class $k$ (i.e. between the predictions and targets of only the nodes of degree $k$). This allows us to quantify the GNN performance for every local structure.

Figures~\ref{fig:generalization}(i--k) confirm that the GNN provides more accurate predictions than the MLE in general and across all degrees. This is especially true in the case of the interacting contagion, where the accuracy of the MLE seems to deteriorate rapidly for large degree nodes.  This is a consequence of how scarce the inputs are for this dynamics compared to both the simple and complex contagion dynamics for training datasets of the same size, and of how fast the size of the set of possible inputs scales, thereby quickly rendering MLE completely ineffective for small training datasets.  The GNN, on the other hand, is less affected by the scarcity of the data, since any sample improves its global performance, as discussed above.

Figure~\ref{fig:generalization} also exposes the crucial role of the network $G$ on which the dynamics evolves in the global performance of the GNN.  Namely, the heterogeneous degree distributions of Barab\'asi-Albert networks (BA)---or any heterogeneous degree distribution---offer a wider range of degrees than those of homogeneous Erd\H{o}s-R\'enyi networks (ER).  We can take advantage of this heterogeneity to train GNN models that generalize well across a larger range of local structures, as seen in Fig.~\ref{fig:generalization}(i--l) (see also Supplementary Information).  However, the predictions on BA networks are not systematically always better for low degrees than those on ER networks, as seen in the interacting and metapopulation cases.  This nonetheless suggests a wide applicability of our approach for real complex systems, whose underlying network structures recurrently exhibit a heterogeneous degree distribution~\cite{Voitalov2019}.

We now test the trained GNN on unseen network structures by recovering the bifurcation diagrams of the four dynamics.  In the infinite-size limit $|\mathcal{V}| \to \infty$, these dynamics have two possible long-term outcomes: the absorbing state where the diseases quickly die out, and the endemic/epidemic state in which a macroscopic fraction of nodes remains (endemic) or has been infected over time (epidemic)~\cite{Kiss2017,Pastor-Satorras2015,Sanz2014}.  These possible long-term outcomes exchange stability during a phase transition which is continuous for the simple contagion and metapopulation dynamics, and discontinuous for the complex and interacting contagion dynamics.  The position of the phase transition depends on the parameters of the dynamics as well as on the topology of the network.  Note that for the interacting contagion dynamics, the stability of absorbing and endemic states do not change at the same point, giving rise to a bistable regime where both states are stable.

Figure~\ref{fig:phasetransition} shows the different bifurcation diagrams obtained by performing numerical simulations with the trained GNN models [using Eq.~\eqref{eq:GNN_outcome}] while varying the average degree of networks, on which the GNN has not been trained.  Quantitatively, the predictions are again strikingly accurate---essentially perfect for the simple and complex contagion dynamics---which is remarkable given that the bifurcation diagrams were obtained on networks the GNN had never seen before. These results illustrate how insights can be gained about the underlying process concerning the existence of phase transitions and their order, among other things.  They also suggest how the GNN can be used for diverse applications, such as predicting the dynamics under various network structures (e.g. designing intervention strategies that affect the way individuals interact and are therefore connected).

\begin{figure*}
    \centering
    \includegraphics[width=1.\textwidth]{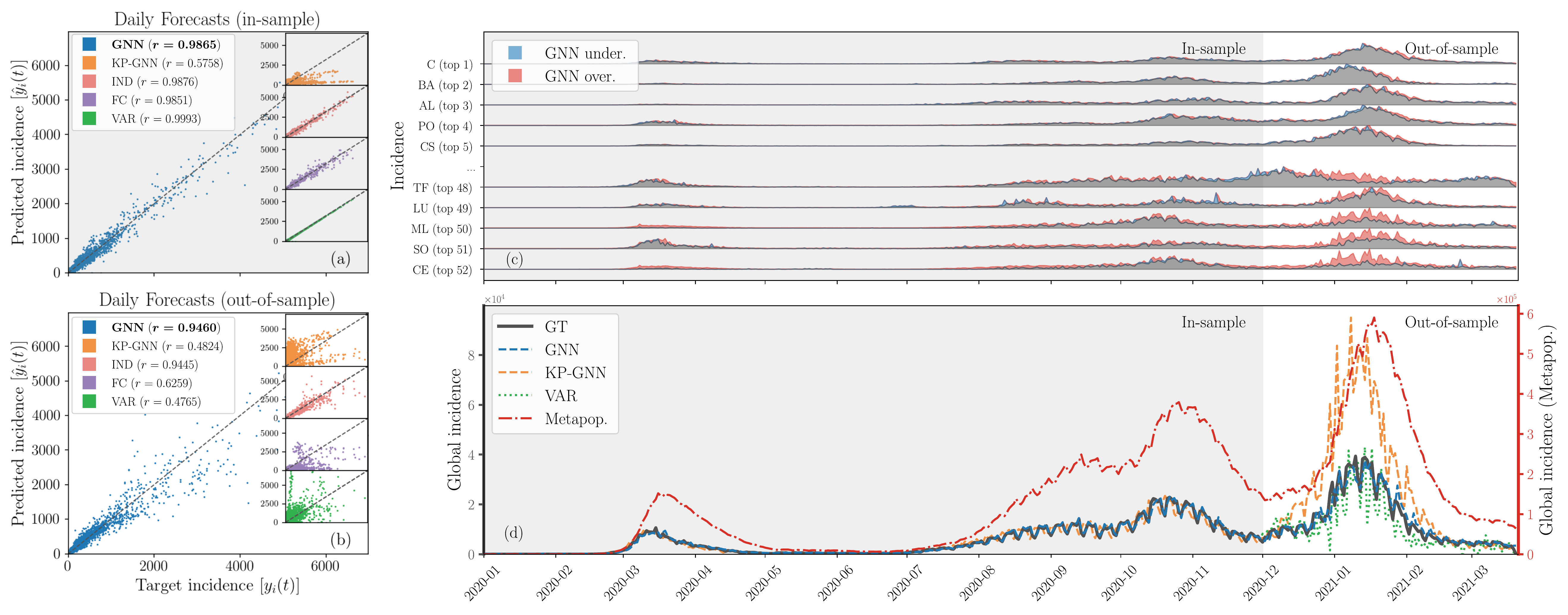}
    \caption{\textbf{Learning the Spain COVID-19 dataset.}
    (a-b) Comparison between the targets and the predictions in the in-sample and the out-of-sample datasets for our GNN model (blue) and for other models (KP-CNN in orange, IND in pink, FC in purple and VAR in green; see main text). The accuracy of the predictions is quantified by the Pearson correlation coefficient provided in the legend.
    (c) Forecasts by our GNN model for individual time series of the provincial daily incidence compared with the ground truth. Underestimation and overestimation are respectively shown in blue and red.  Each time series has been rescaled as in Fig.~\ref{fig:covid-data}(b) and are ordered according to mean square error of the GNN's predictions.
    (d) Forecasts for the global incidence (sum of the daily incidence in every province).  The solid grey line indicates the ground truth (GT); the dashed blue line, the dashed orange line and dotted green line show the forecast of our GNN model, of KP-CNN and of VAR, respectively.  We also show the forecast of an equivalent metapopulation model (red dash-dotted line) which has its own scale (red axis on the right) to improve the visualization; the other lines share the same axis on the left.  Similarly to Fig.~\ref{fig:covid-data}, we differentiate the in-sample from the out-of-sample forecasts using a shaded background.
    }
    \label{fig:covid-results}
\end{figure*}

Finally, we illustrate the applicability of our approach by training our GNN model using the evolution of COVID-19 in Spain between January~1\textsuperscript{st} 2020 and March~27\textsuperscript{th} 2021 (see Fig.~\ref{fig:covid-data}).  This dataset consists in the daily number of new cases (i.e. incidence) for each of the 50 provinces of Spain as well as Ceuta and Melilla~\cite{SpainCOVID}.  We also use a network of the mobility flow recorded in 2018~\cite{SpainMobility} as a proxy to model the interaction network between these 52 regions.  This network is multiplex---each layer corresponds to a different mode of transportation---, directed and weighted (average daily mobility flow).

We compare the performance of our approach with that of different baselines: Four data-driven techniques---three competing neural network architectures~\cite{Tian2020,Kapoor2020} and a linear vector autoregressive model (VAR)~\cite{Wei2018,Rustam2020}---, and an equivalent mechanistic metapopulation model (Metapop.) driven by a simple contagion mechanism~\cite{colizza2007modeling}.  Among the three neural network architectures, we used the model of Ref.~\cite{Kapoor2020} (KP-GNN) that has been used to predict the evolution of COVID-19 in the US.  In a way of an ablation study, the other two GNN architectures embody the assumptions that the nodes of the networks are mutually independent (IND), or that the nodes are arbitrarily codependent (FC) in a way that is learned by the neural network. Note that there exists a wide variety of GNN architectures designed to handle dynamics \emph{of} networks~\cite{skarding2020foundations}---networks whose topology evolves over time---but that these architectures are not typically adapted for learning dynamics \emph{on} networks (see Sec.~III F of the Supplementary Information). Finally, we used the parameters of Ref.~\cite{Aleta2020} for the metapopulation model.  Section~\ref{sub:covid} provides more details on the baselines.

Figure~\ref{fig:covid-results} shows that all data-driven models can generate highly accurate in-sample predictions, with the exception of the KP-GNN model which appears to have a hard time leaning the dynamics, possibly because of its aggregation mechanism (see Sec.~III F of the Supplementary Information). This further substantiates the idea that many GNN architectures designed for structure learning, like the graph convolutional network~\cite{Kipf2016} at the core the KP-GNN model, are suboptimal for dynamics learning problems. However, the other architectures do not appear to have the same capability to generalize the dynamics out-of-sample: The FC and the VAR models, especially, tend to overfit more than the GNN and the IND models.  While this was expected for the linear VAR model, the FC model overfits because it is granted too much freedom in the way it learns how the nodes interact with one another.  Interestingly, the IND and the GNN models seem to perform similarly, which hints to the possibility that the specifics of the mobility network might not have contributed significantly to the dynamics.  This is perhaps not surprising since social distancing and confinement measures were in place during the period covered by the dataset.  Indeed, our results indicate that the global effective dynamics was mostly driven by internal processes within each individual province, rather than driven by the interaction between them.  This last observation suggests that our GNN model is robust to spurious connections in the interaction network.

Finally, Fig.~\ref{fig:covid-results}(d) shows that the metapopulation model is systematically overestimating the incidence by over an order of magnitude.  Again, this is likely due to the confinement measures in place during that period which were not reflected in the original parameters of the dynamics~\cite{Aleta2020}.  Additional mechanisms accounting for this interplay between social restrictions and the prevalence of the disease---e.g. complex contagion mechanisms~\cite{hebert-dufresne2020macroscopic} or time-dependent parameters~\cite{wang2020survivalconvolution}---would therefore be in order to extend the validity of the metapopulation model to the full length of the dataset.  Interestingly, a signature of this interplay is encoded in the daily incidence data and our GNN model appears to be able to capture it to some extent.

\section*{Discussion}
We introduced a data-driven approach that learns effective mechanisms governing the propagation of diverse dynamics on complex networks.  We proposed a reliable training protocol, and we validated the projections of our GNN architecture on simple, complex, interacting contagion and metapopulation dynamics using synthetic networks. Interestingly, we found that many standard GNN architectures do not handle correctly the problem of learning contagion dynamics from time series. Also, we found that our approach performs better when trained on data whose underlying network structure is heterogeneous, which could prove useful in real-world applications of our method given the ubiquitousness of scale-free networks~\cite{voitalov2019scalefree}.

By recovering the bifurcation diagram of various dynamics, we illustrated how our approach can leverage time series from an unknown dynamical process to gain insights about its properties---e.g. the existence of a phase transition and its order.  We have also shown how to use this framework on real datasets, which in turn could then be used to help build better effective models.  In a way, we see this approach as the equivalent of a numerical Petri dish---offering a new way to experiment and gain insights about an unknown dynamics---that is complementary to traditional mechanistic modeling to design better intervention procedures, containment countermeasures and to perform model selection.

Although we focused the presentation of our method on contagion dynamics, its potential applicability reaches many other realms of complex systems modeling where intricate mechanisms are at play.  We believe this work establishes solid foundations for the use of deep learning in the design of realistic effective models of complex systems.

Gathering detailed epidemiological datasets is a complex and labor-intensive process, meaning that datasets suitable for our approach are currently the exception rather than the norm.  The current COVID-19 pandemic has, however, shown how an adequate international reaction to an emerging infectious pathogen critically depends on the free flow of information.  New initiatives like Golbal.health~\cite{GlobalHealth} are good examples on how the international epidemiological community is coming together to share data more openly and to make available comprehensive datasets to all researchers.  Thanks to such initiatives, it is likely that future pandemics will see larger amount of data available to the scientific community in real time.  It is therefore crucial for the community to start developing tools, such as the one presented here, to leverage these datasets so that we are ready for the next pandemic.

\section*{Methods}

\subsection{Graph neural network and training details}
\label{sub:gnn}
In this section, we briefly present our GNN architecture, the training settings, the synthetic data generation procedure and the hyperparmeters used in our experiments.

\begin{figure*}
    \includegraphics[scale=1]{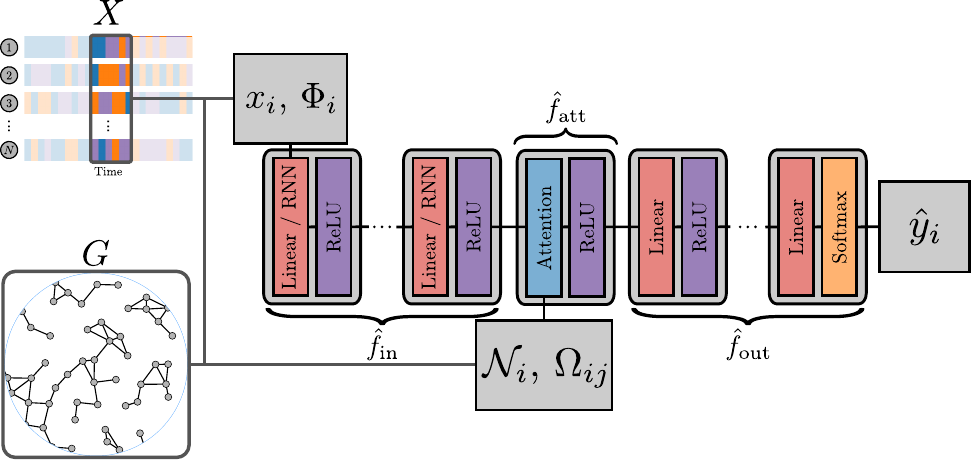}
    \caption{\textbf{Visualization of the GNN architecture}. The blocks of different colors represent mathematical operations. The red blocks correspond to  trainable affine transformation parametrized by weights and biases. The purple blocks represent activation functions between each layer. The core of the model is the attention module~\cite{Velickovic2017}, which is represented in blue. The orange block at the end is an exponential Softmax activation that transforms the output into properly normalized outcomes.}
    \label{fig:architecture}
\end{figure*}

\subsubsection{Architecture}
We use the GNN architecture shown in Fig.~\ref{fig:architecture} and detailed in Tab.~\ref{tab:gnn_layers}. First, we transform the state $x_i$ of every node with a shared multilayer perception (MLP), denoted $\hat{f}_\mathrm{in}:\mathcal{S}\to \mathbb{R}^d$  where $d$ is the resulting number of node features, such that
\begin{equation}
    \xi_i = \hat{f}_\mathrm{in}(x_i)\,.
\end{equation}
We concatenate the node attributes $\Phi_i$ to $x_i$, when these attributes are available, in which case $\hat{f}_\mathrm{in}:\mathcal{S}\times \mathbb{R}^{Q}\to \mathbb{R}^d$. At this point, $\xi_i$ is a vector of features representing the state (and attributes) of node $v_i$. Then, we aggregate the features of the first neighbors using a modified attention mechanism $\hat{f}_\mathrm{att}$, inspired by Ref.~\cite{Velickovic2017} (see Sec.~\ref{sub:attention}),
\begin{equation}
    \nu_i = \hat{f}_\mathrm{att}(\xi_i, \xi_{\mathcal{N}_i})\,,
\end{equation}
where we recall that $\mathcal{N}_i=\set{v_j|e_{ij}\in\mathcal{E}}$ is the set of nodes connected to node $v_i$. We also include the edge attributes $\Omega_{ij}$ into the attention mechanism, when they are available. To do so, we transform the edge attributes $\Omega_{ij}$ into abstract edge features, such that $\psi_{ij} = \hat{f}_\mathrm{edge}(\Omega_{ij})$ where $\hat{f}_\mathrm{edge}:\mathbb{R}^P\to \mathbb{R}^{d_\mathrm{edge}}$ is also a MLP, before they are used in the aggregation. Finally, we compute the outcome $\hat{y}_i$ of each node $v_i$ with another MLP $\hat{f}_\mathrm{out}:\mathbb{R}^d\to \mathcal{R}$ such that
\begin{equation}
    \hat{y}_i = \hat{f}_\mathrm{out}(\nu_i)\,.
\end{equation}

\begin{table*}
    \begin{tabular}{c | c c c c c}
        \hline Dynamics $\quad$&$\quad$ Simple $\quad$&$\quad$ Complex $\quad$&$\quad$ Interacting $\quad$&$\quad$ Metapopulation $\quad$&$\quad$ COVID-19 \\ \hline\hline
        Input layers &
        $\begin{array}{l} \mathrm{Linear}(1, 32) \\ \mathrm{ReLU} \\ \mathrm{Linear}(32, 32) \\ \mathrm{ReLU} \end{array}$ &
        $\begin{array}{l} \mathrm{Linear}(1, 32) \\ \mathrm{ReLU} \\ \mathrm{Linear}(32, 32)\\ \mathrm{ReLU}\end{array}$ &
        $\begin{array}{l} \mathrm{Linear}(1, 32) \\ \mathrm{ReLU} \\ \mathrm{Linear}(32, 32) \\ \mathrm{ReLU} \\ \mathrm{Linear}(32, 32) \\ \mathrm{ReLU} \end{array}$ &
        $\begin{array}{l} \mathrm{Linear}(4, 32)^* \\ \mathrm{ReLU} \\ \mathrm{Linear}(32, 32) \\ \mathrm{ReLU} \\ \mathrm{Linear}(32, 32) \\ \mathrm{ReLU} \\ \mathrm{Linear}(32, 32) \\ \mathrm{ReLU} \end{array}$ &
        $\begin{array}{l} \mathrm{RNN}(2, 4;L)^* \\ \mathrm{ReLU} \\ \mathrm{RNN}(4, 8;L) \\ \mathrm{ReLU} \\ \mathrm{RNN}(8, 16;L)^{**} \\ \mathrm{ReLU} \\ \mathrm{Linear}(16, 32) \\ \mathrm{ReLU} \end{array}$
        \\ \hline
        Number of attention layers & 2 & 2 & 4 & 8\textsuperscript{\dag} & 5\textsuperscript{\dag\dag} \\ \hline
        Output layers &
        $\begin{array}{l} \mathrm{Linear}(32, 32) \\ \mathrm{ReLU} \\ \mathrm{Linear}(32, 2) \\  \mathrm{Softmax} \end{array}$ &
        $\begin{array}{l} \mathrm{Linear}(32, 32) \\ \mathrm{ReLU} \\ \mathrm{Linear}(32, 2) \\  \mathrm{Softmax} \end{array}$ &
        $\begin{array}{l} \mathrm{Linear}(32, 32) \\ \mathrm{ReLU} \\ \mathrm{Linear}(32, 32) \\ \mathrm{ReLU} \\ \mathrm{Linear}(32, 4) \\ \mathrm{Softmax}\end{array}$ &
        $\begin{array}{l} \mathrm{Linear}(32, 32) \\ \mathrm{ReLU} \\ \mathrm{Linear}(32, 32) \\ \mathrm{ReLU}\\ \mathrm{Linear}(32, 32) \\ \mathrm{ReLU} \\ \mathrm{Linear}(32, 3) \\ \mathrm{Softmax}\end{array}$ &
        $\begin{array}{l} \mathrm{Linear}(32, 16)^* \\ \mathrm{ReLU} \\ \mathrm{Linear}(16, 8) \\ \mathrm{ReLU} \\ \mathrm{Linear}(8, 4) \\ \mathrm{ReLU} \\ \mathrm{Linear}(4, 1) \end{array}$

        \\ \hline
        Number of parameters & 6\,698 & 6\,698 & 11\,188 &99\,883 & 7\,190 \\ \hline\hline
    \end{tabular}
    \caption{\textbf{Layer by layer description of the GNN models for each dynamics.} For each sequence, the operations are applied from top to bottom. The operations represented by $\mathrm{Linear}(m, n)$ correspond to linear (or affine) transformations of the form $f(\vec{x}) = \vec{W}\vec{x} + \vec{b}$, where $\vec{x}\in \mathbb{R}^m$ is the input, $\vec{W}\in \mathbb{R}^{n\times m}$ and $\vec{b}\in \mathbb{R}^n$ are  trainable parameters. The operation $\mathrm{RNN}(m,n;L)$ corresponds to an Elman recurrent neural network module~\cite{goodfellow2016deep} with $m$ input features and $n$ output features applied on sequences of length $L$~\cite{goodfellow2016deep}. The operations $\mathrm{ReLU}$ and $\mathrm{Softmax}$ are activation functions given by $\mathrm{ReLU}(x) = \max\set{x, 0}$ and $\mathrm{Softmax}(\vec{x}) = \frac{\exp(\vec{x})}{\sum_i \exp(x_i)}$.
    (*) Here, the dimension of the input is increased by one, because we aggregated to the state of the nodes $x_i$ their rescaled and centered population size $N_i$.
    (**) Here, only the features of the last element of the sequence---those corresponding to the state $X_t$---are kept to proceed further into the architecture.
    (\dag) Because the networks are weighted for the metapopulation dynamics, we initially transform the edge weights into abstract feature representations using a sequence of layers, i.e. $\big(\mathrm{Linear}(1, 4), \mathrm{ReLU}, \mathrm{Linear}(4, 4)\big)$ applied from left to right, before using them in the attention modules. These layers are trained alongside all the other layers.
    (\dag\dag) The network is also weighted in this case, hence we used the same set up as for the metapopulation GNN model to transform the edge weights. Also, note that the five attention modules are each associated to a different layer in the multiplex network.
    }
    \label{tab:gnn_layers}
\end{table*}

\subsubsection{Attention Mechanism} \label{sub:attention}
We use an attention mechanism inspired by the graph attention network architecture (GAT)~\cite{Velickovic2017}. The attention mechanism consists of three trainable functions $\mathcal{A}:\mathbb{R}^{d}\to\mathbb{R}$, $\mathcal{B}:\mathbb{R}^{d}\to\mathbb{R}$ and $\mathcal{C}:\mathbb{R}^{d_{\mathrm{edge}}}\to\mathbb{R}$, that combine the feature vectors $\xi_i$, $\xi_j$ and $\psi_{ij}$ of a connected pair of nodes $v_i$ and $v_j$, where we recall that $d$ and $d_{\mathrm{edge}}$ are the number of node and edge features, respectively. Then, the attention coefficient $a_{ij}$ is computed as follows
\begin{equation}\label{eq:attention_mechanism}
    a_{ij} = \sigma\Big[\mathcal{A}\big(\xi_i\big) + \mathcal{B}\big(\xi_j\big) + \mathcal{C}\big(\psi_{ij}\big)\Big]
\end{equation}
where $\sigma(x) = [1 + e^{-x}]^{-1}$ is the logistic function. Notice that, by using this logistic function, the value of the attention coefficients is constrained to the open interval $(0, 1)$, where $a_{ij} = 0$ implies that the feature $\xi_j$ do not change the value of $\nu_i$, and $a_{ij} = 1$ implies that it maximally changes the value of $\nu_i$. In principle, $a_{ij}$ quantifies the influence of the state of node $v_j$ over the outcome of node $v_i$. In reality, the representation learned by the GNN can be non-sparse, meaning that the neighbor features $\xi_{\mathcal{N}_i}$ can be combined in such a way that their noncontributing parts are canceled out without having $a_{ij}$ being necessarily zero. This can result in the failure of this interpretation of this attention coefficients (see the Supplementary Information for further details). Nevertheless, the attention coefficients can be used to assess how connected nodes interact together.

We compute the aggregated feature vectors $\nu_i$ of node $v_i$ as
\begin{equation}\label{eq:attention_aggregation}
    \nu_i = \hat{f}_\mathrm{att}(\xi_i, \xi_{\mathcal{N}_i}) = \xi_i + \sum_{v_j\in\mathcal{N}_i} a_{ij} \xi_j\,.
\end{equation}
It is important to stress that, at this point, $\nu_i$ contains some information about $v_i$ and all of its neighbors in a pairwise manner. In all our experiments, we fix $\mathcal{A}$, $\mathcal{B}$, and $\mathcal{C}$ to be affine transformations with trainable weight matrix and bias vector. Also, we use multiple attention modules in parallel to increase the expressive power of the GNN architecture, as suggested by Ref.~\cite{Velickovic2017}.

The attention mechanism described by Eq.~\eqref{eq:attention_mechanism} is slightly different from the vanilla version of Ref.~\cite{Velickovic2017}. Similarly to other well-known GNN architectures~\cite{Kipf2016,Hamilton2017,Morris2018}, the aggregation scheme of the vanilla GAT is designed as an average of the feature vectors of the neighbors---where, by definition, $\sum_{v_j\in\mathcal{N}_i}a_{ij} = 1$ for all $v_i$---rather than as a general weighted sum like for Eq.~\eqref{eq:attention_aggregation}. This is often reasonable in the context of structure learning, where the node features represent some coordinates in a metric space where connected nodes are likely to be close~\cite{Hamilton2017}. Yet, in the general case, this type of constraint was shown to lessen dramatically the expressive power of the GNN architecture~\cite{Xu2018}. We also reached the same conclusion while using average-like GNN architectures (see the Supplementary Information). By contrast, the aggregation scheme described by Eq.~\eqref{eq:attention_aggregation} allows our architecture to represent various dynamic processes on networks accurately.

\subsubsection{Training settings}
\label{sub:training}
In all experiments on synthetic data, we use the cross entropy loss as the local loss function,
\begin{equation}
    L\big(y_i, \hat{y}_i\big) = -\sum_{m} y_{i,m} \log \hat{y}_{i,m}\,,
\end{equation}
where $y_{i,m}$ corresponds to the $m$-th element of the outcome vector of node $v_i$, which either is a transition probability for the stochastic contagion dynamics or a fraction of people for the metapopulation dynamics. For the simple, complex and interacting contagion dynamics, we used the observed outcomes $\tilde{y}_i$, corresponding to the stochastic state of node $v_i$ at the next time step, as the target in the loss function. While we noticed a diminished performance when using the observed outcomes as opposed to the true transition probabilities (see Supplementary Information), this setting is more realistic and shows what happens when the targets are noisy. The effect of noise can be tempered by increasing the size of the dataset (see the Supplementary Information). For the metapopulation dynamics, since this model is deterministic, we used the true targets without adding noise.

\subsubsection{Performance measures}
We use the Pearson correlation coefficient $r$ as a global performance measure defined on a set of targets $Y$ and predictions $\hat{Y}$ as
\begin{equation}
r = \frac{\mathbb{E}\big[(Y - \mathbb{E}[Y]) (\hat{Y} - \mathbb{E}[\hat{Y}])\big]}{\sqrt{\mathbb{E}\big[(Y - \mathbb{E}[Y])^2\big]\,\mathbb{E}\big[(\hat{Y} - \mathbb{E}[\hat{Y}])^2\big]}}
\end{equation}
where $\mathbb{E}[W]$ denotes the expectation of $W$. Also, because the maximum correlation occurs at $r = 1$, we also define $1 - r$ as the global error on the set of target-prediction pairs.

\subsubsection{Synthetic data generation}
We generate data from each dynamics using the following algorithm:
\begin{enumerate}
    \item Sample a network $G$ from a given generative model (e.g. the Erd\H{o}s-R\'enyi $G(N,M)$ or the Barab\'asi-Albert network models).
    \item Initialize the state of the system $X(0) = \big(x_i(0)\big)_{i=1..N}$. For the simple, complex and interacting contagion dynamics, sample uniformly the number of nodes in each state. For the metapopulation dynamics, sample the population size for each node from a Poisson distribution of average $10^4$ and then sample the number of infected people within each node from a binomial distribution of parameter $10^{-5}$. For instance, a network of $|\mathcal{V}|=10^3$ nodes will be initialized with a total of 100 infected people, on average, distributed among the nodes.
    \item At time $t$, compute the observed outcome---$Y_t$ for the metapopulation dynamics, and $\tilde{Y}_t$ for the three stochastic dynamics. Then, record the states $X_t$ and $Y_t$ (or $\tilde{Y}_t$).
    \item Repeat step 3 until $(t \mod t_s)= 0$, where $t_s$ is a resampling time. At this moment, apply step 2 to reinitialize the states $X_t$ and repeat step 3.
    \item Stop when $t = T$, where $T$ is the targeted number of samples.
\end{enumerate}
The resampling step parametrized by $t_s$ indirectly controls the diversity of the training dataset.  We allow $t_s$ to be small for the contagion dynamics $(t_s = 2)$ and larger for the metapopulation dynamics $(t_s = 100)$ to emphasize on the performance of the GNN rather than the quality of the training dataset, while acknowledging that different values of $t_s$ could lead to poor training (see Supplementary Information).

We trained the simple, complex and interacting contagion GNN models on networks of size $|\mathcal{V}|=10^3$ nodes and on time series of length $T=10^4$. To generate the networks, we either used Erd\H{o}s-R\'enyi (ER) random networks $G(N,M)$ or Barab\'asi-Albert (BA) random networks. In both cases, the parameters of the generative network models are chosen such that the average degree is fixed to $\avg{k}=4$.

To train our models on the metapopulation dynamics, we generated $10$ networks of $|\mathcal{V}|=100$ nodes and generated for each of them time series of $t_s=100$ time steps. This number of time steps roughly corresponds to the moment where the epidemic dies out. Simiarly to the previous experiments, we used the ER and the BA models to generate the networks, where the parameters were chosen such that $\avg{k}=4$. However, because this dynamics is not stochastic, we varied the average degree of the networks to increase the variability in the time series. This was done by randomly removing a fraction $p = 1 - \ln\big(1 - \mu + e\mu\big)$ of their edges, where $\mu$ was sampled for each network uniformly between 0 and 1. In this scenario, the networks were directed and weighted, with each edge weight $e_{ij}$ being uniformly distributed between 0 and 1.

\subsubsection{Hyperparameters}
The optimization of the parameters was performed using the rectified Adam algorithm~\cite{Liu2019d}, which is hyperparameterized by $b_1=0.9$ and $b_2=0.999$, as suggested in Ref.~\cite{Liu2019d}.

To build a validation dataset, we selected a fraction of the node states randomly for each time step. More specifically, we chose node $v_i$ at time $t$ proportionally to its importance weight $w_i(t)$. For all experiments on synthetic dynamics, we randomly selected 10 nodes to be part of the validation set, on average. For all experiments, the learning rate $\epsilon$ was reduced by a factor 2 every 10 epochs with initial value $\epsilon_0=0.001$. A weight decay of $10^{-4}$ was used as well to help regularize the training. We trained all models for 30 epochs, and selected the GNN model with the lowest loss on validation datasets. We fixed the importance sampling bias exponents for the training to $\lambda=0.5$ in the simple, complex and interacting contagion cases, and fixed it to $\lambda=1$ in the metapopulation case.

\subsection{Importance weights}
\label{sub:weights}
In this section, we show how to implement the importance weights in the different cases. Other versions of the importance weights are also available in the Supplementary Information.

\subsubsection{Discrete state stochastic dynamics}
When $\mathcal{S}$ is a finite countable set, the importance weights can be computed exactly using Eq.~\eqref{eq:ideal_weights},
\begin{equation}
    w_i(t) \propto \Big[\rho\big(k_i, x_i(t), x_{\mathcal{N}_i}(t)\big)\Big]^{-\lambda}\,
\end{equation}
where $\rho\big(k, x, x_{\mathcal{N}}\big)$ is the probability to observe a node of degree $k$ in state $x$ with a neighborhood in state $x_\mathcal{N}$ in the complete dataset $D$. The inputs can be simplified from $(k, x, x_\mathcal{N})$ to $(k, x, \vec{\ell})$ without loss of generality, where $\vec{\ell}$ is a vector whose entries are the number of neighbors in each state.  The distribution is then estimated from the complete dataset $D$ by computing the fraction of inputs that are in every configuration
\begin{align}
    \label{eq:exact_weights}
    \rho(k, x, \vec{\ell}) = &\frac{1}{|\mathcal{V}|T}\sum_{i=1}^{|\mathcal{V}|} I\Big(k_i=k\Big) \nonumber\\
    &\qquad\times \sum_{t=1}^T I\Big(x_i(t)=x\Big)\,\, I\Big(\vec{\ell}_i(t)=\vec{\ell}\Big)\,
\end{align}
where $I(\cdot)$ is the indicator function.

\subsubsection{Continuous state deterministic dynamics}
The case of continuous states---e.g. for metapopulation dynamics---is more challenging than its discrete counterpart, especially if the node and edge attributes, $\Phi_i$ and $\Omega_{ij}$, need to be accounted for. One of the challenges is that we cannot count the inputs like in the previous case. 
As a result, the estimated distribution $\rho$ cannot be estimated directly using Eq.~\eqref{eq:exact_weights}, and we use instead
%
\begin{align}
  w_i(t) = \big[ P(k_i)\ \Sigma(\Phi_i, \Omega_i|k_i)\ \Pi\big(\bar{x}(t)\big) \big]^{-\lambda}
\end{align}
where $P(k_i)$ is the fraction of nodes with degree $k_i$, $\Sigma(\Phi_i, \Omega_i|k_i)$ is the joint probability density function (pdf) conditioned on the degree $k_i$ for the node attributes $\Phi_i$ and the sum of the edge attributes $\Omega_i \equiv \sum_{v_j\in\mathcal{N}_i}\Omega_{ij}$, and where $\Pi\big(\bar{x}(t)\big)$ is the pdf for the average of node states at time $t$ $\bar{x}(t) = \frac{1}{|\mathcal{V}|} \sum_{v_i\in\mathcal{V}} x_i(t)$.  The pdf are obtained using nonparametric Gaussian kernel density estimators (KDE)~\cite{conover1998}. Provided that the density values of the KDE are unbounded above, we normalize the pdf such that the density of each sample used to construct the KDE sum to one. Further details on how we designed the importance weights are provided in the Supplementary Information.

\subsection{Dynamics}
\label{sub:dynamics}
In what follows, we describe in detail the contagion dynamics used for our experiments. We specify the node outcome function $f$ introduced in Eq.~\eqref{eq:locality} and the parameters of the dynamics.

\subsubsection{Simple contagion}
We consider the simple contagion dynamics called the susceptible-infected-susceptible (SIS) dynamics for which $\mathcal{S} = \set{S, I}=\set{0,1}$---we use these two representations of $\mathcal{S}$ interchangeably. Because this dynamics is stochastic, we let $\mathcal{R}=[0,1]^2$. We define the infection function $\alpha(\ell)$ as the probability that a susceptible node becomes infected given its number of infected neighbors $\ell$
\begin{equation}
    \label{eq:infection_sis}
    \mathrm{Pr}(S\to I| \ell) = \alpha(\ell) = 1 - (1 - \gamma)^\ell\ ,
\end{equation}
where $\gamma\in[0, 1]$ is the disease transmission probability. In other words, a node can be infected by any of its infected neighbors independently with probability $\gamma$. We also define the constant recovery probability as
\begin{equation}
    \mathrm{Pr}(I\to S) = \beta\,.
\end{equation}
The node outcome function for the SIS dynamics is therefore
\begin{equation}
    \label{eq:outcome_sis}
    f(x_i, x_{\mathcal{N}_i}) =
    \begin{cases}
        \Big(1-\alpha(\ell_i),\, \alpha(\ell_i)\Big) & \text{if $x_i=0$,} \\
        (\beta,\, 1 - \beta) & \text{if $x_i=1$,}
    \end{cases}
\end{equation}
where
\begin{equation}
    \ell_i = \sum_{v_j\in\mathcal{N}_i} \delta(x_j,1)\,
\end{equation}
is the number of infected neighbors of $v_i$ and $\delta(x,y)$ is the Kronecker delta. Note that for each case in Eq.~\eqref{eq:outcome_sis}, the outcome is a two-dimensional probability vector, where the first entry is the probability that node $v_i$ becomes/remains susceptible at the following time step, and the second entry is the probability that it becomes/remains infected. We used $(\gamma, \beta)=(0.04, 0.08)$ in all experiments involving this simple contagion dynamics.

\subsubsection{Complex contagion}
To lift the independent transmission assumption of the SIS dynamics, we consider a complex contagion dynamics for which the node outcome function has a similar form as Eq.~\eqref{eq:outcome_sis}, but where the infection function $\alpha(\ell)$ has the nonmonotonic form
\begin{equation}
    \label{eq:infection_planck}
    \alpha(\ell) = \frac{1}{z(\eta)}\frac{\ell^3}{e^{\ell/\eta} - 1}
\end{equation}
where  $z(\eta)$ normalizes the infection function such that $\alpha(\ell^*) = 1$ at its global maximum $\ell^*$ and $\eta>0$ is a parameter controlling the position of $\ell^*$. This function is inspired by the Planck distribution for the black-body radiation, although it was chosen for its general shape rather than for any physical meaning whatsoever. We used $(\eta, \beta)=(8, 0.06)$ in all experiments involving this complex contagion dynamics.

\subsubsection{Interacting contagion}
We define the interacting contagion as two SIS dynamics that are interacting and denote it as the SIS-SIS dynamics. In this case, we have $\mathcal{S} = \set{S_1S_2, I_1S_2, S_1I_2, I_1I_2} = \set{0, 1, 2, 3}$. Simiarly to the SIS dynamics, we have $\mathcal{R} = [0,1]^4$ and we define the infection probability functions
\begin{subequations}
    \begin{align}
        \alpha_g(\ell_g) & = 1 - (1 - \gamma_g)^{\ell_g}        && \text{if } x = 0 \\
        \alpha_g^*(\ell_g) & = 1 - (1 - \zeta\gamma_g)^{\ell_g} && \text{if } x = 1,2 \ ,
    \end{align}
\end{subequations}
where $\zeta\geq0$ is a coupling constant and $\ell_g$ is the number of neighbors infected by disease $g$, and also define the recovery probabilities $\beta_g$ for each disease ($g=1,2$).  The case where $\zeta > 1$ corresponds to the situation in which the diseases are synergistic (i.e. being infected by one increases the probability of getting infected by the other), whereas competition is introduced if $\zeta < 1$ (being already infected by one decreases the probability of getting infected by the other). The case $\zeta=1$ falls back on two independent SIS dynamics that evolve simultaneously on the network. The outcome function is composed of 16 entries that are expressed as follows
\begin{widetext}
    \begin{align}
        f(x_i, x_{\mathcal{N}_i}) &=
        \begin{cases}
            \Big(\big[1 - \alpha_1(\ell_{i,1}) \big]\big[1 - \alpha_2(\ell_{i,2})\big],\,
            \alpha_1(\ell_{i,1})\big[1 - \alpha_2(\ell_{i,2})\big],\,
            \big[1 - \alpha_1(\ell_{i,1}) \big]\,\alpha_2(\ell_{i,2}),\,
            \alpha_1(\ell_{i,1})\alpha_2(\ell_{i,2})\Big)
             & \text{if $x_i = 0$,} \\
            \Big(\beta_1 \big[1 - \alpha_2^*(\ell_{i,2})\big],\,
            \big[1 - \beta_1\big] \big[1 - \alpha_2^*(\ell_{i,2})\big],\,
            \beta_1 \alpha_2^*(\ell_{i,2}),\,
            \big[1 - \beta_1\big] \alpha_2^*(\ell_{i,2})\Big)
            & \text{if $x_i=1$,} \\
            \Big(\big[1 - \alpha_1^*(\ell_{i,1})\big]\,\beta_2,\,
            \alpha_1^*(\ell_{i,1})\beta_2,\,
            \big[1 - \alpha_1^*(\ell_{i,1})\big]\,\big[1 - \beta_2\big],\,
            \alpha_1^*(\ell_{i,1})\big[1 - \beta_2\big]\Big)
            & \text{if $x_i=2$,} \\
            \Big(\beta_1\beta_2,\,
            [1 - \beta_1]\beta_2,\,
            \beta_1[1 - \beta_2],\,
            [1 - \beta_1][1 - \beta_2]\Big)
            & \text{if $x_i=3$.
            }
        \end{cases}
    \end{align}
\end{widetext}
where we define $\ell_{i,g}$ as the number of neighbors of $v_i$ that are infected by disease $g$. We used $(\gamma_1,\gamma_2, \beta_1, \beta_2, \zeta) = (0.01, 0.012, 0.19, 0.22, 50)$ in all experiments involving this interacting contagion dynamics.

\subsubsection{Metapopulation}
\label{sub:metapopulation}
The metapopulation dynamics considered is a deterministic version of the susceptible-infection-recovered (SIR) metapopulation model~\cite{Colizza2007,Balcan2010,Ajelli2018,soriano-panos2018spreading}. We consider that the nodes are populated by a fixed number of people $N_i$, which can be in three states---susceptible (S), infected (I) or recovered (R). We therefore track the number of people in every state at each time. Furthermore, we let the network $G$ be weighted, with the weights describing the mobility flow of people between regions. In this case, $\Omega_{ij}\in\mathbb{R}$ is the average number of people that are traveling from node $v_j$ to node $v_i$. Finally, because we assume that the population size is on average steady, we let $\Phi_i = N_i$ be a node attribute and work with the fraction of people in every epidemiological state. More precisely, we define the state of node $v_j$ by $x_j = (s_j, i_j, r_j)$, where $s_j$, $i_j$ and $r_j$ are the fractions of susceptible, infected and recovered people, respectively. From these definitions, we define the node outcome function of this dynamics as
\begin{equation}
    f(x_j, x_{\mathcal{N}_j}, G) = \left(\begin{array}{c}
    s_j - s_j \tilde{\alpha}_j\\
    i_j -\frac{i_j}{\tau_r} + s_j \tilde{\alpha}_j\\
    r_j + \frac{i_j}{\tau_r}
    \end{array}\right)
\end{equation}
where
\begin{equation}\label{eq:metapop_agg}
    \tilde{\alpha}_j = \alpha(i_j,N_j) + \sum_{v_l\in\mathcal{N}_j}\frac{k_j\Omega_{jl}\alpha(i_l, N_l)}{\sum_{v_n\in\mathcal{N}_j} \Omega_{jn}}\,,
\end{equation}
and $k_j$ is the degree of node $v_j$. The function $\alpha(i, N)$ corresponds to the infection rate, per day, at which an individual is infected by someone visiting from a neighboring region with $iN$ infected people in it, and is equal to
\begin{equation}\label{eq:metapop_infection}
    \alpha(i, N) = 1 - \left(1 - \frac{R_0}{\tau_r N}\right)^{iN} \approx 1 - e^{-\frac{R_0}{\tau_r}i}\,.
\end{equation}
where $R_0$ corresponds to the reproduction number and, $\tau_r$ is the average recovery time in days. In all experiments with this metapopulation dynamics, we used $(R_0, \tau_r) = (8.31, 7.5)$.

\subsection{COVID-19 outbreak in Spain}
\label{sub:covid}

\subsubsection{Dataset}
The dataset is composed of the daily incidence of the 52 Spanish provinces (including Ceuta and Melilla) monitored for 450 days between January 1\textsuperscript{st} 2020 and March 27\textsuperscript{th} 2021~\cite{SpainCOVID}. The dataset is augmented with the origin-destination (OD) network of individual mobility~\cite{SpainMobility}. This mobility network is multiplex, directed and weighted, where the weight of each edge $e_{ij}^\nu$ represents mobility flow from province $v_j$ and to province $v_j$ using transportation $\nu$. The metadata associated to each node is the population of province $v_i$~\cite{SpainPopulation}, noted $\Phi_i = N_i$. The metadata associated to each edge, $\Omega_{ij}^\nu$, corresponds to the average number of people that moved from $v_j$ to $v_i$ using $\nu$ as the main means of transportation.

\subsubsection{Models}
The GNN model used in Fig.~\ref{fig:covid-results} is very similar to the metapopulation GNN model---with node and edge attributes---, with the exception that different attention modules are used to model the different OD edge types (plane, car, coach, train and boat, see Tab.~\ref{tab:gnn_layers}). To combine the features of each layer of the multiplex network, we average pooled the output features of the attention modules. We also generalize our model to take in input a sequence of $L$ states of the system, that is
\begin{equation}
    \hat{Y}_t = \hat{\mathcal{M}}(X_{t:t-L+1}, G;\param)
\end{equation}
where $X_{t:t-L+1} = (X_t, X_{t-1},\cdots, X_{t-L+1})$ and $L$ is a lag. At the local level, it reads
\begin{align}
    \hat{y}_i(t) & = \hat{f}\Big(x_i(t:t-L+1), \nonumber \\
                 & \hspace{0.15\linewidth} \Phi_i, x_{\mathcal{N}_i}(t:t-L+1), \Phi_{\mathcal{N}_i}, \Omega_{i\mathcal{N}_i}; \param\Big)
\end{align}
where $x_{i}(t:t-L+1)$ corresponds to the $L$ previous state of node $i$ from time $t$ to time $t-L+1$. As we now feed sequences of node states to the GNN, we use Elman recurrent neural networks~\cite{goodfellow2016deep} to transform these sequences of states before aggregating them instead of linear layers, as shown in Fig.~\ref{fig:architecture} and Tab.~\ref{tab:gnn_layers}. Additionally, because the outputs of the models are not probability vectors, like for the dynamics of Sec.~\ref{sub:dynamics}, but real numbers, we use the mean square error (MSE) loss to train the model:
\begin{equation}
    L(y_i, \hat{y}_i) = (y_{i} - \hat{y}_{i})^2\,.
\end{equation}

We use five different baseline models to compare with the performance of our GNN: Three additional neural network architectures, a vector autoregressive model (VAR)~\cite{Wei2018} and an equivalent metapopulation model driven by a simple contagion mechanism.  The first neural network architecture, denoted the KP-GNN model, was used in Ref.~\cite{Kapoor2020} to forecast the evolution COVID-19 in the US using a similar strategy as ours with respect to the mobility network.  As described in Ref.~\cite{Kapoor2020}, we used a single-layered MLP with 64 hidden units to transform the input, and then we used two graph convolutional networks (GCN) in series, each with 32 hidden units, to perform the feature aggregation.  Finally, we computed the output of the model using another single-layered MLP with 32 hidden units.  The layers of this model are separated by ReLU activation functions and are sampled with a dropout rate of 0.5.  Because this model is not directly adapted to multiplex networks, we merged all layers together into a single network and summed the weights of the edges.  Then, as prescribed in Ref.~\cite{Kapoor2020}, we thresholded the merged network by keeping at most 32 neighbors with the highest edge weight for each node.  We did not use our importance sampling procedure to train the KP-GNN model---letting $\lambda = 0$---to remain as close as possible to the original model.

The other two neural network architectures are very similar to the GNN model we presented in Tab.~\ref{tab:gnn_layers}: The only different component is their aggregation mechanism.  The IND model, where the nodes are assumed to be mutually independent, does not aggregate the features of the neighbors.  It therefore acts like a univariate model, where the time series of each node are processed like different elements of a minibatch.  In the FC model, the nodes interact via a single-layered MLP connecting all nodes together.  The parameters of this MLP are learnable, which effectively allows the model to express any interaction patterns.  Because the number of parameters of this MLP scales with $d|\mathcal{V}|^2$, where $d$ is the number of node features after the input layers, we introduce another layer of 8 hidden units to compress the input features before aggregating.

The VAR model is a linear generative model adapted for multivariate time series forecasting
\begin{equation}
    \hat{Y}_t = \hat{\mathcal{M}}(X_t, X_{t-1}, \cdots, X_{t-L+1}) = \sum_{l=0}^{L - 1
    } \vec{A}_l X_{t - l} + \vec{b} + \vec{\epsilon}_t
\end{equation}
where $\vec{A}_l\in\mathbb{R}^{|\mathcal{V}|\times |\mathcal{V}|}$ are weight matrices, $\vec{b}\in\mathbb{R}^{|\mathcal{V}|}$ is a trend vector and $\vec{\epsilon}_t$ is an error term with $\mathbb{E}[\vec{\epsilon}_t] = \vec{0}$ and $\mathbb{E}[\vec{\epsilon}_t \vec{\epsilon}_s] = \delta_{t, s}\vec{\Sigma}$, with $\vec{\Sigma}$ being a positive-semidefinite covariance matrix. While autoregressive models are often used to predict stock markets~\cite{Sims1980}, they have also been used recently to forecast diverse COVID-19 outbreaks~\cite{Rustam2020}. This model is fitted to the COVID-19 time series dataset also by minimizing the MSE.

The metapopulation model is essentially identical to the model presented in Sec.~\ref{sub:dynamics}.  However, because we track the incidence, i.e. the number of newly infectious cases $\chi_i(t)$ in each province $i$, instead of the complete state $(S_i, I_i, R_i)$ representing the number of individuals in each state, we allow the model to internally track the complete state based on the ground truth.  At first, the whole population is susceptible, i.e. $S_i(1) = N_i$.  Then, at each time step, we subtract the number of newly infectious cases in each node from $S_i(t)$, and add to $I_i$.  Finally, the model allows a fraction $\frac{1}{\tau_r}$ of its infected people to recover. The evolution equations of this model are as follows
\begin{subequations}
    \begin{align}
        S_i(t+1) &= S_i(t) - \chi_i(t)\,,\\
        I_i(t+1) &= I_i(t) + \chi_i(t) - \frac{1}{\tau_r}I_i(t)\,,\\
        R_i(t+1) &= R_i(t) + \frac{1}{\tau_r}I_i(t)\,.
    \end{align}
\end{subequations}
Finally, we computed the incidence $\hat{\chi}_i(t)$ predicted by the metapopulation model using the current internal state as follows:
\begin{equation}
    \hat{\chi}_i(t) = S_i \tilde{\alpha}_i\,,
\end{equation}
where $\tilde{\alpha}_i$ is given by Eq.~\eqref{eq:metapop_agg}, using the mobility network $G$, Eq.~\eqref{eq:metapop_infection} for $\alpha(i, N)$ and $i_j=\frac{I_j}{N_j}$. Since the mobility weights $\Omega_{ij}^\nu$ represent the average number of people traveling from province $v_j$ to province $v_i$, we assumed all layers to be equivalent and aggregated each layer into a single typeless network where $\Omega_{ij} = \sum_\nu \Omega_{ij}^\nu$. We fixed the parameters of the model to $R_0 = 2.5$ and $\tau_r = 7.5$, as these values were used in other contexts for modeling the propagation of COVID-19 in Spain~\cite{Aleta2020}.

\subsubsection{Training}
We trained the GNN and other neural networks for 200 epochs, while decreasing the learning rate by a factor of 2 every 20 epochs with an initial value of $10^{-3}$. For our GNN, the IND and the FC models, we fixed the importance sampling bias exponent to $\lambda=0.5$ and, like the models trained on synthetic data, we used a weight decay of $10^{-4}$ (see Sec.~\ref{sub:training}). We fixed the lag of these models, including the VAR model, to $L=5$.  The KP-GNN model was trained using a weight decay of $10^{-5}$ following Ref.~\cite{Kapoor2020}, and we chose a lag of $L=7$. For all models, we constructed the validation dataset by randomly selecting a partition of the nodes at each time step proportionally to their importance weights $w_i(t)$: 20\% of the nodes are used for validation in this case. The test dataset was constructed by selecting the last 100 time steps of the time series of all nodes, which rough corresponds to the third wave of the outbreak in Spain.\\


\begin{thebibliography}{72}%
\makeatletter
\providecommand \@ifxundefined [1]{%
 \@ifx{#1\undefined}
}%
\providecommand \@ifnum [1]{%
 \ifnum #1\expandafter \@firstoftwo
 \else \expandafter \@secondoftwo
 \fi
}%
\providecommand \@ifx [1]{%
 \ifx #1\expandafter \@firstoftwo
 \else \expandafter \@secondoftwo
 \fi
}%
\providecommand \natexlab [1]{#1}%
\providecommand \enquote  [1]{``#1''}%
\providecommand \bibnamefont  [1]{#1}%
\providecommand \bibfnamefont [1]{#1}%
\providecommand \citenamefont [1]{#1}%
\providecommand \href@noop [0]{\@secondoftwo}%
\providecommand \href [0]{\begingroup \@sanitize@url \@href}%
\providecommand \@href[1]{\@@startlink{#1}\@@href}%
\providecommand \@@href[1]{\endgroup#1\@@endlink}%
\providecommand \@sanitize@url [0]{\catcode `\\12\catcode `\$12\catcode
  `\&12\catcode `\#12\catcode `\^12\catcode `\_12\catcode `\%12\relax}%
\providecommand \@@startlink[1]{}%
\providecommand \@@endlink[0]{}%
\providecommand \url  [0]{\begingroup\@sanitize@url \@url }%
\providecommand \@url [1]{\endgroup\@href {#1}{\urlprefix }}%
\providecommand \urlprefix  [0]{URL }%
\providecommand \Eprint [0]{\href }%
\providecommand \doibase [0]{http://dx.doi.org/}%
\providecommand \selectlanguage [0]{\@gobble}%
\providecommand \bibinfo  [0]{\@secondoftwo}%
\providecommand \bibfield  [0]{\@secondoftwo}%
\providecommand \translation [1]{[#1]}%
\providecommand \BibitemOpen [0]{}%
\providecommand \bibitemStop [0]{}%
\providecommand \bibitemNoStop [0]{.\EOS\space}%
\providecommand \EOS [0]{\spacefactor3000\relax}%
\providecommand \BibitemShut  [1]{\csname bibitem#1\endcsname}%
\let\auto@bib@innerbib\@empty
\bibitem [{\citenamefont {Kermack}\ and\ \citenamefont
  {McKendrick}(1927)}]{kermack1927contribution}%
  \BibitemOpen
  \bibfield  {author} {\bibinfo {author} {\bibfnamefont {W.~O.}\ \bibnamefont
  {Kermack}}\ and\ \bibinfo {author} {\bibfnamefont {A.~G.}\ \bibnamefont
  {McKendrick}},\ }\bibfield  {title} {\enquote {\bibinfo {title} {A
  {{Contribution}} to the {{Mathematical Theory}} of {{Epidemics}}},}\ }\href
  {\doibase 10.1098/rspa.1927.0118} {\bibfield  {journal} {\bibinfo  {journal}
  {Proc. R. Soc. A}\ }\textbf {\bibinfo {volume} {115}},\ \bibinfo {pages}
  {700--721} (\bibinfo {year} {1927})}\BibitemShut {NoStop}%
\bibitem [{\citenamefont {Hethcote}(2000)}]{hethcote2000mathematics}%
  \BibitemOpen
  \bibfield  {author} {\bibinfo {author} {\bibfnamefont {H.~W.}\ \bibnamefont
  {Hethcote}},\ }\bibfield  {title} {\enquote {\bibinfo {title} {The
  {{Mathematics}} of {{Infectious Diseases}}},}\ }\href {\doibase
  10.1137/S0036144500371907} {\bibfield  {journal} {\bibinfo  {journal} {SIAM
  Rev.}\ }\textbf {\bibinfo {volume} {42}},\ \bibinfo {pages} {599--653}
  (\bibinfo {year} {2000})}\BibitemShut {NoStop}%
\bibitem [{\citenamefont {Siettos}\ and\ \citenamefont
  {Russo}(2013)}]{siettos2013mathematical}%
  \BibitemOpen
  \bibfield  {author} {\bibinfo {author} {\bibfnamefont {C.~I.}\ \bibnamefont
  {Siettos}}\ and\ \bibinfo {author} {\bibfnamefont {L.}~\bibnamefont
  {Russo}},\ }\bibfield  {title} {\enquote {\bibinfo {title} {Mathematical
  modeling of infectious disease dynamics},}\ }\href {\doibase
  10.4161/viru.24041} {\bibfield  {journal} {\bibinfo  {journal} {Virulence}\
  }\textbf {\bibinfo {volume} {4}},\ \bibinfo {pages} {295--306} (\bibinfo
  {year} {2013})}\BibitemShut {NoStop}%
\bibitem [{\citenamefont {Kiss}\ \emph {et~al.}(2017)\citenamefont {Kiss},
  \citenamefont {Miller},\ and\ \citenamefont {Simon}}]{Kiss2017}%
  \BibitemOpen
  \bibfield  {author} {\bibinfo {author} {\bibfnamefont {I.~Z.}\ \bibnamefont
  {Kiss}}, \bibinfo {author} {\bibfnamefont {J.~C.}\ \bibnamefont {Miller}}, \
  and\ \bibinfo {author} {\bibfnamefont {P.~L.}\ \bibnamefont {Simon}},\ }\href
  {\doibase 10.1007/978-3-319-50806-1} {\emph {\bibinfo {title} {{Mathematics
  of Epidemics on Networks}}}}\ (\bibinfo  {publisher} {Springer},\ \bibinfo
  {year} {2017})\ p.\ \bibinfo {pages} {598}\BibitemShut {NoStop}%
\bibitem [{\citenamefont {Brauer}\ \emph {et~al.}(2019)\citenamefont {Brauer},
  \citenamefont {{Castillo-Chavez}},\ and\ \citenamefont
  {Feng}}]{brauer2019mathematical}%
  \BibitemOpen
  \bibfield  {author} {\bibinfo {author} {\bibfnamefont {F.}~\bibnamefont
  {Brauer}}, \bibinfo {author} {\bibfnamefont {C.}~\bibnamefont
  {{Castillo-Chavez}}}, \ and\ \bibinfo {author} {\bibfnamefont
  {Z.}~\bibnamefont {Feng}},\ }\href {\doibase 10.1007/978-1-4939-9828-9}
  {\emph {\bibinfo {title} {Mathematical {{Models}} in {{Epidemiology}}}}}\
  (\bibinfo  {publisher} {{Springer}},\ \bibinfo {year} {2019})\BibitemShut
  {NoStop}%
\bibitem [{\citenamefont {Grassly}\ and\ \citenamefont
  {Fraser}(2008)}]{grassly2008mathematical}%
  \BibitemOpen
  \bibfield  {author} {\bibinfo {author} {\bibfnamefont {N.~C.}\ \bibnamefont
  {Grassly}}\ and\ \bibinfo {author} {\bibfnamefont {C.}~\bibnamefont
  {Fraser}},\ }\bibfield  {title} {\enquote {\bibinfo {title} {Mathematical
  models of infectious disease transmission},}\ }\href {\doibase
  10.1038/nrmicro1845} {\bibfield  {journal} {\bibinfo  {journal} {Nat. Rev.
  Microbiol.}\ }\textbf {\bibinfo {volume} {6}},\ \bibinfo {pages} {477--487}
  (\bibinfo {year} {2008})}\BibitemShut {NoStop}%
\bibitem [{\citenamefont {{Pastore y Piontti}}\ \emph
  {et~al.}(2019)\citenamefont {{Pastore y Piontti}}, \citenamefont {Perra},
  \citenamefont {Rossi}, \citenamefont {Samay},\ and\ \citenamefont
  {Vespignani}}]{pastoreypiontti2019charting}%
  \BibitemOpen
  \bibfield  {author} {\bibinfo {author} {\bibfnamefont {A.}~\bibnamefont
  {{Pastore y Piontti}}}, \bibinfo {author} {\bibfnamefont {N.}~\bibnamefont
  {Perra}}, \bibinfo {author} {\bibfnamefont {L.}~\bibnamefont {Rossi}},
  \bibinfo {author} {\bibfnamefont {N.}~\bibnamefont {Samay}}, \ and\ \bibinfo
  {author} {\bibfnamefont {A.}~\bibnamefont {Vespignani}},\ }\href {\doibase
  10.1007/978-3-319-93290-3} {\emph {\bibinfo {title} {Charting the {{Next
  Pandemic}}: {{Modeling Infectious Disease Spreading}} in the {{Data Science
  Age}}}}}\ (\bibinfo  {publisher} {{Springer}},\ \bibinfo {year}
  {2019})\BibitemShut {NoStop}%
\bibitem [{\citenamefont {Viboud}\ and\ \citenamefont
  {Vespignani}(2019)}]{viboud2019future}%
  \BibitemOpen
  \bibfield  {author} {\bibinfo {author} {\bibfnamefont {C.}~\bibnamefont
  {Viboud}}\ and\ \bibinfo {author} {\bibfnamefont {A.}~\bibnamefont
  {Vespignani}},\ }\bibfield  {title} {\enquote {\bibinfo {title} {The future
  of influenza forecasts},}\ }\href {\doibase 10.1073/pnas.1822167116}
  {\bibfield  {journal} {\bibinfo  {journal} {Proc. Natl. Acad. Sci. U.S.A.}\
  }\textbf {\bibinfo {volume} {116}},\ \bibinfo {pages} {2802--2804} (\bibinfo
  {year} {2019})}\BibitemShut {NoStop}%
\bibitem [{\citenamefont {Morens}\ \emph {et~al.}(2008)\citenamefont {Morens},
  \citenamefont {Taubenberger},\ and\ \citenamefont
  {Fauci}}]{morens2008predominant}%
  \BibitemOpen
  \bibfield  {author} {\bibinfo {author} {\bibfnamefont {D.~M.}\ \bibnamefont
  {Morens}}, \bibinfo {author} {\bibfnamefont {J.~K.}\ \bibnamefont
  {Taubenberger}}, \ and\ \bibinfo {author} {\bibfnamefont {A.~S.}\
  \bibnamefont {Fauci}},\ }\bibfield  {title} {\enquote {\bibinfo {title}
  {Predominant {{Role}} of {{Bacterial Pneumonia}} as a {{Cause}} of {{Death}}
  in {{Pandemic Influenza}}: {{Implications}} for {{Pandemic Influenza
  Preparedness}}},}\ }\href {\doibase 10.1086/591708} {\bibfield  {journal}
  {\bibinfo  {journal} {J. Infect. Dis.}\ }\textbf {\bibinfo {volume} {198}},\
  \bibinfo {pages} {962--970} (\bibinfo {year} {2008})}\BibitemShut {NoStop}%
\bibitem [{\citenamefont {Sanz}\ \emph {et~al.}(2014)\citenamefont {Sanz},
  \citenamefont {Xia}, \citenamefont {Meloni},\ and\ \citenamefont
  {Moreno}}]{Sanz2014}%
  \BibitemOpen
  \bibfield  {author} {\bibinfo {author} {\bibfnamefont {J.}~\bibnamefont
  {Sanz}}, \bibinfo {author} {\bibfnamefont {C.-Y}\ \bibnamefont {Xia}},
  \bibinfo {author} {\bibfnamefont {S.}~\bibnamefont {Meloni}}, \ and\ \bibinfo
  {author} {\bibfnamefont {Y.}~\bibnamefont {Moreno}},\ }\bibfield  {title}
  {\enquote {\bibinfo {title} {{Dynamics of Interacting Diseases}},}\ }\href
  {\doibase 10.1103/PhysRevX.4.041005} {\bibfield  {journal} {\bibinfo
  {journal} {Phys. Rev. X}\ }\textbf {\bibinfo {volume} {4}},\ \bibinfo {pages}
  {041005} (\bibinfo {year} {2014})}\BibitemShut {NoStop}%
\bibitem [{\citenamefont {Nickbakhsh}\ \emph {et~al.}(2019)\citenamefont
  {Nickbakhsh}, \citenamefont {Mair}, \citenamefont {Matthews}, \citenamefont
  {Reeve}, \citenamefont {Johnson}, \citenamefont {Thorburn}, \citenamefont
  {{von Wissmann}}, \citenamefont {Reynolds}, \citenamefont {McMenamin},
  \citenamefont {Gunson},\ and\ \citenamefont {Murcia}}]{nickbakhsh2019virus}%
  \BibitemOpen
  \bibfield  {author} {\bibinfo {author} {\bibfnamefont {S.}~\bibnamefont
  {Nickbakhsh}}, \bibinfo {author} {\bibfnamefont {C.}~\bibnamefont {Mair}},
  \bibinfo {author} {\bibfnamefont {L.}~\bibnamefont {Matthews}}, \bibinfo
  {author} {\bibfnamefont {R.}~\bibnamefont {Reeve}}, \bibinfo {author}
  {\bibfnamefont {P.~C.~D.}\ \bibnamefont {Johnson}}, \bibinfo {author}
  {\bibfnamefont {F.}~\bibnamefont {Thorburn}}, \bibinfo {author}
  {\bibfnamefont {B.}~\bibnamefont {{von Wissmann}}}, \bibinfo {author}
  {\bibfnamefont {A.}~\bibnamefont {Reynolds}}, \bibinfo {author}
  {\bibfnamefont {J.}~\bibnamefont {McMenamin}}, \bibinfo {author}
  {\bibfnamefont {R.~N.}\ \bibnamefont {Gunson}}, \ and\ \bibinfo {author}
  {\bibfnamefont {P.~R.}\ \bibnamefont {Murcia}},\ }\bibfield  {title}
  {\enquote {\bibinfo {title} {Virus\textendash virus interactions impact the
  population dynamics of influenza and the common cold},}\ }\href {\doibase
  10.1073/pnas.1911083116} {\bibfield  {journal} {\bibinfo  {journal} {Proc.
  Natl. Acad. Sci. U.S.A.}\ }\textbf {\bibinfo {volume} {116}},\ \bibinfo
  {pages} {27142--27150} (\bibinfo {year} {2019})}\BibitemShut {NoStop}%
\bibitem [{\citenamefont {{H{\'e}bert-Dufresne}}\ \emph
  {et~al.}(2020)\citenamefont {{H{\'e}bert-Dufresne}}, \citenamefont
  {Scarpino},\ and\ \citenamefont {Young}}]{hebert-dufresne2020macroscopic}%
  \BibitemOpen
  \bibfield  {author} {\bibinfo {author} {\bibfnamefont {L.}~\bibnamefont
  {{H{\'e}bert-Dufresne}}}, \bibinfo {author} {\bibfnamefont {S.~V.}\
  \bibnamefont {Scarpino}}, \ and\ \bibinfo {author} {\bibfnamefont {J.-G.}\
  \bibnamefont {Young}},\ }\bibfield  {title} {\enquote {\bibinfo {title}
  {Macroscopic patterns of interacting contagions are indistinguishable from
  social reinforcement},}\ }\href {\doibase 10.1038/s41567-020-0791-2}
  {\bibfield  {journal} {\bibinfo  {journal} {Nat. Phys.}\ }\textbf {\bibinfo
  {volume} {16}},\ \bibinfo {pages} {426--431} (\bibinfo {year}
  {2020})}\BibitemShut {NoStop}%
\bibitem [{\citenamefont {Centola}(2010)}]{centola2010spread}%
  \BibitemOpen
  \bibfield  {author} {\bibinfo {author} {\bibfnamefont {D.}~\bibnamefont
  {Centola}},\ }\bibfield  {title} {\enquote {\bibinfo {title} {The {{Spread}}
  of {{Behavior}} in an {{Online Social Network Experiment}}},}\ }\href
  {\doibase 10.1126/science.1185231} {\bibfield  {journal} {\bibinfo  {journal}
  {Science}\ }\textbf {\bibinfo {volume} {329}},\ \bibinfo {pages} {1194--1197}
  (\bibinfo {year} {2010})}\BibitemShut {NoStop}%
\bibitem [{\citenamefont {Lehmann}\ and\ \citenamefont
  {Ahn}(2018)}]{lehmann2018complex}%
  \BibitemOpen
  \bibinfo {editor} {\bibfnamefont {S.}~\bibnamefont {Lehmann}}\ and\ \bibinfo
  {editor} {\bibfnamefont {Y.-Y.}\ \bibnamefont {Ahn}},\ eds.,\ \href {\doibase
  10.1007/978-3-319-77332-2} {\emph {\bibinfo {title} {Complex {{Spreading
  Phenomena}} in {{Social Systems}}}}},\ Computational {{Social Sciences}}\
  (\bibinfo  {publisher} {{Springer}},\ \bibinfo {year} {2018})\BibitemShut
  {NoStop}%
\bibitem [{\citenamefont {Iacopini}\ \emph {et~al.}(2019)\citenamefont
  {Iacopini}, \citenamefont {Petri}, \citenamefont {Barrat},\ and\
  \citenamefont {Latora}}]{iacopini2019simplicial}%
  \BibitemOpen
  \bibfield  {author} {\bibinfo {author} {\bibfnamefont {I.}~\bibnamefont
  {Iacopini}}, \bibinfo {author} {\bibfnamefont {G.}~\bibnamefont {Petri}},
  \bibinfo {author} {\bibfnamefont {A.}~\bibnamefont {Barrat}}, \ and\ \bibinfo
  {author} {\bibfnamefont {V.}~\bibnamefont {Latora}},\ }\bibfield  {title}
  {\enquote {\bibinfo {title} {Simplicial models of social contagion},}\ }\href
  {\doibase 10.1038/s41467-019-10431-6} {\bibfield  {journal} {\bibinfo
  {journal} {Nat. Commun.}\ }\textbf {\bibinfo {volume} {10}},\ \bibinfo
  {pages} {2485} (\bibinfo {year} {2019})}\BibitemShut {NoStop}%
\bibitem [{\citenamefont {Biggerstaff}\ \emph {et~al.}(2018)\citenamefont
  {Biggerstaff}, \citenamefont {Johansson}, \citenamefont {Alper},
  \citenamefont {Brooks}, \citenamefont {Chakraborty}, \citenamefont {Farrow},
  \citenamefont {Hyun}, \citenamefont {Kandula}, \citenamefont {McGowan},
  \citenamefont {Ramakrishnan}, \citenamefont {Rosenfeld}, \citenamefont
  {Shaman}, \citenamefont {Tibshirani}, \citenamefont {Tibshirani},
  \citenamefont {Vespignani}, \citenamefont {Yang}, \citenamefont {Zhang},\
  and\ \citenamefont {Reed}}]{biggerstaff2018results}%
  \BibitemOpen
  \bibfield  {author} {\bibinfo {author} {\bibfnamefont {M.}~\bibnamefont
  {Biggerstaff}}, \bibinfo {author} {\bibfnamefont {M.}~\bibnamefont
  {Johansson}}, \bibinfo {author} {\bibfnamefont {D.}~\bibnamefont {Alper}},
  \bibinfo {author} {\bibfnamefont {L.~C.}\ \bibnamefont {Brooks}}, \bibinfo
  {author} {\bibfnamefont {P.}~\bibnamefont {Chakraborty}}, \bibinfo {author}
  {\bibfnamefont {D.~C.}\ \bibnamefont {Farrow}}, \bibinfo {author}
  {\bibfnamefont {S.}~\bibnamefont {Hyun}}, \bibinfo {author} {\bibfnamefont
  {S.}~\bibnamefont {Kandula}}, \bibinfo {author} {\bibfnamefont
  {C.}~\bibnamefont {McGowan}}, \bibinfo {author} {\bibfnamefont
  {N.}~\bibnamefont {Ramakrishnan}}, \bibinfo {author} {\bibfnamefont
  {R.}~\bibnamefont {Rosenfeld}}, \bibinfo {author} {\bibfnamefont
  {J.}~\bibnamefont {Shaman}}, \bibinfo {author} {\bibfnamefont
  {R.}~\bibnamefont {Tibshirani}}, \bibinfo {author} {\bibfnamefont {R.~J.}\
  \bibnamefont {Tibshirani}}, \bibinfo {author} {\bibfnamefont
  {A.}~\bibnamefont {Vespignani}}, \bibinfo {author} {\bibfnamefont
  {W.}~\bibnamefont {Yang}}, \bibinfo {author} {\bibfnamefont {Q.}~\bibnamefont
  {Zhang}}, \ and\ \bibinfo {author} {\bibfnamefont {C.}~\bibnamefont {Reed}},\
  }\bibfield  {title} {\enquote {\bibinfo {title} {Results from the second year
  of a collaborative effort to forecast influenza seasons in the {{United
  States}}},}\ }\href {\doibase 10.1016/j.epidem.2018.02.003} {\bibfield
  {journal} {\bibinfo  {journal} {Epidemics}\ }\textbf {\bibinfo {volume}
  {24}},\ \bibinfo {pages} {26--33} (\bibinfo {year} {2018})}\BibitemShut
  {NoStop}%
\bibitem [{\citenamefont {Brunton}\ \emph {et~al.}(2016)\citenamefont
  {Brunton}, \citenamefont {Proctor},\ and\ \citenamefont
  {Kutz}}]{Brunton2016}%
  \BibitemOpen
  \bibfield  {author} {\bibinfo {author} {\bibfnamefont {S.~L.}\ \bibnamefont
  {Brunton}}, \bibinfo {author} {\bibfnamefont {J.~L.}\ \bibnamefont
  {Proctor}}, \ and\ \bibinfo {author} {\bibfnamefont {J.~N.}\ \bibnamefont
  {Kutz}},\ }\bibfield  {title} {\enquote {\bibinfo {title} {{Discovering
  governing equations from data by sparse identification of nonlinear dynamical
  systems}},}\ }\href {\doibase 10.1073/pnas.1517384113} {\bibfield  {journal}
  {\bibinfo  {journal} {Proc. Natl. Acad. Sci. USA}\ }\textbf {\bibinfo
  {volume} {113}},\ \bibinfo {pages} {3932--3937} (\bibinfo {year}
  {2016})}\BibitemShut {NoStop}%
\bibitem [{\citenamefont {Kutz}(2017)}]{Kutz2017}%
  \BibitemOpen
  \bibfield  {author} {\bibinfo {author} {\bibfnamefont {J.~N.}\ \bibnamefont
  {Kutz}},\ }\bibfield  {title} {\enquote {\bibinfo {title} {{Deep learning in
  fluid dynamics}},}\ }\href {\doibase 10.1017/jfm.2016.803} {\bibfield
  {journal} {\bibinfo  {journal} {J. Fluid Mech.}\ }\textbf {\bibinfo {volume}
  {814}},\ \bibinfo {pages} {1--4} (\bibinfo {year} {2017})}\BibitemShut
  {NoStop}%
\bibitem [{\citenamefont {Pathak}\ \emph {et~al.}(2017)\citenamefont {Pathak},
  \citenamefont {Lu}, \citenamefont {Hunt}, \citenamefont {Girvan},\ and\
  \citenamefont {Ott}}]{Pathak2017}%
  \BibitemOpen
  \bibfield  {author} {\bibinfo {author} {\bibfnamefont {J.}~\bibnamefont
  {Pathak}}, \bibinfo {author} {\bibfnamefont {Z.}~\bibnamefont {Lu}}, \bibinfo
  {author} {\bibfnamefont {B.~R.}\ \bibnamefont {Hunt}}, \bibinfo {author}
  {\bibfnamefont {M.}~\bibnamefont {Girvan}}, \ and\ \bibinfo {author}
  {\bibfnamefont {E.}~\bibnamefont {Ott}},\ }\bibfield  {title} {\enquote
  {\bibinfo {title} {{Using machine learning to replicate chaotic attractors
  and calculate Lyapunov exponents from data}},}\ }\href {\doibase
  10.1063/1.5010300} {\bibfield  {journal} {\bibinfo  {journal} {Chaos}\
  }\textbf {\bibinfo {volume} {27}},\ \bibinfo {pages} {121102} (\bibinfo
  {year} {2017})}\BibitemShut {NoStop}%
\bibitem [{\citenamefont {Lusch}\ \emph {et~al.}(2018)\citenamefont {Lusch},
  \citenamefont {Kutz},\ and\ \citenamefont {Brunton}}]{Lusch2018}%
  \BibitemOpen
  \bibfield  {author} {\bibinfo {author} {\bibfnamefont {B.}~\bibnamefont
  {Lusch}}, \bibinfo {author} {\bibfnamefont {J.~N.}\ \bibnamefont {Kutz}}, \
  and\ \bibinfo {author} {\bibfnamefont {S.~L.}\ \bibnamefont {Brunton}},\
  }\bibfield  {title} {\enquote {\bibinfo {title} {{Deep learning for universal
  linear embeddings of nonlinear dynamics}},}\ }\href {\doibase
  10.1038/s41467-018-07210-0} {\bibfield  {journal} {\bibinfo  {journal} {Nat.
  Commun.}\ }\textbf {\bibinfo {volume} {9}},\ \bibinfo {pages} {1--10}
  (\bibinfo {year} {2018})}\BibitemShut {NoStop}%
\bibitem [{\citenamefont {Pathak}\ \emph {et~al.}(2018)\citenamefont {Pathak},
  \citenamefont {Hunt}, \citenamefont {Girvan}, \citenamefont {Lu},\ and\
  \citenamefont {Ott}}]{Pathak2018}%
  \BibitemOpen
  \bibfield  {author} {\bibinfo {author} {\bibfnamefont {J.}~\bibnamefont
  {Pathak}}, \bibinfo {author} {\bibfnamefont {B.}~\bibnamefont {Hunt}},
  \bibinfo {author} {\bibfnamefont {M.}~\bibnamefont {Girvan}}, \bibinfo
  {author} {\bibfnamefont {Z.}~\bibnamefont {Lu}}, \ and\ \bibinfo {author}
  {\bibfnamefont {E.}~\bibnamefont {Ott}},\ }\bibfield  {title} {\enquote
  {\bibinfo {title} {{Model-Free Prediction of Large Spatiotemporally Chaotic
  Systems from Data: A Reservoir Computing Approach}},}\ }\href {\doibase
  10.1103/PhysRevLett.120.024102} {\bibfield  {journal} {\bibinfo  {journal}
  {Phys. Rev. Lett.}\ }\textbf {\bibinfo {volume} {120}},\ \bibinfo {pages}
  {024102} (\bibinfo {year} {2018})}\BibitemShut {NoStop}%
\bibitem [{\citenamefont {{de Silva}}\ \emph {et~al.}(2020)\citenamefont {{de
  Silva}}, \citenamefont {Higdon}, \citenamefont {Brunton},\ and\ \citenamefont
  {Kutz}}]{desilva2020discovery}%
  \BibitemOpen
  \bibfield  {author} {\bibinfo {author} {\bibfnamefont {B.~M.}\ \bibnamefont
  {{de Silva}}}, \bibinfo {author} {\bibfnamefont {D.~M.}\ \bibnamefont
  {Higdon}}, \bibinfo {author} {\bibfnamefont {S.~L.}\ \bibnamefont {Brunton}},
  \ and\ \bibinfo {author} {\bibfnamefont {J.~N.}\ \bibnamefont {Kutz}},\
  }\bibfield  {title} {\enquote {\bibinfo {title} {Discovery of {{Physics From
  Data}}: {{Universal Laws}} and {{Discrepancies}}},}\ }\href {\doibase
  10.3389/frai.2020.00025} {\bibfield  {journal} {\bibinfo  {journal} {Front.
  Artif. Intell.}\ }\textbf {\bibinfo {volume} {3}},\ \bibinfo {pages} {25}
  (\bibinfo {year} {2020})}\BibitemShut {NoStop}%
\bibitem [{\citenamefont {Chen}\ \emph {et~al.}(2020)\citenamefont {Chen},
  \citenamefont {Weng}, \citenamefont {Yang}, \citenamefont {Gu}, \citenamefont
  {Zhang},\ and\ \citenamefont {Small}}]{Chen2020a}%
  \BibitemOpen
  \bibfield  {author} {\bibinfo {author} {\bibfnamefont {X.}~\bibnamefont
  {Chen}}, \bibinfo {author} {\bibfnamefont {T.}~\bibnamefont {Weng}}, \bibinfo
  {author} {\bibfnamefont {H.}~\bibnamefont {Yang}}, \bibinfo {author}
  {\bibfnamefont {C.}~\bibnamefont {Gu}}, \bibinfo {author} {\bibfnamefont
  {J.}~\bibnamefont {Zhang}}, \ and\ \bibinfo {author} {\bibfnamefont
  {M.}~\bibnamefont {Small}},\ }\bibfield  {title} {\enquote {\bibinfo {title}
  {{Mapping topological characteristics of dynamical systems into neural
  networks: A reservoir computing approach}},}\ }\href {\doibase
  10.1103/PhysRevE.102.033314} {\bibfield  {journal} {\bibinfo  {journal}
  {Phys. Rev. E}\ }\textbf {\bibinfo {volume} {102}},\ \bibinfo {pages}
  {033314} (\bibinfo {year} {2020})}\BibitemShut {NoStop}%
\bibitem [{\citenamefont {Dutta}\ \emph {et~al.}(2018)\citenamefont {Dutta},
  \citenamefont {Mira},\ and\ \citenamefont {Onnela}}]{Dutta2018}%
  \BibitemOpen
  \bibfield  {author} {\bibinfo {author} {\bibfnamefont {R.}~\bibnamefont
  {Dutta}}, \bibinfo {author} {\bibfnamefont {A.}~\bibnamefont {Mira}}, \ and\
  \bibinfo {author} {\bibfnamefont {J.-P.}\ \bibnamefont {Onnela}},\ }\bibfield
   {title} {\enquote {\bibinfo {title} {{Bayesian inference of spreading
  processes on networks}},}\ }\href {\doibase 10.1098/rspa.2018.0129}
  {\bibfield  {journal} {\bibinfo  {journal} {Proc. R. Soc. A}\ }\textbf
  {\bibinfo {volume} {474}},\ \bibinfo {pages} {20180129} (\bibinfo {year}
  {2018})}\BibitemShut {NoStop}%
\bibitem [{\citenamefont {Shah}\ \emph {et~al.}(2020)\citenamefont {Shah},
  \citenamefont {Dehmamy}, \citenamefont {Perra}, \citenamefont {Chinazzi},
  \citenamefont {Barab{\'{a}}si}, \citenamefont {Vespignani},\ and\
  \citenamefont {Yu}}]{Shah2020}%
  \BibitemOpen
  \bibfield  {author} {\bibinfo {author} {\bibfnamefont {C.}~\bibnamefont
  {Shah}}, \bibinfo {author} {\bibfnamefont {N.}~\bibnamefont {Dehmamy}},
  \bibinfo {author} {\bibfnamefont {N.}~\bibnamefont {Perra}}, \bibinfo
  {author} {\bibfnamefont {M.}~\bibnamefont {Chinazzi}}, \bibinfo {author}
  {\bibfnamefont {A.-L.}\ \bibnamefont {Barab{\'{a}}si}}, \bibinfo {author}
  {\bibfnamefont {A.}~\bibnamefont {Vespignani}}, \ and\ \bibinfo {author}
  {\bibfnamefont {R.}~\bibnamefont {Yu}},\ }\bibfield  {title} {\enquote
  {\bibinfo {title} {{Finding Patient Zero: Learning Contagion Source with
  Graph Neural Networks}},}\ }\href {http://arxiv.org/abs/2006.11913} {\
  (\bibinfo {year} {2020})},\ \Eprint {http://arxiv.org/abs/2006.11913}
  {arXiv:2006.11913} \BibitemShut {NoStop}%
\bibitem [{\citenamefont {Rodrigues}\ \emph {et~al.}(2019)\citenamefont
  {Rodrigues}, \citenamefont {Peron}, \citenamefont {Connaughton},
  \citenamefont {Kurths},\ and\ \citenamefont {Moreno}}]{Rodrigues2019}%
  \BibitemOpen
  \bibfield  {author} {\bibinfo {author} {\bibfnamefont {F.~A.}\ \bibnamefont
  {Rodrigues}}, \bibinfo {author} {\bibfnamefont {T.}~\bibnamefont {Peron}},
  \bibinfo {author} {\bibfnamefont {C.}~\bibnamefont {Connaughton}}, \bibinfo
  {author} {\bibfnamefont {J.}~\bibnamefont {Kurths}}, \ and\ \bibinfo {author}
  {\bibfnamefont {Y.}~\bibnamefont {Moreno}},\ }\bibfield  {title} {\enquote
  {\bibinfo {title} {A machine learning approach to predicting dynamical
  observables from network structure},}\ }\href
  {https://arxiv.org/abs/1910.00544} {\  (\bibinfo {year} {2019})},\ \Eprint
  {http://arxiv.org/abs/1910.00544} {arXiv:1910.00544} \BibitemShut {NoStop}%
\bibitem [{\citenamefont {Salova}\ \emph {et~al.}(2019)\citenamefont {Salova},
  \citenamefont {Emenheiser}, \citenamefont {Rupe}, \citenamefont
  {Crutchfield},\ and\ \citenamefont {D'Souza}}]{Salova2019}%
  \BibitemOpen
  \bibfield  {author} {\bibinfo {author} {\bibfnamefont {A.}~\bibnamefont
  {Salova}}, \bibinfo {author} {\bibfnamefont {J.}~\bibnamefont {Emenheiser}},
  \bibinfo {author} {\bibfnamefont {A.}~\bibnamefont {Rupe}}, \bibinfo {author}
  {\bibfnamefont {J.~P.}\ \bibnamefont {Crutchfield}}, \ and\ \bibinfo {author}
  {\bibfnamefont {R.~M.}\ \bibnamefont {D'Souza}},\ }\bibfield  {title}
  {\enquote {\bibinfo {title} {{Koopman operator and its approximations for
  systems with symmetries}},}\ }\href {\doibase 10.1063/1.5099091} {\bibfield
  {journal} {\bibinfo  {journal} {Chaos}\ }\textbf {\bibinfo {volume} {29}},\
  \bibinfo {pages} {093128} (\bibinfo {year} {2019})}\BibitemShut {NoStop}%
\bibitem [{\citenamefont {Laurence}\ \emph {et~al.}(2020)\citenamefont
  {Laurence}, \citenamefont {Murphy}, \citenamefont {St-Onge}, \citenamefont
  {Roy-Pomerleau},\ and\ \citenamefont {Thibeault}}]{Laurence2020}%
  \BibitemOpen
  \bibfield  {author} {\bibinfo {author} {\bibfnamefont {E.}~\bibnamefont
  {Laurence}}, \bibinfo {author} {\bibfnamefont {C.}~\bibnamefont {Murphy}},
  \bibinfo {author} {\bibfnamefont {G.}~\bibnamefont {St-Onge}}, \bibinfo
  {author} {\bibfnamefont {X.}~\bibnamefont {Roy-Pomerleau}}, \ and\ \bibinfo
  {author} {\bibfnamefont {V.}~\bibnamefont {Thibeault}},\ }\bibfield  {title}
  {\enquote {\bibinfo {title} {{Detecting structural perturbations from time
  series with deep learning}},}\ }\href {http://arxiv.org/abs/2006.05232} {\
  (\bibinfo {year} {2020})},\ \Eprint {http://arxiv.org/abs/2006.05232}
  {arXiv:2006.05232} \BibitemShut {NoStop}%
\bibitem [{\citenamefont {Zhang}\ \emph {et~al.}(2018)\citenamefont {Zhang},
  \citenamefont {Cui},\ and\ \citenamefont {Zhu}}]{Zhang2018i}%
  \BibitemOpen
  \bibfield  {author} {\bibinfo {author} {\bibfnamefont {Z.}~\bibnamefont
  {Zhang}}, \bibinfo {author} {\bibfnamefont {P.}~\bibnamefont {Cui}}, \ and\
  \bibinfo {author} {\bibfnamefont {W.}~\bibnamefont {Zhu}},\ }\bibfield
  {title} {\enquote {\bibinfo {title} {{Deep Learning on Graphs: A Survey}},}\
  }\href {http://arxiv.org/abs/1812.04202} {\  (\bibinfo {year} {2018})},\
  \Eprint {http://arxiv.org/abs/1812.04202} {arXiv:1812.04202} \BibitemShut
  {NoStop}%
\bibitem [{\citenamefont {Zhou}\ \emph {et~al.}(2018)\citenamefont {Zhou},
  \citenamefont {Cui}, \citenamefont {Zhang}, \citenamefont {Yang},
  \citenamefont {Liu}, \citenamefont {Wang}, \citenamefont {Li},\ and\
  \citenamefont {Sun}}]{Zhou2018a}%
  \BibitemOpen
  \bibfield  {author} {\bibinfo {author} {\bibfnamefont {J.}~\bibnamefont
  {Zhou}}, \bibinfo {author} {\bibfnamefont {G{\'{e}}}~\bibnamefont {Cui}},
  \bibinfo {author} {\bibfnamefont {Z.}~\bibnamefont {Zhang}}, \bibinfo
  {author} {\bibfnamefont {C.}~\bibnamefont {Yang}}, \bibinfo {author}
  {\bibfnamefont {Z.}~\bibnamefont {Liu}}, \bibinfo {author} {\bibfnamefont
  {L.}~\bibnamefont {Wang}}, \bibinfo {author} {\bibfnamefont {C.}~\bibnamefont
  {Li}}, \ and\ \bibinfo {author} {\bibfnamefont {M.}~\bibnamefont {Sun}},\
  }\bibfield  {title} {\enquote {\bibinfo {title} {{Graph Neural Networks: A
  Review of Methods and Applications}},}\ }\href
  {http://arxiv.org/abs/1812.08434} {\  (\bibinfo {year} {2018})},\ \Eprint
  {http://arxiv.org/abs/1812.08434} {arXiv:1812.08434} \BibitemShut {NoStop}%
\bibitem [{\citenamefont {Xu}\ \emph {et~al.}(2018)\citenamefont {Xu},
  \citenamefont {Hu}, \citenamefont {Leskovec},\ and\ \citenamefont
  {Jegelka}}]{Xu2018}%
  \BibitemOpen
  \bibfield  {author} {\bibinfo {author} {\bibfnamefont {K.}~\bibnamefont
  {Xu}}, \bibinfo {author} {\bibfnamefont {W.}~\bibnamefont {Hu}}, \bibinfo
  {author} {\bibfnamefont {J.}~\bibnamefont {Leskovec}}, \ and\ \bibinfo
  {author} {\bibfnamefont {S.}~\bibnamefont {Jegelka}},\ }\bibfield  {title}
  {\enquote {\bibinfo {title} {{How Powerful are Graph Neural Networks?}}}\
  }\href {http://arxiv.org/abs/1810.00826} {\  (\bibinfo {year} {2018})},\
  \Eprint {http://arxiv.org/abs/1810.00826} {arXiv:1810.00826} \BibitemShut
  {NoStop}%
\bibitem [{\citenamefont {Perozzi}\ \emph {et~al.}(2014)\citenamefont
  {Perozzi}, \citenamefont {Al-Rfou},\ and\ \citenamefont
  {Skiena}}]{perozzi2014deepwalk}%
  \BibitemOpen
  \bibfield  {author} {\bibinfo {author} {\bibfnamefont {B.}~\bibnamefont
  {Perozzi}}, \bibinfo {author} {\bibfnamefont {R.}~\bibnamefont {Al-Rfou}}, \
  and\ \bibinfo {author} {\bibfnamefont {S.}~\bibnamefont {Skiena}},\
  }\bibfield  {title} {\enquote {\bibinfo {title} {Deepwalk: Online learning of
  social representations},}\ }in\ \href {\doibase 10.1145/2623330.2623732}
  {\emph {\bibinfo {booktitle} {Proceedings of the 20th ACM SIGKDD
  International Conference on Knowledge Discovery and Data Mining}}}\ (\bibinfo
  {address} {New York, NY, USA},\ \bibinfo {year} {2014})\ p.\ \bibinfo {pages}
  {701–710}\BibitemShut {NoStop}%
\bibitem [{\citenamefont {Hamilton}\ \emph {et~al.}(2017)\citenamefont
  {Hamilton}, \citenamefont {Ying},\ and\ \citenamefont
  {Leskovec}}]{Hamilton2017}%
  \BibitemOpen
  \bibfield  {author} {\bibinfo {author} {\bibfnamefont {W.~L.}\ \bibnamefont
  {Hamilton}}, \bibinfo {author} {\bibfnamefont {R.}~\bibnamefont {Ying}}, \
  and\ \bibinfo {author} {\bibfnamefont {J.}~\bibnamefont {Leskovec}},\
  }\bibfield  {title} {\enquote {\bibinfo {title} {{Representation Learning on
  Graphs: Methods and Applications}},}\ }\href
  {http://arxiv.org/abs/1709.05584} {\  (\bibinfo {year} {2017})},\ \Eprint
  {http://arxiv.org/abs/1709.05584} {arXiv:1709.05584} \BibitemShut {NoStop}%
\bibitem [{\citenamefont {Zhang}\ \emph {et~al.}(2019)\citenamefont {Zhang},
  \citenamefont {Zhao}, \citenamefont {Liu}, \citenamefont {Wang},
  \citenamefont {Tao}, \citenamefont {Xin},\ and\ \citenamefont
  {Zhang}}]{zhang2019general}%
  \BibitemOpen
  \bibfield  {author} {\bibinfo {author} {\bibfnamefont {Z.}~\bibnamefont
  {Zhang}}, \bibinfo {author} {\bibfnamefont {Y.}~\bibnamefont {Zhao}},
  \bibinfo {author} {\bibfnamefont {J.}~\bibnamefont {Liu}}, \bibinfo {author}
  {\bibfnamefont {S.}~\bibnamefont {Wang}}, \bibinfo {author} {\bibfnamefont
  {R.}~\bibnamefont {Tao}}, \bibinfo {author} {\bibfnamefont {R.}~\bibnamefont
  {Xin}}, \ and\ \bibinfo {author} {\bibfnamefont {J.}~\bibnamefont {Zhang}},\
  }\bibfield  {title} {\enquote {\bibinfo {title} {A general deep learning
  framework for network reconstruction and dynamics learning},}\ }\href
  {\doibase 10.1007/s41109-019-0194-4} {\bibfield  {journal} {\bibinfo
  {journal} {Appl. Netw. Sci.}\ }\textbf {\bibinfo {volume} {4}},\ \bibinfo
  {pages} {110} (\bibinfo {year} {2019})}\BibitemShut {NoStop}%
\bibitem [{\citenamefont {Fout}\ \emph {et~al.}(2017)\citenamefont {Fout},
  \citenamefont {Byrd}, \citenamefont {Shariat},\ and\ \citenamefont
  {Ben-Hur}}]{Fout2017}%
  \BibitemOpen
  \bibfield  {author} {\bibinfo {author} {\bibfnamefont {A.}~\bibnamefont
  {Fout}}, \bibinfo {author} {\bibfnamefont {J.}~\bibnamefont {Byrd}}, \bibinfo
  {author} {\bibfnamefont {B.}~\bibnamefont {Shariat}}, \ and\ \bibinfo
  {author} {\bibfnamefont {A.}~\bibnamefont {Ben-Hur}},\ }\bibfield  {title}
  {\enquote {\bibinfo {title} {{Protein Interface Prediction using Graph
  Convolutional Networks}},}\ }in\ \href
  {http://papers.nips.cc/paper/7231-protein-interface-prediction-using-graph-convolutional-networks}
  {\emph {\bibinfo {booktitle} {Adv. Neural Inf. Process. Syst. 30}}}\
  (\bibinfo {year} {2017})\ pp.\ \bibinfo {pages} {6530--6539}\BibitemShut
  {NoStop}%
\bibitem [{\citenamefont {Zitnik}\ \emph {et~al.}(2018)\citenamefont {Zitnik},
  \citenamefont {Agrawal},\ and\ \citenamefont {Leskovec}}]{Zitnik2018}%
  \BibitemOpen
  \bibfield  {author} {\bibinfo {author} {\bibfnamefont {M.}~\bibnamefont
  {Zitnik}}, \bibinfo {author} {\bibfnamefont {M.}~\bibnamefont {Agrawal}}, \
  and\ \bibinfo {author} {\bibfnamefont {J.}~\bibnamefont {Leskovec}},\
  }\bibfield  {title} {\enquote {\bibinfo {title} {Modeling polypharmacy side
  effects with graph convolutional networks},}\ }\href {\doibase
  10.1093/bioinformatics/bty294} {\bibfield  {journal} {\bibinfo  {journal}
  {Bioinformatics}\ }\textbf {\bibinfo {volume} {34}},\ \bibinfo {pages}
  {i457--i466} (\bibinfo {year} {2018})}\BibitemShut {NoStop}%
\bibitem [{\citenamefont {Kapoor}\ \emph {et~al.}(2020)\citenamefont {Kapoor},
  \citenamefont {Ben}, \citenamefont {Liu}, \citenamefont {Perozzi},
  \citenamefont {Barnes}, \citenamefont {Blais},\ and\ \citenamefont
  {O'Banion}}]{Kapoor2020}%
  \BibitemOpen
  \bibfield  {author} {\bibinfo {author} {\bibfnamefont {A.}~\bibnamefont
  {Kapoor}}, \bibinfo {author} {\bibfnamefont {X.}~\bibnamefont {Ben}},
  \bibinfo {author} {\bibfnamefont {L.}~\bibnamefont {Liu}}, \bibinfo {author}
  {\bibfnamefont {B.}~\bibnamefont {Perozzi}}, \bibinfo {author} {\bibfnamefont
  {M.}~\bibnamefont {Barnes}}, \bibinfo {author} {\bibfnamefont
  {M.}~\bibnamefont {Blais}}, \ and\ \bibinfo {author} {\bibfnamefont
  {S.}~\bibnamefont {O'Banion}},\ }\bibfield  {title} {\enquote {\bibinfo
  {title} {Examining covid-19 forecasting using spatio-temporal graph neural
  networks},}\ }\href@noop {} {\  (\bibinfo {year} {2020})},\ \Eprint
  {http://arxiv.org/abs/2007.03113} {arXiv:2007.03113} \BibitemShut {NoStop}%
\bibitem [{\citenamefont {Skarding}\ \emph {et~al.}(2020)\citenamefont
  {Skarding}, \citenamefont {Gabrys},\ and\ \citenamefont
  {K.}}]{skarding2020foundations}%
  \BibitemOpen
  \bibfield  {author} {\bibinfo {author} {\bibfnamefont {J.}~\bibnamefont
  {Skarding}}, \bibinfo {author} {\bibfnamefont {B.}~\bibnamefont {Gabrys}}, \
  and\ \bibinfo {author} {\bibfnamefont {Musial}\ \bibnamefont {K.}},\
  }\href@noop {} {\enquote {\bibinfo {title} {Foundations and modelling of
  dynamic networks using dynamic graph neural networks: A survey},}\ }
  (\bibinfo {year} {2020}),\ \Eprint {http://arxiv.org/abs/2005.07496}
  {arXiv:2005.07496} \BibitemShut {NoStop}%
\bibitem [{\citenamefont {Fritz}\ \emph {et~al.}(2021)\citenamefont {Fritz},
  \citenamefont {Dorigatti},\ and\ \citenamefont {R{\"u}gamer}}]{Fritz2021}%
  \BibitemOpen
  \bibfield  {author} {\bibinfo {author} {\bibfnamefont {C.}~\bibnamefont
  {Fritz}}, \bibinfo {author} {\bibfnamefont {E.}~\bibnamefont {Dorigatti}}, \
  and\ \bibinfo {author} {\bibfnamefont {D.}~\bibnamefont {R{\"u}gamer}},\
  }\bibfield  {title} {\enquote {\bibinfo {title} {Combining graph neural
  networks and spatio-temporal disease models to predict covid-19 cases in
  germany},}\ }\href@noop {} {\  (\bibinfo {year} {2021})},\ \Eprint
  {http://arxiv.org/abs/2101.00661} {arXiv:2101.00661} \BibitemShut {NoStop}%
\bibitem [{\citenamefont {Gao}\ \emph {et~al.}(2021)\citenamefont {Gao},
  \citenamefont {Sharma}, \citenamefont {Qian}, \citenamefont {Glass},
  \citenamefont {Spaeder}, \citenamefont {Romberg}, \citenamefont {Sun},\ and\
  \citenamefont {Xiao}}]{Gao2021}%
  \BibitemOpen
  \bibfield  {author} {\bibinfo {author} {\bibfnamefont {J.}~\bibnamefont
  {Gao}}, \bibinfo {author} {\bibfnamefont {R.}~\bibnamefont {Sharma}},
  \bibinfo {author} {\bibfnamefont {C.}~\bibnamefont {Qian}}, \bibinfo {author}
  {\bibfnamefont {L.~M}\ \bibnamefont {Glass}}, \bibinfo {author}
  {\bibfnamefont {J.}~\bibnamefont {Spaeder}}, \bibinfo {author} {\bibfnamefont
  {J.}~\bibnamefont {Romberg}}, \bibinfo {author} {\bibfnamefont
  {J.}~\bibnamefont {Sun}}, \ and\ \bibinfo {author} {\bibfnamefont
  {C.}~\bibnamefont {Xiao}},\ }\bibfield  {title} {\enquote {\bibinfo {title}
  {{STAN: spatio-temporal attention network for pandemic prediction using
  real-world evidence}},}\ }\href {\doibase 10.1093/jamia/ocaa322} {\bibfield
  {journal} {\bibinfo  {journal} {J. Am. Med. Inform. Assoc}\ }\textbf
  {\bibinfo {volume} {28}},\ \bibinfo {pages} {733--743} (\bibinfo {year}
  {2021})}\BibitemShut {NoStop}%
\bibitem [{\citenamefont {Veličković}\ \emph {et~al.}(2018)\citenamefont
  {Veličković}, \citenamefont {Cucurull}, \citenamefont {Casanova},
  \citenamefont {Romero}, \citenamefont {Liò},\ and\ \citenamefont
  {Bengio}}]{Velickovic2017}%
  \BibitemOpen
  \bibfield  {author} {\bibinfo {author} {\bibfnamefont {P.}~\bibnamefont
  {Veličković}}, \bibinfo {author} {\bibfnamefont {G.}~\bibnamefont
  {Cucurull}}, \bibinfo {author} {\bibfnamefont {A.}~\bibnamefont {Casanova}},
  \bibinfo {author} {\bibfnamefont {A.}~\bibnamefont {Romero}}, \bibinfo
  {author} {\bibfnamefont {P.}~\bibnamefont {Liò}}, \ and\ \bibinfo {author}
  {\bibfnamefont {Y.}~\bibnamefont {Bengio}},\ }\bibfield  {title} {\enquote
  {\bibinfo {title} {{Graph Attention Networks}},}\ }\href
  {http://arxiv.org/abs/1710.10903} {\  (\bibinfo {year} {2018})},\ \Eprint
  {http://arxiv.org/abs/1710.10903} {arXiv:1710.10903} \BibitemShut {NoStop}%
\bibitem [{\citenamefont {Barab{\'{a}}si}(2016)}]{Barabasi2016}%
  \BibitemOpen
  \bibfield  {author} {\bibinfo {author} {\bibfnamefont {A.-L.}\ \bibnamefont
  {Barab{\'{a}}si}},\ }\href {http://networksciencebook.com} {\emph {\bibinfo
  {title} {{Network Science}}}}\ (\bibinfo  {publisher} {Cambridge University
  Press},\ \bibinfo {year} {2016})\ p.\ \bibinfo {pages} {474}\BibitemShut
  {NoStop}%
\bibitem [{\citenamefont {Rubinstein}\ and\ \citenamefont
  {Kroese}(2016)}]{Rubinstein2016}%
  \BibitemOpen
  \bibfield  {author} {\bibinfo {author} {\bibfnamefont {R.~Y.}\ \bibnamefont
  {Rubinstein}}\ and\ \bibinfo {author} {\bibfnamefont {D.~P.}\ \bibnamefont
  {Kroese}},\ }\href {\doibase 10.1002/9781118631980} {\emph {\bibinfo {title}
  {Simulation and the Monte Carlo Method}}},\ \bibinfo {edition} {3rd}\ ed.\
  (\bibinfo  {publisher} {Wiley},\ \bibinfo {year} {2016})\ p.\ \bibinfo
  {pages} {414}\BibitemShut {NoStop}%
\bibitem [{\citenamefont {Isella}\ \emph {et~al.}(2011)\citenamefont {Isella},
  \citenamefont {Stehl{\'e}}, \citenamefont {Barrat}, \citenamefont {Cattuto},
  \citenamefont {Pinton},\ and\ \citenamefont {{Van den Broeck}}}]{Isella2011}%
  \BibitemOpen
  \bibfield  {author} {\bibinfo {author} {\bibfnamefont {L.}~\bibnamefont
  {Isella}}, \bibinfo {author} {\bibfnamefont {J.}~\bibnamefont {Stehl{\'e}}},
  \bibinfo {author} {\bibfnamefont {A.}~\bibnamefont {Barrat}}, \bibinfo
  {author} {\bibfnamefont {C.}~\bibnamefont {Cattuto}}, \bibinfo {author}
  {\bibfnamefont {J.-F.}\ \bibnamefont {Pinton}}, \ and\ \bibinfo {author}
  {\bibfnamefont {W.}~\bibnamefont {{Van den Broeck}}},\ }\bibfield  {title}
  {\enquote {\bibinfo {title} {{What's in a crowd? Analysis of face-to-face
  behavioral networks}},}\ }\href {\doibase 10.1016/j.jtbi.2010.11.033}
  {\bibfield  {journal} {\bibinfo  {journal} {J. Theor. Biol.}\ }\textbf
  {\bibinfo {volume} {271}},\ \bibinfo {pages} {166--180} (\bibinfo {year}
  {2011})}\BibitemShut {NoStop}%
\bibitem [{\citenamefont {Mastrandrea}\ \emph {et~al.}(2015)\citenamefont
  {Mastrandrea}, \citenamefont {Fournet},\ and\ \citenamefont
  {Barrat}}]{Mastrandrea2015}%
  \BibitemOpen
  \bibfield  {author} {\bibinfo {author} {\bibfnamefont {R.}~\bibnamefont
  {Mastrandrea}}, \bibinfo {author} {\bibfnamefont {J.}~\bibnamefont
  {Fournet}}, \ and\ \bibinfo {author} {\bibfnamefont {A.}~\bibnamefont
  {Barrat}},\ }\bibfield  {title} {\enquote {\bibinfo {title} {{Contact
  Patterns in a High School: A Comparison between Data Collected Using Wearable
  Sensors, Contact Diaries and Friendship Surveys}},}\ }\href {\doibase
  10.1371/journal.pone.0136497} {\bibfield  {journal} {\bibinfo  {journal}
  {PLOS One}\ }\textbf {\bibinfo {volume} {10}},\ \bibinfo {pages} {e0136497}
  (\bibinfo {year} {2015})}\BibitemShut {NoStop}%
\bibitem [{\citenamefont {G{\'{e}}nois}\ and\ \citenamefont
  {Barrat}(2018)}]{Genois2018}%
  \BibitemOpen
  \bibfield  {author} {\bibinfo {author} {\bibfnamefont {M.}~\bibnamefont
  {G{\'{e}}nois}}\ and\ \bibinfo {author} {\bibfnamefont {A.}~\bibnamefont
  {Barrat}},\ }\bibfield  {title} {\enquote {\bibinfo {title} {{Can co-location
  be used as a proxy for face-to-face contacts?}}}\ }\href {\doibase
  10.1140/epjds/s13688-018-0140-1} {\bibfield  {journal} {\bibinfo  {journal}
  {EPJ Data Sci.}\ }\textbf {\bibinfo {volume} {7}},\ \bibinfo {pages} {11}
  (\bibinfo {year} {2018})}\BibitemShut {NoStop}%
\bibitem [{\citenamefont {Colizza}\ \emph
  {et~al.}(2007{\natexlab{a}})\citenamefont {Colizza}, \citenamefont
  {Pastor-Satorras},\ and\ \citenamefont {Vespignani}}]{Colizza2007}%
  \BibitemOpen
  \bibfield  {author} {\bibinfo {author} {\bibfnamefont {V.}~\bibnamefont
  {Colizza}}, \bibinfo {author} {\bibfnamefont {R.}~\bibnamefont
  {Pastor-Satorras}}, \ and\ \bibinfo {author} {\bibfnamefont {A.}~\bibnamefont
  {Vespignani}},\ }\bibfield  {title} {\enquote {\bibinfo {title}
  {{Reaction–diffusion processes and metapopulation models in heterogeneous
  networks}},}\ }\href {\doibase 10.1038/nphys560} {\bibfield  {journal}
  {\bibinfo  {journal} {Nat. Phys.}\ }\textbf {\bibinfo {volume} {3}},\
  \bibinfo {pages} {276--282} (\bibinfo {year}
  {2007}{\natexlab{a}})}\BibitemShut {NoStop}%
\bibitem [{\citenamefont {Balcan}\ \emph {et~al.}(2010)\citenamefont {Balcan},
  \citenamefont {Gon{\c{c}}alves}, \citenamefont {Hu}, \citenamefont {Ramasco},
  \citenamefont {Colizza},\ and\ \citenamefont {Vespignani}}]{Balcan2010}%
  \BibitemOpen
  \bibfield  {author} {\bibinfo {author} {\bibfnamefont {D.}~\bibnamefont
  {Balcan}}, \bibinfo {author} {\bibfnamefont {B.}~\bibnamefont
  {Gon{\c{c}}alves}}, \bibinfo {author} {\bibfnamefont {H.}~\bibnamefont {Hu}},
  \bibinfo {author} {\bibfnamefont {J.~J.}\ \bibnamefont {Ramasco}}, \bibinfo
  {author} {\bibfnamefont {V.}~\bibnamefont {Colizza}}, \ and\ \bibinfo
  {author} {\bibfnamefont {A.}~\bibnamefont {Vespignani}},\ }\bibfield  {title}
  {\enquote {\bibinfo {title} {{Modeling the spatial spread of infectious
  diseases: The global epidemic and mobility computational model}},}\ }\href
  {\doibase 10.1016/j.jocs.2010.07.002} {\bibfield  {journal} {\bibinfo
  {journal} {J. Comput. Sci.}\ }\textbf {\bibinfo {volume} {1}},\ \bibinfo
  {pages} {132--145} (\bibinfo {year} {2010})}\BibitemShut {NoStop}%
\bibitem [{\citenamefont {Ajelli}\ \emph {et~al.}(2018)\citenamefont {Ajelli},
  \citenamefont {Zhang}, \citenamefont {Sun}, \citenamefont {Merler},
  \citenamefont {Fumanelli}, \citenamefont {Chowell}, \citenamefont {Simonsen},
  \citenamefont {Viboud},\ and\ \citenamefont {Vespignani}}]{Ajelli2018}%
  \BibitemOpen
  \bibfield  {author} {\bibinfo {author} {\bibfnamefont {M.}~\bibnamefont
  {Ajelli}}, \bibinfo {author} {\bibfnamefont {Q.}~\bibnamefont {Zhang}},
  \bibinfo {author} {\bibfnamefont {K.}~\bibnamefont {Sun}}, \bibinfo {author}
  {\bibfnamefont {S.}~\bibnamefont {Merler}}, \bibinfo {author} {\bibfnamefont
  {L.}~\bibnamefont {Fumanelli}}, \bibinfo {author} {\bibfnamefont
  {G.}~\bibnamefont {Chowell}}, \bibinfo {author} {\bibfnamefont
  {L.}~\bibnamefont {Simonsen}}, \bibinfo {author} {\bibfnamefont
  {C.}~\bibnamefont {Viboud}}, \ and\ \bibinfo {author} {\bibfnamefont
  {A.}~\bibnamefont {Vespignani}},\ }\bibfield  {title} {\enquote {\bibinfo
  {title} {{The RAPIDD Ebola forecasting challenge: Model description and
  synthetic data generation}},}\ }\href {\doibase 10.1016/J.EPIDEM.2017.09.001}
  {\bibfield  {journal} {\bibinfo  {journal} {Epidemics}\ }\textbf {\bibinfo
  {volume} {22}},\ \bibinfo {pages} {3--12} (\bibinfo {year}
  {2018})}\BibitemShut {NoStop}%
\bibitem [{\citenamefont {{Soriano-Pa{\~n}os}}\ \emph
  {et~al.}(2018)\citenamefont {{Soriano-Pa{\~n}os}}, \citenamefont {Lotero},
  \citenamefont {Arenas},\ and\ \citenamefont
  {{G{\'o}mez-Garde{\~n}es}}}]{soriano-panos2018spreading}%
  \BibitemOpen
  \bibfield  {author} {\bibinfo {author} {\bibfnamefont {D.}~\bibnamefont
  {{Soriano-Pa{\~n}os}}}, \bibinfo {author} {\bibfnamefont {L.}~\bibnamefont
  {Lotero}}, \bibinfo {author} {\bibfnamefont {A.}~\bibnamefont {Arenas}}, \
  and\ \bibinfo {author} {\bibfnamefont {J.}~\bibnamefont
  {{G{\'o}mez-Garde{\~n}es}}},\ }\bibfield  {title} {\enquote {\bibinfo {title}
  {Spreading {{Processes}} in {{Multiplex Metapopulations Containing Different
  Mobility Networks}}},}\ }\href {\doibase 10.1103/PhysRevX.8.031039}
  {\bibfield  {journal} {\bibinfo  {journal} {Phys. Rev. X}\ }\textbf {\bibinfo
  {volume} {8}},\ \bibinfo {pages} {031039} (\bibinfo {year}
  {2018})}\BibitemShut {NoStop}%
\bibitem [{Spa(2018)}]{SpainMobility}%
  \BibitemOpen
  \href@noop {} {\enquote {\bibinfo {title} {{Observatorio del Transporte y la
  Logística en Espa\~na}},}\ }\bibinfo {howpublished}
  {\url{https://observatoriotransporte.mitma.gob.es/estudio-experimental}}
  (\bibinfo {year} {2018}),\ \bibinfo {note} {[Accessed:
  11-August-2020]}\BibitemShut {NoStop}%
\bibitem [{Spa(2020{\natexlab{a}})}]{SpainCOVID}%
  \BibitemOpen
  \href@noop {} {\enquote {\bibinfo {title} {{COVID-19 en Espa\~na}},}\
  }\bibinfo {howpublished} {\url{https://cnecovid.isciii.es}} (\bibinfo {year}
  {2020}{\natexlab{a}}),\ \bibinfo {note} {[Accessed:
  27-March-2021]}\BibitemShut {NoStop}%
\bibitem [{\citenamefont {Eichelsbacher}\ and\ \citenamefont
  {Ganesh}(2002)}]{Eichelsbacher2002}%
  \BibitemOpen
  \bibfield  {author} {\bibinfo {author} {\bibfnamefont {P.}~\bibnamefont
  {Eichelsbacher}}\ and\ \bibinfo {author} {\bibfnamefont {A.}~\bibnamefont
  {Ganesh}},\ }\bibfield  {title} {\enquote {\bibinfo {title} {{Bayesian
  Inference for Markov Chains}},}\ }\href {\doibase 10.1239/jap/1019737990}
  {\bibfield  {journal} {\bibinfo  {journal} {J. Appl. Probab.}\ }\textbf
  {\bibinfo {volume} {39}},\ \bibinfo {pages} {91--99} (\bibinfo {year}
  {2002})}\BibitemShut {NoStop}%
\bibitem [{\citenamefont {Voitalov}\ \emph
  {et~al.}(2019{\natexlab{a}})\citenamefont {Voitalov}, \citenamefont {van~der
  Hoorn}, \citenamefont {van~der Hofstad},\ and\ \citenamefont
  {Krioukov}}]{Voitalov2019}%
  \BibitemOpen
  \bibfield  {author} {\bibinfo {author} {\bibfnamefont {I.}~\bibnamefont
  {Voitalov}}, \bibinfo {author} {\bibfnamefont {P.}~\bibnamefont {van~der
  Hoorn}}, \bibinfo {author} {\bibfnamefont {R.}~\bibnamefont {van~der
  Hofstad}}, \ and\ \bibinfo {author} {\bibfnamefont {D.}~\bibnamefont
  {Krioukov}},\ }\bibfield  {title} {\enquote {\bibinfo {title} {{Scale-free
  networks well done}},}\ }\href {\doibase 10.1103/PhysRevResearch.1.033034}
  {\bibfield  {journal} {\bibinfo  {journal} {Phys. Rev. Research}\ }\textbf
  {\bibinfo {volume} {1}},\ \bibinfo {pages} {033034} (\bibinfo {year}
  {2019}{\natexlab{a}})}\BibitemShut {NoStop}%
\bibitem [{\citenamefont {Pastor-Satorras}\ \emph {et~al.}(2015)\citenamefont
  {Pastor-Satorras}, \citenamefont {Castellano}, \citenamefont {{Van
  Mieghem}},\ and\ \citenamefont {Vespignani}}]{Pastor-Satorras2015}%
  \BibitemOpen
  \bibfield  {author} {\bibinfo {author} {\bibfnamefont {R.}~\bibnamefont
  {Pastor-Satorras}}, \bibinfo {author} {\bibfnamefont {C.}~\bibnamefont
  {Castellano}}, \bibinfo {author} {\bibfnamefont {P.}~\bibnamefont {{Van
  Mieghem}}}, \ and\ \bibinfo {author} {\bibfnamefont {A.}~\bibnamefont
  {Vespignani}},\ }\bibfield  {title} {\enquote {\bibinfo {title} {{Epidemic
  processes in complex networks}},}\ }\href {\doibase
  10.1103/RevModPhys.87.925} {\bibfield  {journal} {\bibinfo  {journal} {Rev.
  Mod. Phys.}\ }\textbf {\bibinfo {volume} {87}},\ \bibinfo {pages} {925--979}
  (\bibinfo {year} {2015})}\BibitemShut {NoStop}%
\bibitem [{\citenamefont {Tian}\ \emph {et~al.}(2020)\citenamefont {Tian},
  \citenamefont {Luthra},\ and\ \citenamefont {Zhang}}]{Tian2020}%
  \BibitemOpen
  \bibfield  {author} {\bibinfo {author} {\bibfnamefont {Y.}~\bibnamefont
  {Tian}}, \bibinfo {author} {\bibfnamefont {Ish.}\ \bibnamefont {Luthra}}, \
  and\ \bibinfo {author} {\bibfnamefont {.}~\bibnamefont {Zhang}},\ }\bibfield
  {title} {\enquote {\bibinfo {title} {Forecasting covid-19 cases using machine
  learning models},}\ }\href {\doibase 10.1101/2020.07.02.20145474} {\
  (\bibinfo {year} {2020}),\ 10.1101/2020.07.02.20145474}\BibitemShut {NoStop}%
\bibitem [{\citenamefont {Wei}(2018)}]{Wei2018}%
  \BibitemOpen
  \bibfield  {author} {\bibinfo {author} {\bibfnamefont {W.~W.~S.}\
  \bibnamefont {Wei}},\ }\href@noop {} {\emph {\bibinfo {title} {Multivariate
  time series analysis and applications}}}\ (\bibinfo  {publisher} {John Wiley
  \& Sons},\ \bibinfo {year} {2018})\BibitemShut {NoStop}%
\bibitem [{\citenamefont {Rustam}\ \emph {et~al.}(2020)\citenamefont {Rustam},
  \citenamefont {Reshi}, \citenamefont {Mehmood}, \citenamefont {Ullah},
  \citenamefont {On}, \citenamefont {Aslam},\ and\ \citenamefont
  {Choi}}]{Rustam2020}%
  \BibitemOpen
  \bibfield  {author} {\bibinfo {author} {\bibfnamefont {F.}~\bibnamefont
  {Rustam}}, \bibinfo {author} {\bibfnamefont {A.~A.}\ \bibnamefont {Reshi}},
  \bibinfo {author} {\bibfnamefont {A.}~\bibnamefont {Mehmood}}, \bibinfo
  {author} {\bibfnamefont {S.}~\bibnamefont {Ullah}}, \bibinfo {author}
  {\bibfnamefont {B.}~\bibnamefont {On}}, \bibinfo {author} {\bibfnamefont
  {W.}~\bibnamefont {Aslam}}, \ and\ \bibinfo {author} {\bibfnamefont {G.~S.}\
  \bibnamefont {Choi}},\ }\bibfield  {title} {\enquote {\bibinfo {title}
  {Covid-19 future forecasting using supervised machine learning models},}\
  }\href {\doibase 10.1109/ACCESS.2020.2997311} {\bibfield  {journal} {\bibinfo
   {journal} {IEEE Access}\ }\textbf {\bibinfo {volume} {8}},\ \bibinfo {pages}
  {101489--101499} (\bibinfo {year} {2020})}\BibitemShut {NoStop}%
\bibitem [{\citenamefont {Colizza}\ \emph
  {et~al.}(2007{\natexlab{b}})\citenamefont {Colizza}, \citenamefont {Barrat},
  \citenamefont {Barthelemy}, \citenamefont {Valleron},\ and\ \citenamefont
  {Vespignani}}]{colizza2007modeling}%
  \BibitemOpen
  \bibfield  {author} {\bibinfo {author} {\bibfnamefont {V.}~\bibnamefont
  {Colizza}}, \bibinfo {author} {\bibfnamefont {A.}~\bibnamefont {Barrat}},
  \bibinfo {author} {\bibfnamefont {M.}~\bibnamefont {Barthelemy}}, \bibinfo
  {author} {\bibfnamefont {A.-J.}\ \bibnamefont {Valleron}}, \ and\ \bibinfo
  {author} {\bibfnamefont {A.}~\bibnamefont {Vespignani}},\ }\bibfield  {title}
  {\enquote {\bibinfo {title} {Modeling the {{Worldwide Spread}} of {{Pandemic
  Influenza}}: {{Baseline Case}} and {{Containment Interventions}}},}\ }\href
  {\doibase 10.1371/journal.pmed.0040013} {\bibfield  {journal} {\bibinfo
  {journal} {PLOS Med.}\ }\textbf {\bibinfo {volume} {4}},\ \bibinfo {pages}
  {e13} (\bibinfo {year} {2007}{\natexlab{b}})}\BibitemShut {NoStop}%
\bibitem [{\citenamefont {Aleta}\ and\ \citenamefont
  {Moreno}(2020)}]{Aleta2020}%
  \BibitemOpen
  \bibfield  {author} {\bibinfo {author} {\bibfnamefont {A.}~\bibnamefont
  {Aleta}}\ and\ \bibinfo {author} {\bibfnamefont {Y.}~\bibnamefont {Moreno}},\
  }\bibfield  {title} {\enquote {\bibinfo {title} {{Evaluation of the potential
  incidence of COVID-19 and effectiveness of containment measures in Spain: A
  data-driven approach}},}\ }\href {\doibase 10.1186/s12916-020-01619-5}
  {\bibfield  {journal} {\bibinfo  {journal} {BMC Med.}\ }\textbf {\bibinfo
  {volume} {18}},\ \bibinfo {pages} {157} (\bibinfo {year} {2020})}\BibitemShut
  {NoStop}%
\bibitem [{\citenamefont {Kipf}\ and\ \citenamefont
  {Welling}(2016)}]{Kipf2016}%
  \BibitemOpen
  \bibfield  {author} {\bibinfo {author} {\bibfnamefont {T.~N.}\ \bibnamefont
  {Kipf}}\ and\ \bibinfo {author} {\bibfnamefont {M.}~\bibnamefont {Welling}},\
  }\bibfield  {title} {\enquote {\bibinfo {title} {{Semi-Supervised
  Classification with Graph Convolutional Networks}},}\ }\href
  {http://arxiv.org/abs/1609.02907} {\  (\bibinfo {year} {2016})},\ \Eprint
  {http://arxiv.org/abs/1609.02907} {arXiv:1609.02907} \BibitemShut {NoStop}%
\bibitem [{\citenamefont {Wang}\ \emph {et~al.}(2020)\citenamefont {Wang},
  \citenamefont {Xie}, \citenamefont {Wang},\ and\ \citenamefont
  {Zeng}}]{wang2020survivalconvolution}%
  \BibitemOpen
  \bibfield  {author} {\bibinfo {author} {\bibfnamefont {Q.}~\bibnamefont
  {Wang}}, \bibinfo {author} {\bibfnamefont {S.}~\bibnamefont {Xie}}, \bibinfo
  {author} {\bibfnamefont {Y.}~\bibnamefont {Wang}}, \ and\ \bibinfo {author}
  {\bibfnamefont {D.}~\bibnamefont {Zeng}},\ }\bibfield  {title} {\enquote
  {\bibinfo {title} {Survival-{{Convolution Models}} for {{Predicting
  COVID}}-19 {{Cases}} and {{Assessing Effects}} of {{Mitigation
  Strategies}}},}\ }\href {\doibase 10.3389/fpubh.2020.00325} {\bibfield
  {journal} {\bibinfo  {journal} {Front. Public Health}\ }\textbf {\bibinfo
  {volume} {8}},\ \bibinfo {pages} {325} (\bibinfo {year} {2020})}\BibitemShut
  {NoStop}%
\bibitem [{\citenamefont {Voitalov}\ \emph
  {et~al.}(2019{\natexlab{b}})\citenamefont {Voitalov}, \citenamefont {{van der
  Hoorn}}, \citenamefont {{van der Hofstad}},\ and\ \citenamefont
  {Krioukov}}]{voitalov2019scalefree}%
  \BibitemOpen
  \bibfield  {author} {\bibinfo {author} {\bibfnamefont {I.}~\bibnamefont
  {Voitalov}}, \bibinfo {author} {\bibfnamefont {P.}~\bibnamefont {{van der
  Hoorn}}}, \bibinfo {author} {\bibfnamefont {R.}~\bibnamefont {{van der
  Hofstad}}}, \ and\ \bibinfo {author} {\bibfnamefont {D.}~\bibnamefont
  {Krioukov}},\ }\bibfield  {title} {\enquote {\bibinfo {title} {Scale-free
  networks well done},}\ }\href {\doibase 10.1103/PhysRevResearch.1.033034}
  {\bibfield  {journal} {\bibinfo  {journal} {Phys. Rev. Research}\ }\textbf
  {\bibinfo {volume} {1}},\ \bibinfo {pages} {033034} (\bibinfo {year}
  {2019}{\natexlab{b}})}\BibitemShut {NoStop}%
\bibitem [{Glo(2020)}]{GlobalHealth}%
  \BibitemOpen
  \href@noop {} {\enquote {\bibinfo {title} {{Global.health: A Data Science
  Initiative}},}\ }\bibinfo {howpublished} {\url{https://global.health}}
  (\bibinfo {year} {2020}),\ \bibinfo {note} {[Accessed:
  21-May-2021]}\BibitemShut {NoStop}%
\bibitem [{\citenamefont {Goodfellow}\ \emph {et~al.}(2016)\citenamefont
  {Goodfellow}, \citenamefont {Bengio},\ and\ \citenamefont
  {Courtville}}]{goodfellow2016deep}%
  \BibitemOpen
  \bibfield  {author} {\bibinfo {author} {\bibfnamefont {I.}~\bibnamefont
  {Goodfellow}}, \bibinfo {author} {\bibfnamefont {Y.}~\bibnamefont {Bengio}},
  \ and\ \bibinfo {author} {\bibfnamefont {A.}~\bibnamefont {Courtville}},\
  }\href@noop {} {\emph {\bibinfo {title} {Deep {{Learning}}}}}\ (\bibinfo
  {publisher} {{MIT Press}},\ \bibinfo {year} {2016})\BibitemShut {NoStop}%
\bibitem [{\citenamefont {Morris}\ \emph {et~al.}(2018)\citenamefont {Morris},
  \citenamefont {Ritzert}, \citenamefont {Fey}, \citenamefont {Hamilton},
  \citenamefont {Lenssen}, \citenamefont {Rattan},\ and\ \citenamefont
  {Grohe}}]{Morris2018}%
  \BibitemOpen
  \bibfield  {author} {\bibinfo {author} {\bibfnamefont {C.}~\bibnamefont
  {Morris}}, \bibinfo {author} {\bibfnamefont {M.}~\bibnamefont {Ritzert}},
  \bibinfo {author} {\bibfnamefont {M.}~\bibnamefont {Fey}}, \bibinfo {author}
  {\bibfnamefont {W.~L.}\ \bibnamefont {Hamilton}}, \bibinfo {author}
  {\bibfnamefont {J.~E.}\ \bibnamefont {Lenssen}}, \bibinfo {author}
  {\bibfnamefont {G.}~\bibnamefont {Rattan}}, \ and\ \bibinfo {author}
  {\bibfnamefont {M.}~\bibnamefont {Grohe}},\ }\bibfield  {title} {\enquote
  {\bibinfo {title} {{Weisfeiler and Leman Go Neural: Higher-order Graph Neural
  Networks}},}\ }\href {http://arxiv.org/abs/1810.02244} {\  (\bibinfo {year}
  {2018})},\ \Eprint {http://arxiv.org/abs/1810.02244} {arXiv:1810.02244}
  \BibitemShut {NoStop}%
\bibitem [{\citenamefont {Liu}\ \emph {et~al.}(2019)\citenamefont {Liu},
  \citenamefont {Jiang}, \citenamefont {He}, \citenamefont {Chen},
  \citenamefont {Liu}, \citenamefont {Gao},\ and\ \citenamefont
  {Han}}]{Liu2019d}%
  \BibitemOpen
  \bibfield  {author} {\bibinfo {author} {\bibfnamefont {L.}~\bibnamefont
  {Liu}}, \bibinfo {author} {\bibfnamefont {H.}~\bibnamefont {Jiang}}, \bibinfo
  {author} {\bibfnamefont {P.}~\bibnamefont {He}}, \bibinfo {author}
  {\bibfnamefont {W.}~\bibnamefont {Chen}}, \bibinfo {author} {\bibfnamefont
  {X.}~\bibnamefont {Liu}}, \bibinfo {author} {\bibfnamefont {J.}~\bibnamefont
  {Gao}}, \ and\ \bibinfo {author} {\bibfnamefont {J.}~\bibnamefont {Han}},\
  }\bibfield  {title} {\enquote {\bibinfo {title} {{On the Variance of the
  Adaptive Learning Rate and Beyond}},}\ }\href
  {http://arxiv.org/abs/1908.03265} {\  (\bibinfo {year} {2019})},\ \Eprint
  {http://arxiv.org/abs/1908.03265} {arXiv:1908.03265} \BibitemShut {NoStop}%
\bibitem [{\citenamefont {Conover}(1998)}]{conover1998}%
  \BibitemOpen
  \bibfield  {author} {\bibinfo {author} {\bibfnamefont {W.~J.}\ \bibnamefont
  {Conover}},\ }\href@noop {} {\emph {\bibinfo {title} {{Practical
  nonparametric statistics}}}}\ (\bibinfo  {publisher} {John Wiley {\&} Sons},\
  \bibinfo {year} {1998})\ p.\ \bibinfo {pages} {350}\BibitemShut {NoStop}%
\bibitem [{Spa(2020{\natexlab{b}})}]{SpainPopulation}%
  \BibitemOpen
  \href@noop {} {\enquote {\bibinfo {title} {{Instituto Nacional de
  Estadística}},}\ }\bibinfo {howpublished} {\url{https://www.ine.es}}
  (\bibinfo {year} {2020}{\natexlab{b}}),\ \bibinfo {note} {[Accessed:
  11-August-2020]}\BibitemShut {NoStop}%
\bibitem [{\citenamefont {Sims}(1980)}]{Sims1980}%
  \BibitemOpen
  \bibfield  {author} {\bibinfo {author} {\bibfnamefont {C.~A.}\ \bibnamefont
  {Sims}},\ }\bibfield  {title} {\enquote {\bibinfo {title} {Macroeconomics and
  reality},}\ }\href {\doibase 10.2307/1912017} {\bibfield  {journal} {\bibinfo
   {journal} {Econometrica}\ ,\ \bibinfo {pages} {1--48}} (\bibinfo {year}
  {1980})}\BibitemShut {NoStop}%
\bibitem [{\citenamefont {Murphy}\ \emph
  {et~al.}(2021{\natexlab{a}})\citenamefont {Murphy}, \citenamefont
  {Laurence},\ and\ \citenamefont {Allard}}]{data-dynalearn2021}%
  \BibitemOpen
  \bibfield  {author} {\bibinfo {author} {\bibfnamefont {C.}~\bibnamefont
  {Murphy}}, \bibinfo {author} {\bibfnamefont {E.}~\bibnamefont {Laurence}}, \
  and\ \bibinfo {author} {\bibfnamefont {A.}~\bibnamefont {Allard}},\ }\href
  {\doibase 10.5281/zenodo.5015063} {\enquote {\bibinfo {title}
  {{DynamicaLab/data-dynalearn}},}\ } (\bibinfo {year}
  {2021}{\natexlab{a}})\BibitemShut {NoStop}%
\bibitem [{\citenamefont {Murphy}\ \emph
  {et~al.}(2021{\natexlab{b}})\citenamefont {Murphy}, \citenamefont
  {Laurence},\ and\ \citenamefont {Allard}}]{code-dynalearn2021}%
  \BibitemOpen
  \bibfield  {author} {\bibinfo {author} {\bibfnamefont {C.}~\bibnamefont
  {Murphy}}, \bibinfo {author} {\bibfnamefont {E.}~\bibnamefont {Laurence}}, \
  and\ \bibinfo {author} {\bibfnamefont {A.}~\bibnamefont {Allard}},\ }\href
  {\doibase 10.5281/zenodo.4974521} {\enquote {\bibinfo {title}
  {{DynamicaLab/code-dynalearn}},}\ } (\bibinfo {year}
  {2021}{\natexlab{b}})\BibitemShut {NoStop}%
\end{thebibliography}
%

\noindent\textbf{Data availability}: The raw COVID-19, mobility and population datasets are publicly available respectively from~\cite{SpainMobility},~\cite{SpainCOVID}, and~\cite{SpainPopulation}.  The processed datasets shown in Fig.~\ref{fig:covid-data} and used in Fig.~\ref{fig:covid-results} are available on Zenodo~\cite{data-dynalearn2021}.

\noindent\textbf{Code availability}: The software developed for this project is available on Zenodo~\cite{code-dynalearn2021}.

\noindent\textbf{Acknowledgements}: The authors are grateful to Emily N. Cyr and Guillaume St-Onge for their comments and to Vincent Thibeault, Xavier Roy-Pomerleau, Fran\c{c}ois Thibault, Patrick Desrosiers, Louis J. Dub\'e, Simon Hardy and Laurent H\'ebert-Dufresne for fruitful discussions.  This work was supported by the Sentinelle Nord initiative from the Fonds d’excellence en recherche Apogée Canada (CM, EL, AA), the Conseil de recherches en sciences naturelles et en g\'enie du Canada (CM, AA) and the Fonds de recherche du Qu\'ebec-Nature et technologie (EL).

\noindent\textbf{Author contributions}: CM, EL and AA designed the study.  CM and EL designed the models and methods.  CM performed the numerical analysis.  CM and AA wrote the paper.

\noindent\textbf{Competing Interests}: The authors declare no competing interests.

\clearpage


\section*{Supplementary material}

# Deep learning of contagion dynamics on complex networks
# — Supplementary Information —

Charles Murphy, Edward Laurence, and Antoine Allard

*Département de Physique, de Génie Physique, et d'Optique,*

*Université Laval, Québec (Québec), Canada G1V 0A6 and*

*Centre interdisciplinaire en modélisation mathématique,*

*Université Laval, Québec (Québec), Canada G1V 0A6*

(Dated: June 23, 2021)

# Contents

# Supplementary Note I — Derivation of the importance weights

## A. Preliminaries

The importance weight $w_i(t)$ quantifies the extent to which the configuration of a node $i$ at a time $t$ weighs in the loss function. In turn, this affects how the parameters are optimized by correcting further for more important configurations. Yet, it is necessary to correctly define what "importance" means in this context, otherwise it can lead to badly trained models.

In the main paper, we considered the idea that the configurations should be weighted by an importance sampling (IS) scheme [1] where the target distribution is assumed uniform over all possible configurations. By doing so, we enforce the assumption that all configurations are equally important. Thus, the weights must be inversely proportional to the distribution of these configurations as they are observed in the training dataset. In the context of dynamical processes with a discrete and finite state set $\mathcal{S}$ on simple networks, this observed distribution is simply $\rho(k, x, x_{\mathcal{N}})$, where $k$ is the degree of a node, $x \in \mathcal{S}$ is its state and $x_{\mathcal{N}} \in \mathcal{S}^k$ is the vector state of its neighbors. The importance weight of a node $i$ at time $t$ is then

$$w_i(t) \propto \left[ \rho(k_i, x_i, x_{\mathcal{N}_i}) \right]^{-\lambda} \tag{1}$$

where we allow $\lambda$ to vary between 0 and 1, corresponding to no IS and pure IS, respectively.

## B. Generalizing the importance weights to continuous states

For the metapopulation dynamics, we need to generalize Eq. (1) because the probability distribution $\rho$ in this form can only be evaluated where $\mathcal{S}$ is a finite and countable set: $\rho$ can be computed by counting. When $\mathcal{S}$ is a subset of $\mathbb{R}$, counting cannot really be done directly and efficiently, which in turn prevents us from evaluating $\rho$. Instead, we must rely on some assumptions in order to evaluate $\rho$ efficiently.

### 1. Neighbor-Dependent Weights

We consider the direct generalization of $\rho(k_i, x_i, x_{\mathcal{N}_i})$ to real numbers. First, we factor $\rho(k_i, x_i, x_{\mathcal{N}_i}) = P(k_i)Q(x_i, x_{\mathcal{N}_i}|k_i)$. By doing so, the dependence of the importance weight with the degree is more conspicuous. Then, we break apart $Q(x_i, x_{\mathcal{N}_i}|k_i)$, because $Q(x, x_{\mathcal{N}}|k)$ must be permutation invariant under the neighbors states. This can be done in various ways, but the simplest one is probably

$$Q(x_i, x_{\mathcal{N}_i}|k_i) = \prod_{j \in \mathcal{N}_i} \left[ q(x_i, x_j|k_i) \right]^{1/k_i}, \tag{2}$$

where $q(x, x'|k)$ is the pairwise state probability conditioned on the degree $k$. Here, the geometric mean ensures that $Q(x, x_{\mathcal{N}}|k)$ does not have an artificially small value for nodes of high degree, as the values of $Q(x, x_{\mathcal{N}}|k)$ should roughly be of the size magnitude, for any degrees. In this context, we interpret $q(x, x'|k)$ as being the probability to observe in the training dataset a node of degree $k$ that is in state $x$ and connected to node in state $x'$. Therefore, its values must be normalized and bounded by the interval $[0, 1]$.

We make use of kernel density estimators (KDE) [2] with a Gaussian kernel to represent $q(x, x'|k)$. For each value of $k$, we simply build a different KDE, denoted $\hat{q}(x, x'|k)$. The function $\hat{q}(x, x'|k)$ returns density values, which can have any positive value. Thereby, we normalize it to obtain a probability value such that

$$q(x, x'|k) = \frac{\hat{q}(x, x'|k)}{z_k} \tag{3}$$

where

$$z_k = \sum_{t=1}^{T} \sum_{i \in \mathcal{V}} \sum_{j \in \mathcal{N}_i} I(k_i = k)\hat{q}\big(x_i(t), x_j(t)\big|k_i\big) \tag{4}$$

where $I(\cdot)$ is the indicator function.

Furthermore, the configurations must also be weighted with all additional information, that is with the node and edge attributes, i.e. $\Phi_i$ and $\Omega_{ij}$, respectively. This is easily achieved with KDE, where we simply concatenate these attributes to the state pairs, i.e. $q(x, x', \phi, \omega|k)$.

## 2. *Reducing the Complexity*

For continuous state dynamics such as the metapopulaton one, one would like to compute the importance weights using Eq. (3). Now, the problem with using Eq. (3) is that, for a given KDE function $q(x, x', \phi, \omega|k)$, a lot of samples are used to build it. The evaluation of standard KDE is known to scale like $\mathcal{O}(nm)$, where $n$ is the number of samples used to build the KDE and $m$ is the number of samples on which we wish to evaluate the KDE. Consequently, we need $\mathcal{O}\big[(Nk_{\max}T)^2\big]$ steps in order to evaluate all the normalization constant $z_k$ and all the importance weights. For reasonably lengthy time series with not too large networks, it renders the evaluation of the importance weights very inefficient. This is especially intensive for scale-free networks whose maximum degree is $k_{\max} = \mathcal{O}(N^{\frac{1}{\nu-1}})$ [3], where $\nu$ is the exponent of the degree distribution.

To reduce the computational burden of evaluating the importance weights, we consider including additional assumptions. First, we assume that the node and edge attributes are conditionally independent from the state pair. This is equivalent to assuming that the degree encodes all the information needed

to describe the state pairs. This allows us to factor them out of the pairwise state probability such that $q(x, x', \Phi, \Omega | k) = \Sigma(\Phi, \Omega | k) q(x, x' | k)$. We additionally assume that the edge attributes are correctly described by their respective mean taken over the neighbors of the node. This is equivalent to using the strength of the node $\Omega_i = \sum_{j \in \mathcal{N}_i} \Omega_{ij}$. Finally, we assume that, at a given time $t$, the state of the nodes are also correctly described by the average taken over all the nodes, i.e. $\bar{x}(t) = \frac{1}{N} \sum_{i \in \mathcal{V}} x_i(t)$. Thus, we obtained the form we used in the main paper,

$$w_i(t) = \Big[ P(k_i) \Sigma\big(\Phi_i, \Omega_i | k_i\big) \ \Pi\big(\bar{x}(t)\big) \Big]^{-\lambda} \tag{5}$$

where $\Sigma$ and $\Pi$ are represented by KDE using a similar strategy to Eq. (3). Those simplifications reduce the complexity of evaluating of the importance weights to $\mathcal{O}(N^2 + T^2)$, a considerable improvement to $\mathcal{O}\big[(N k_{\max} T)^2\big]$.

# Supplementary Note II — Loss descent patterns

In the case of the simple, complex and interacting contagion dynamics, we address a problem similar to a classification problem: For a given input, the model learns to assign it the correct label, i.e. the discrete state to which the node transition to. However, contrary to more standard classification problems, the label that the graph neural network (GNN) model learns to assign is not deterministic. Instead, it is assigned stochastically with a transition probability distribution provided by the dynamical process. This dramatically changes how the cross entropy loss decreases as the training goes on, because it is no longer expected to descend to zero (see Fig. 1(a–c)). What is expected to descend to zero is hopefully the difference between the ground truth transition probabilities and those of the GNN. Hence, choosing an objective function such as the Kullback-Liebler divergence (KLD), denoted $\mathcal{D}$, which is intimately related to the cross entropy loss and that quantifies the difference between two probability distributions, should shed some light as to why the cross entropy loss drops to a non-zero constant value. Consider the KLD between two discrete probability distributions $p$ and $q$,

$$\mathcal{D}(p||q) = \mathcal{H}(p, q) - \mathcal{H}(p) \tag{6}$$

where $\mathcal{H}(p, q) = -\sum_i p_i \log q_i$ is the cross entropy of $p$ and $q$, and $\mathcal{H}(p) = -\sum_i p_i \log p_i$ is the entropy of $p$. It is well known that minimizing $\mathcal{D}(p||q)$ with respect to $q$ yields $\mathcal{D}(p||q) = 0$ when $q = p$, which in turn leads to

$$\min_q \left[ \mathcal{D}(p||q) \right] = \min_q \left[ \mathcal{H}(p, q) - \mathcal{H}(p) \right] = \min_q \left[ \mathcal{H}(p, q) \right] - \mathcal{H}(p) = 0.$$

From this expression, we readily obtain the minimum expected value of the cross entropy loss,

$$\min_q \left[ \mathcal{H}(p, q) \right] = \mathcal{H}(p) \,. \tag{7}$$

In a realistic scenario, where the ground truth probabilities are not accessible directly, the entropy $\mathcal{H}(p)$ is also not accessible directly, which prevents the use of the KLD as the objective function altogether. For this reason, we use the cross entropy loss in our experiments involving stochastic dynamics. It is also worth monitoring the average entropy of the GNN outcome, which should, in the event that it is close to the ground truth, be of the same magnitude as the minimum of cross entropy loss.

In Fig. 1, we show an example of loss descent for each stochastic dynamics case. We also show the entropy of the GNN outcome averaged over the training dataset, and show the average Jensen-Shannon distance [4] (JSD), a symmetric version of the KLD, between the ground truth transition probabilities and the GNN predictions.

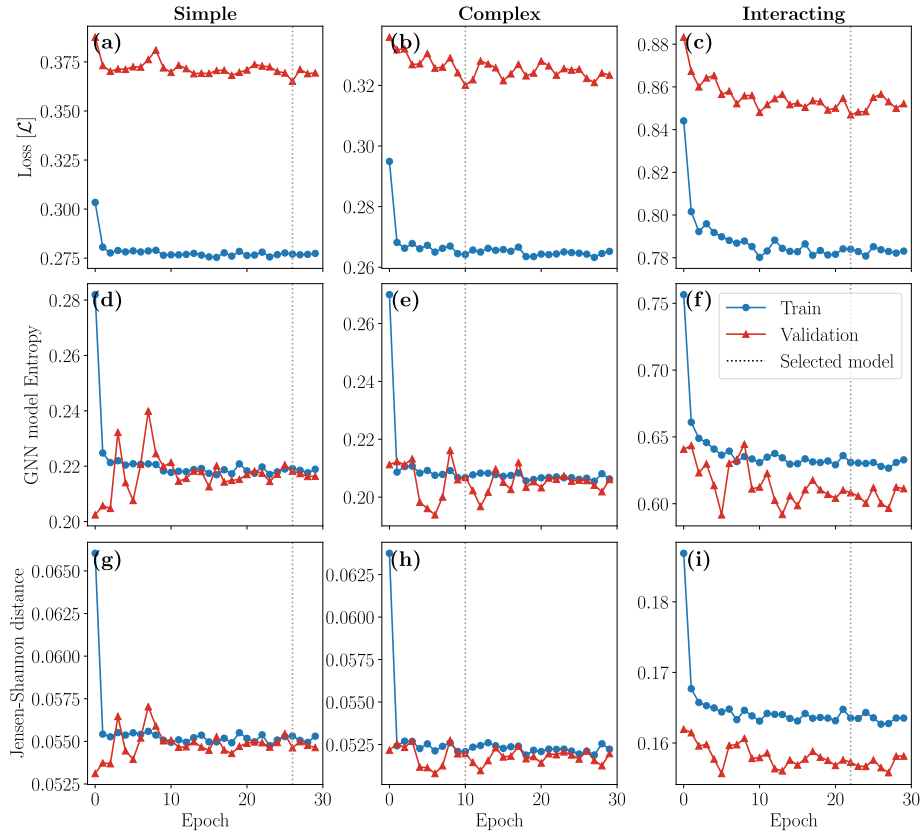

**Supplementary Figure** 1. **Loss optimization patterns during training.** (a–c) Loss as expressed by Eq. (3) in the main text, (d–f) average entropy of the GNN model predictions, (g–i) average Jensen-Shannon distance (JSD) between the GNN predicted LTPs and the ones given by the MLE. We show the results obtained when using Barabási-Albert networks to generate the data; similar conclusions are obtained when using data generated with Erdős-Rényi networks. All measures shown by these plots are approximated using the importance sampling scheme used to compute the loss. The vertical dotted lines show the minimum value of the validation loss, corresponding to our criterion for the model selection.

# Supplementary Note III — Impact of hyperparameters

### A. Performance measures and other metrics

In this section, we investigate the impact of several hyperparameters. To quantify the performance of our models, we used two kinds of metrics. The first one is similar to the one we used in the main paper, which is the Pearson error $1 - r$ computed from the Pearson correlation coefficient $r$ between all target-prediction pairs in the dataset. For completeness, the exact definition of $r$ is provided in the Material and Methods section of the main paper. The second one corresponds to the log Jensen-Shannon distance [4] averaged over all target-prediction pairs. The two metrics provide a similar picture of the global performance of the GNN model. Also, because we use discrete state dynamics in this context, we are allowed to evaluate the effective sample size (ESS) in the following way:

$$n_{\text{eff}} = \frac{\left[\sum_{x \in \mathcal{S}} \sum_{\boldsymbol{\ell}} n(x, \boldsymbol{\ell})\right]^2}{\sum_{x \in \mathcal{S}} \sum_{\boldsymbol{\ell}} \left[n(x, \boldsymbol{\ell})\right]^2} \tag{8}$$

where

$$n(x, \boldsymbol{\ell}) = \sum_{i \in \mathcal{V}} \sum_{t=1}^{T} I\Big(x_i(t) = x \,\wedge\, \boldsymbol{\ell}_i(t) = \boldsymbol{\ell}\Big) \tag{9}$$

is the number of nodes at any times in the dataset that were in state $x$ and that had a neighborhood state vector $\boldsymbol{\ell}$. To better appreciate the relationship between the performance metrics and the ESS, we center and rescale the ESSs with the mean and standard deviation ESS across the experiments which varies the same hyperparameter.

### B. Time Series Length

The time series length, denoted by $T$, corresponds to the number of time steps in the training dataset. It also affects the length of an epoch. We investigate the values $T = \{100, 500, 1000, 5000, 10000\}$.

Figure 2 shows the accuracy diagrams of GNN models trained using different time series lengths. As we can expect, longer time series tend to yield better models. This is unsurprising for two reasons. First, because the targets with which the models are trained are noisy, it generally helps to have larger a training dataset. Using noisy targets yields a noisy objective function as well, for which the noise can be reduced by increasing the number of samples. Second, using larger datasets means that we train the model for a longer period of time. We also note that, because the gradient descent is performed using a stochastic technique, the results can be a bit inconsistent with our previous observations. This is likely to also affect our next

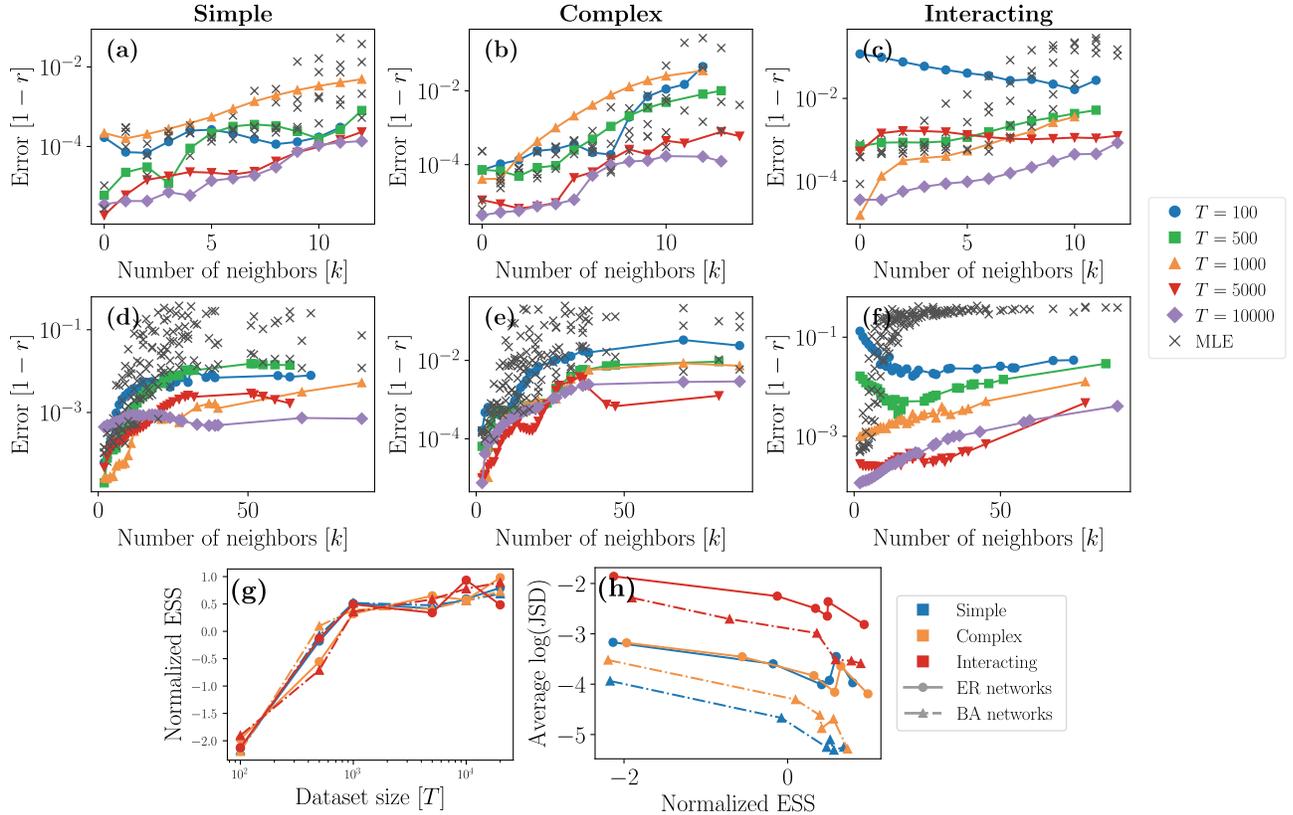

**Supplementary Figure** 2. **Accuracy diagrams for different time series lengths** $T$: We show the accuracy diagrams, that is the error as a function of the degree of the nodes, of GNNs trained on the simple (left column), complex (middle column) and interacting contagion dynamics evolving on Erdős-Rényi (ER, top row) and Barabási-Albert (BA, bottom row) networks. In every panel, we indicate the value of the changing hyperparameter, namely the time series length, with the symbols and the colors according to the legend. The maximum likelihood estimators (MLE), computed from the procedure specified in the main paper, is indicated as a reference. Panel (g) shows the normalized effective sample size (ESS) as a function of the hyperparameter. Finally, panel (h) shows the relationship between the error—the average log-JSD error to be more precise—as a function of the ESS. In panels (g, h), the symbols and line style encode the type of networks used to generate the training dataset and the colors indicate the dynamics.

results, hence we need to keep it in mind. A time-consuming way of addressing this issue would be to train multiple GNNs in the same configurations, and to then average their errors together.

## C. Network size

The network size, denoted by $N$, is the number of nodes in the networks on which the dynamics evolved to generate the training dataset. We investigate the values $N = \{100, 500, 1000, 5000\}$.

Similarly, Fig. 3 shows the accuracy diagrams when changing the network size. At first, increasing

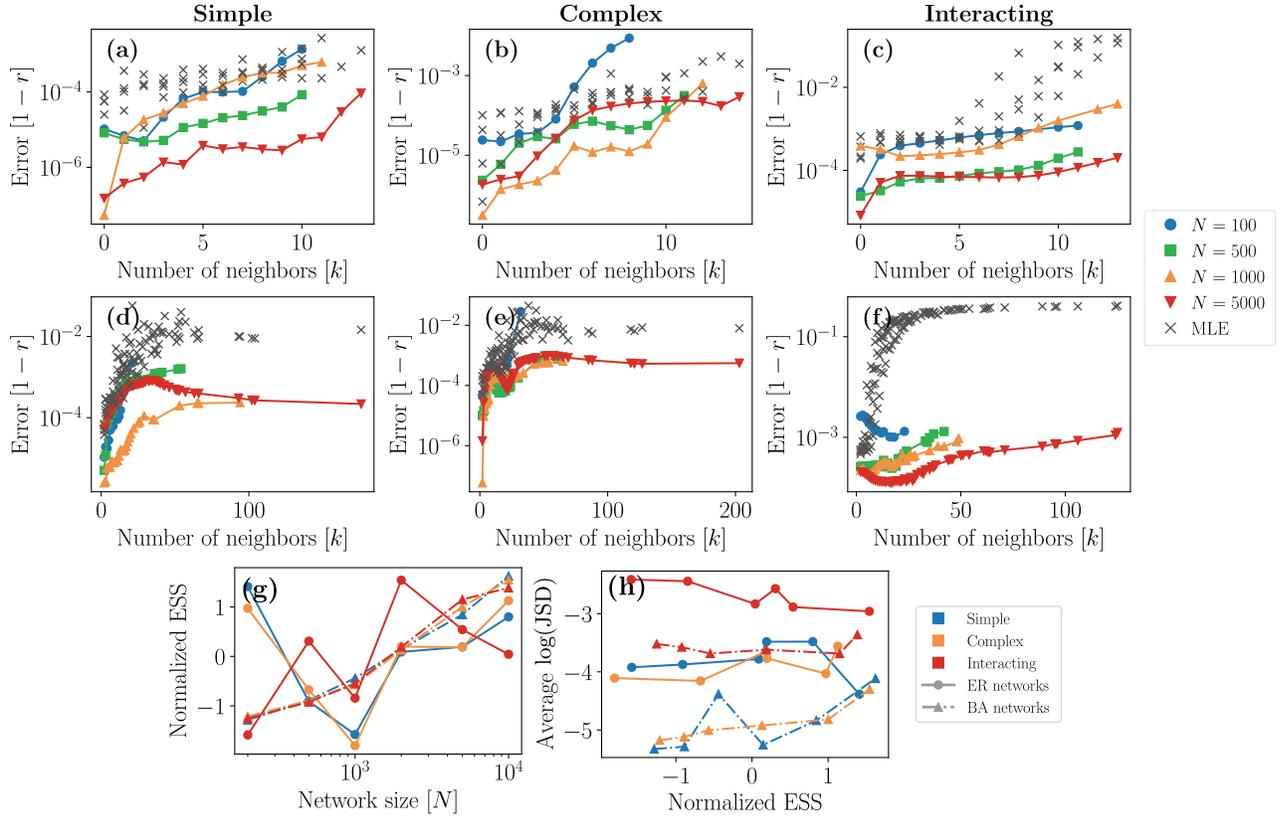

**Supplementary Figure** 3. **Accuracy diagrams for different network sizes** $N$**:** We refer to Fig. 2 for the organization of the panels.

$N$ seems to affect the performance of the models differently depending on the type of networks used. First, for Erdős-Rényi (ER) networks, increasing $N$ does not tend to increase the ESS. This is expected because the maximum degree only slighting increase when the number of nodes is increased, for fixed the average degree $\langle k \rangle$. Hence, we do not observe additional degree classes when $N$ is marginally increased and the training dataset variety remains similar. For Barabási-Albert (BA) networks, we observe something different: While the increase in $N$ leads to higher ESS, there is still no substantial gain in performance. This can be explained by looking at the degree distribution. As more nodes are added to the network, the degree classes get more populated, resulting in increased ESS. However, because the degree distribution is scale-free (with exponent $-3$), these are not populated evenly and more degree classes are created as $N$ increases.

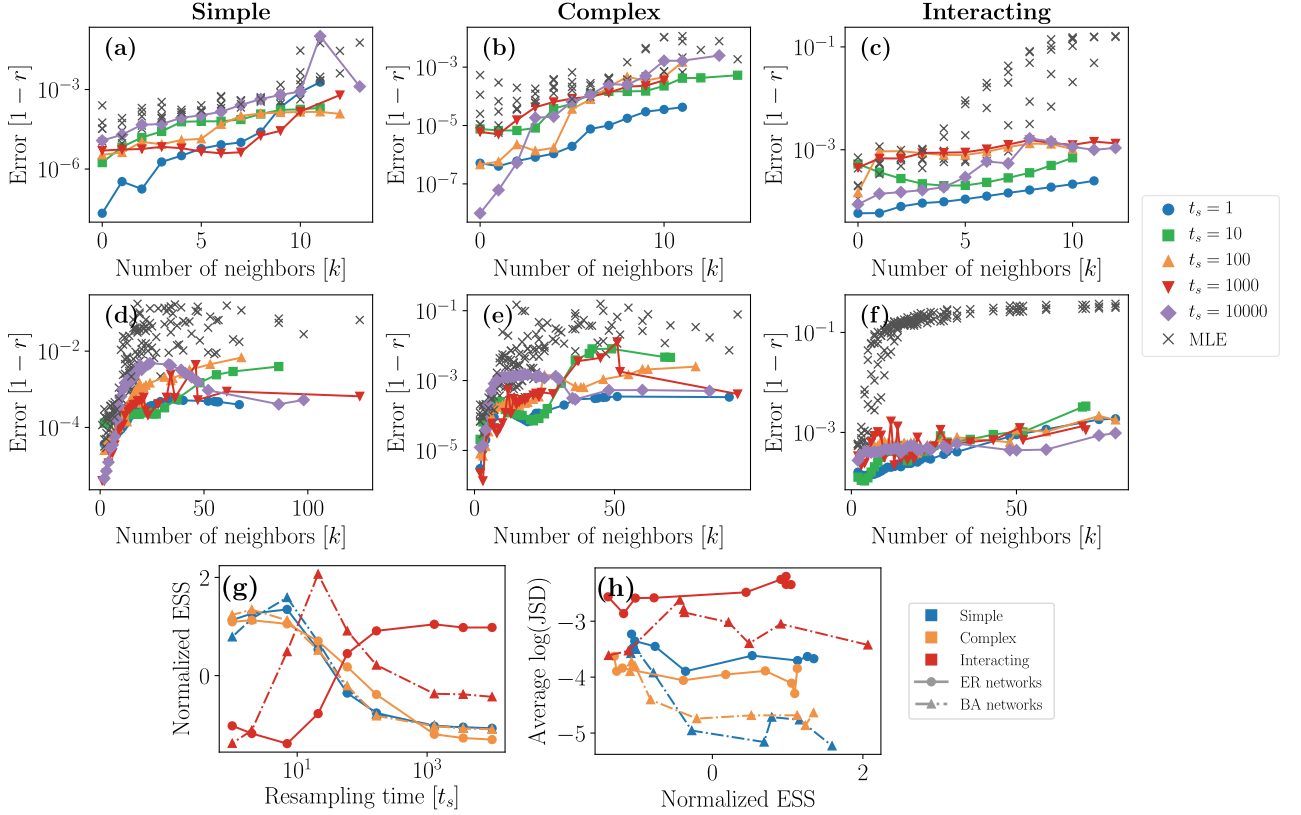

**Supplementary Figure** 4. **Accuracy diagrams for different resampling times** $t_s$**:** We refer to Fig. 2 for the organization of the panels.

### D.   Resampling time

The resampling time, denoted by $t_s$, corresponds to the number of time steps before the states of the nodes are reinitialized when generating the training dataset. Recall that this hyperparameter was introduced in the main paper to improve the variability in the dataset, where small values of $t_s$ is expected to increase the ESS. We investigate the values $t_s = \{1, 10, 100, 1000, 10000\}$.

In Fig. 4, we show the accuracy diagrams when the resampling time is changed. It is clear from Fig. 4 that decreasing the resampling time increases the ESS, thus we tend to train better models. However, in most cases, the gain seems to be marginal.

### E.   Importance Sampling Bias

The role of the importance sampling bias, denoted $\lambda$, is to modulate the influence of the importance weights, where $\lambda = 1$ corresponds to the ideal case, which is a standard IS scheme, and $\lambda = 0$ correspond

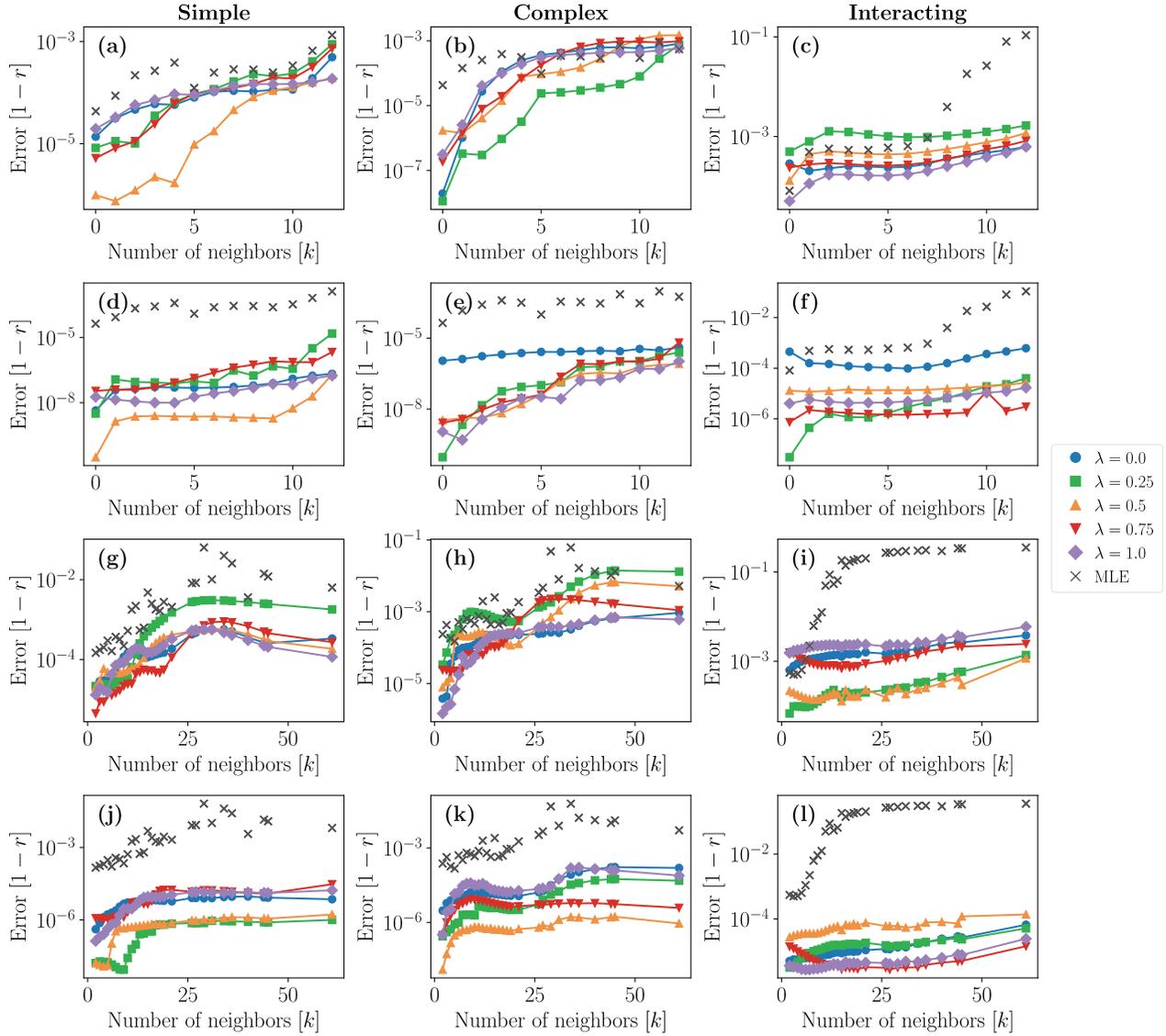

**Supplementary Figure** 5. **Accuracy diagrams for different important sampling bias exponents** $\lambda$**:** Similarly to Fig. 2, we show the accuracy diagrams of GNN models trained on (left column) simple, (middle column) complex and (right column) interacting contagion dynamics propagating on (a–f) Erdős-Rényi (ER) and (g–l) Barabási-Albert (BA) networks. We also show the maximum likelihood estimators (MLE) for comparison. Additionally, the panels (a–c) and (g–i) correspond to GNN models trained using the observed outcome, denoted $\tilde{y}_i(t)$ in the main paper, which corresponds to the state of the node at the next time step: the labels are noisy in this case. Conversely, the GNNs corresponding to panels (d–f) and (j–l) used the true transition probabilities, denoted $y_i(t)$ in the main paper: the labels are deterministic in this case. On all panels, the symbols and colors indicate the value of $\lambda$ as specified by the legend.

to a uniform sampling scheme, that is without IS. There are multiple reasons why it would be preferable to use an exponent $\lambda < 1$. First, it is possible to poorly define the importance weights $w_i(t)$ by a bad choice of assumptions (see Sec. I). This is in part due to the fact that we rely on some statistics to represent the training dataset, which can either contain false assumptions or be poorly estimated due to a small sample size. Second, in the case of stochastic dynamics, letting $\lambda = 1$ is analogous to putting a strong emphasis on rare configurations, which in turn are likely to suffer from a small sample size. This will lead to a poor estimation of the objective function, which is likely to reduce the overall performance of the model. That being said, we investigate the values $\lambda \in \{0, 0.25, 0.5, 0.75, 1\}$.

In Fig. 5, we show the accuracy diagrams when the IS bias exponent $\lambda$ is changed. To better appreciate the comparison between these different training settings, we used the same dataset, networks and training settings. We also trained our models in two different scenarios: we considered using the observed outcome $\tilde{y}_i(t)$, defined as the state of node $i$ at the next time step, to evaluate the objective function. We also used the true transition probabilities, that is the outcome $y_i(t)$, to evaluate the loss. The difference between these two scenarios is that, in the first case, the targets $\tilde{y}_i(t)$ are noisy, and in the other, the targets $y_i(t)$ are deterministic. As it was mentioned earlier, we argue that the choice of $\lambda$ will be dependent on the stochastic nature of the underlying dynamics.

First, from Fig. 5, we can see that choosing $\lambda = 1$ rarely leads to the most proficient models, for models trained on both the ER and the BA networks. Only the case of the interacting contagion dynamics does it seem to improve the performance, but as we discussed before, this is conditioned on the fact that either the sample size is large enough or the dynamic is deterministic. Interestingly, the case $\lambda = 0$ often has similar performance as the case $\lambda = 1$ in the noisy scenarios. In general, the best models seem to be obtained when $\lambda$ is somewhat in the middle, as if the pure IS and the no IS cases are both too strong assumptions.

### F. Graph Neural Network Architecture

We investigate the accuracy diagrams of the models when we use different GNN architectures. To be more specific, we consider six additional GNN architectures that has been shown to perform well in the context of structure learning [5–8].

#### 1. Models

We label the 6 models as follows: We call our architecture the *Att-GNN*, which is described in detail in the main paper. We also consider the architecture from Ref. [8] with multiple aggregation scheme. This

class of architectures aggregate the neighbors' features as follows:

$$\nu_i = \mathcal{A}(\xi_i) + f_{\mathrm{AGG}}(\{\!\{\mathcal{B}(\xi_j) | j \in \mathcal{N}_i\}\!\}) \tag{10}$$

where $\{\!\{\cdot\}\!\}$ denoted a multiset and we recall that $\mathcal{A}$ and $\mathcal{B}$ are linear transformations with a trainable weight matrix and bias vector. Also, we need to specify the $f_{\mathrm{AGG}}$ function, which is a differentiable and permutation-invariant function that aggregates the neighbors' features. We consider three cases for the $f_{\mathrm{AGG}}$ function: the mean pooling case, denoted *Mean-GNN*, where

$$f_{\mathrm{AGG}}(\{\!\{x_1, \cdots, x_k\}\!\}) = \sum_{i=1}^{k} \frac{x_i}{k} \,, \tag{11}$$

the max pooling case, denoted *Max-GNN*, where the $\mu^{\mathrm{th}}$ feature is aggregated such that

$$[f_{\mathrm{AGG}}(\{\!\{x_1, \cdots, x_k\}\!\})]_\mu = \max\{x_{\mu,1}, \cdots, x_{\mu,k}\} \,, \tag{12}$$

and the sum pooling case, denoted *Sum-GNN*, where similar to the mean pooling case,

$$f_{\mathrm{AGG}}(\{\!\{x_1, \cdots, x_k\}\!\}) = \sum_{i=1}^{k} x_i \,, \tag{13}$$

Then, we consider four additional standard architectures: the *GraphSage* architecture from Ref. [9], the graph convolution network (denoted *GCN*) from Ref. [5], the original graph attention network (denoted *GAT*) from Ref. [7] and a GNN architecture used to forecast COVID-19 [10] (denoted *Kapoor-GNN*). The GraphSage aggregates the neighbors' features similarly to the Mean-GNN:

$$\nu_i = \boldsymbol{W}_1 \xi_i + \boldsymbol{W}_2 \sum_{j \in \mathcal{N}_i} \xi_j \,, \tag{14}$$

which, in turn is similar to the GCN,

$$\nu_i = \boldsymbol{W} \sum_{j \in \mathcal{N}_i \vee \{i\}} \frac{\xi_j}{(k_i+1)(k_j+1)} \,. \tag{15}$$

Here, $\boldsymbol{W}$ and $\boldsymbol{W}_i$ are a trainable weight matrix. The GAT architecture aggregates the neighbors' features as follows:

$$\nu_i = \boldsymbol{W} \sum_{j \in \mathcal{N}_i \vee \{i\}} a_{ij} \xi_j \tag{16}$$

where

$$a_{ij} = \frac{e^{\theta_{ij}}}{\sum_{j \in \mathcal{N}_i \vee \{i\}} e^{\theta_{ij}}} \tag{17}$$

and

$$\theta_{ij} = \text{LeakyReLU}_\alpha \left( \boldsymbol{a}^T \boldsymbol{W} \xi_i + \boldsymbol{b}^T \boldsymbol{W} \xi_j \right). \tag{18}$$

In this case, $\boldsymbol{a}$ and $\boldsymbol{b}$ are weight vectors and $\text{LeakyReLU}_\alpha$ is an activation function such that

$$\text{LeakyReLU}_\alpha(x) = \begin{cases} x & \text{if } x > 0 \\ \alpha x & \text{otherwise} \end{cases}, \tag{19}$$

and $\alpha$, the negative slope, is generally fixed to $0.2$. Finally, we consider the Kapoor-GNN architecture which was used to forecast COVID-19 in this US at the county level based on the mobility flow [10] between counties. This architecture is composed of a sequence of two GCN layers in series. As the first layer aggregates the features of the first neighbor, the second adds that of the second neighbors as well.

### 2. Results

In Fig. 6, we show the predicted transition probabilities for the simple and complex contagion dynamics. We see that, in general, the standard GNN architectures yield poor performance in predicting the infection probabilities, even though they have been trained using the same dataset, networks and hyperparameters. We believe this is due to the fact that they internally use a non-extensive aggregation operator—for instance mean-pooling and max-pooling, whose output does not scale with the size of the input. To be clearer, let us assume a node of degree $k$ of which we wish to aggregate the features of its $k$ neighbors. By using a non-extensive aggregator, the output is expected to be of a similar scale as that of any other node of degree $k'$. Hence, the GNN model is likely to have a hard time distinguishing the vector features of nodes of different degrees—a structural feature that we know has a huge impact on most of the dynamical processes on networks. This is in part why almost all GNN architectures described above fail at learning and representing contagion dynamics. The only ones that perform similarly are the Att-GNN and Sum-GNN, which both use extensive aggregators.

From Figs. 6 and 7, we can also appreciate how some GNN architectures are better than others at predicting the recovery probability, this is independent from the neighbors' states unlike the infection probability. Specifically, the GraphSage and GCN architectures have a hard time predicting the recovery probabilities. They may be due to the fact that their aggregator does not distinguish the different values of neighbor features and accept all contributions equally, as opposed to for instance the GAT which is expected to weigh the neighbors' features before aggregating them. The same principle applies to the other architectures that predict correctly the recovery probability. Also, while the Kapoor-GNN, which we recall is composed of

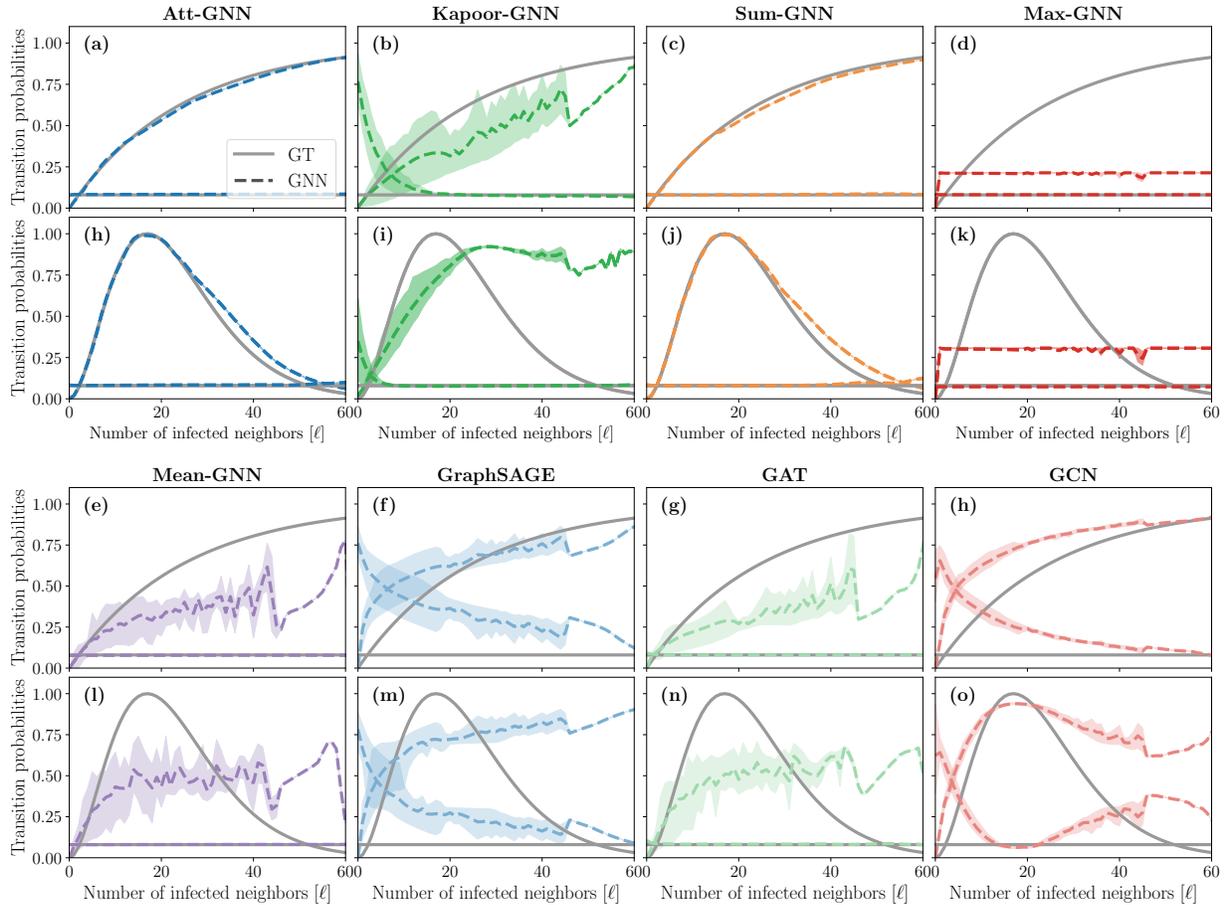

**Supplementary Figure** 6. **Prediction of different GNN architectures on (a–h) simple and (i–o) complex contagion dynamics on Barabási-Albert Networks:** We show the infection and recovery probabilities as predicted by the trained GNNs (dashed lines), and given by the ground truth (GT, solid lines). Each column corresponds to a different architecture. In the top and bottom rows, all models have been trained on the same training dataset and networks. The training settings and parameters of the dynamics are the same as described in the main paper. Also, we used the same training dataset and networks to train each GNN architecture.

two GCN layers in series, seem to perform better than the single GCN architecture, it still struggles overall in comparison with the Att-GNN. Hence, it suggests that increasing the depth of the GNN aggregation scheme may be insufficient to improve the accuracy of the models.

In summary, not all GNN architectures are capable of learning a dynamical process on networks, which also supports the idea, presented in Ref. [11], that most GNN architectures in fact do not have a high expressive power. Then, we can ask if having an extensive aggregator will always be sufficient in the context of dynamical process learning. From our work, it seems to be the case, but the few examples we provide in this paper are far from conclusive in that regard. However, in the case where extensive aggregator would be insufficient, one could consider new strategies such that which is presented in Ref. [12], where

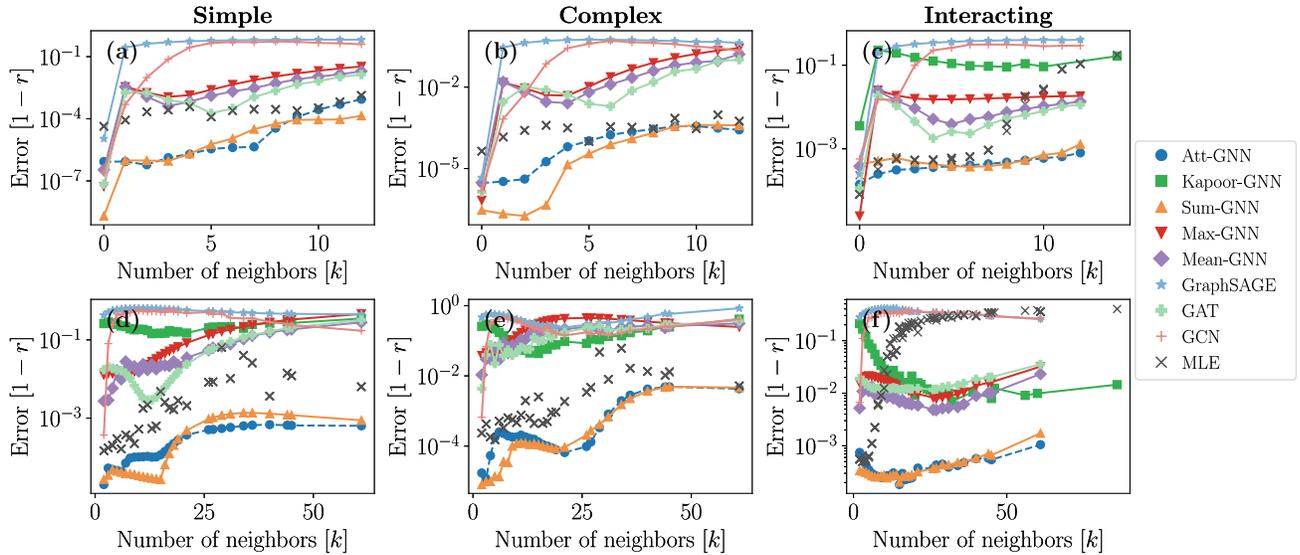

**Supplementary Figure** 7. **Accuracy diagrams for different GNN architectures:** On each panel, the different GNN architectures were trained on the same dataset with the same training settings and hyperparameters. For further details, we refer to Fig. 2.

multiple aggregators are used in parallel.

### 3. Graph neural networks on dynamic networks

It is worth mentioning that there exists a wide variety of GNN architectures designed specifically to handle dynamic networks [13], i.e. networks whose topology evolves over time. These architectures are typically composed of a sequence of GNN layers each of which being applied to one element of a sequence of temporally ordered networks $(\cdots, G_{t-1}, G_t, G_{t+1}, \cdots)$. These learned representations are then combined to obtain some network and/or node embeddings containing temporal information. While useful for temporal networks, this class of models could hardly be adapted to the task of dynamics learning on *static* networks where it is solely the states of nodes that evolve over time—such as in Figs. 6 and 7. They could, nevertheless, be useful in the context of adaptive systems, where the topology of the network changes over time according to the dynamics of the nodes [14]. However, we suspect that the GNN layers at their core should also be chosen carefully to avoid the complications caused by standard GNN architectures in the context of dynamics learning.

## Supplementary Note IV — Interpretability of the models

Like most deep learning models, ours suffer from an interpretability problem in that the parameters learned during training can hardly be associated with specific mechanisms guiding the dynamics. This is in contrast with mechanistic models where the parameters are chosen beforehand to emulate a specific, interpretable behavior. The interpretability problem of our models surely is a drawback of the method, but it can be slightly alleviated. First, as we have demonstrated in the main paper (see Fig. 3), our models can be used to recover the phase transition bifurcation diagrams. As a result, even though the learned parameters of the model are non-interpretable, the behavior of the model, and more specifically the influence of the structure on the dynamics, can still be investigated.

Second, we argue in the Material and Methods section of the main paper that the parameters of our attention mechanism are partially interpretable [15, 16]. Recall that the attention mechanism computes attention coefficients $a_{ij}$ that weigh the interaction between some node features $\xi_i$ and that of one of its neighbors, $\xi_j$. Hence, we should expect $a_{ij}$ to be high—close to 1—when two nodes are expected to interact, and it should be low—close to 0—otherwise. For example, in simple contagion dynamics, two nodes interact strictly when the target node is susceptible and the source node is infected. Otherwise, the transition probabilities of a target node $v_i$ are invariant with respect to the states of its neighbors, $x_{\mathcal{N}_i}$. We say that the state of a node is neighbor invariant when the transition probability of this node is independent of the state of its neighbors.

In reality, it is not exactly what happens, as we can see in Fig. 8. In fact, when the target nodes are in a neighbor-invariant state ($I$ for the simple and complex contagion dynamics, and $I_1 I_2$ for the interacting contagion dynamics), the attention coefficients are correctly close to zero. However, when they are not (e.g. state $S$ for the simple and complex contagion dynamics), the attention mechanism can be non-zero regardless of the state of the neighbors. We think this is due to the possibility that the representations combined by the attention mechanism are not sparse: the features of the neighbors of a node are combined in such a way that they cancel out the contributions of the features of the noncontributing neighbors. This way, the attention coefficients are not necessarily constrained to be zero even though the nodes are effectively not interacting together. This degeneracy seems to be amplified with a greater number of parallel attention modules, i.e. the number of different available representations learned by the model. This phenomenon is analogous to the sparsity problem in under-determined linear regression models, where multiple parameters can fit the same training dataset depending on the loss function [17]. The addition to the loss function of a L1-norm regularization on the attention coefficients is therefore a promising avenue to increase the

**Supplementary Figure** 8. **Attention coefficients as a function of the source-target states for (a-d-g) simple contagion, (b-e-h) complex contagion and (c-f-i) interacting contagion dynamics**. We show the attention coefficients of different models: (a–c) are models with one attention layer, (d–f) have two attention layers and (g–i) have four. The values of the different attention layers are shown by the increasingly lighter colored bars. For the source-target states, we indicate the type of node using the directionality of the arrows: for $X \rightarrow Y$, $X$ is state of the source node and $Y$ is the state of the target node. We also highlight the source-target states where we expect the transition probability of the target node to be non neighbor invariant. The other hyperparameters are given in Tab. 1 of the main paper and in Sec. A6.

interpretability of our approach.

# Supplementary References

[1] R. Y. Rubinstein and D. P. Kroese, *Simulation and the Monte Carlo Method*, 3rd ed. (Wiley, 2016) p. 414.

[2] W. J. Conover, *Practical nonparametric statistics* (John Wiley & Sons, 1998) p. 350.

[3] M. Boguñá, R. Pastor-Satorras, and A. Vespignani, "Cut-offs and finite size effects in scale-free networks," Eur. Phys. J. B **38**, 205–209 (2004).

[4] T. M. Cover and J. A. Thomas, *Elements of Information Theory*, 2nd ed. (Wiley-Interscience, 2005) p. 776.

[5] T. N. Kipf and M. Welling, "Semi-Supervised Classification with Graph Convolutional Networks," (2016), arXiv:1609.02907.

[6] W. L. Hamilton, R. Ying, and J. Leskovec, "Representation Learning on Graphs: Methods and Applications," (2017), arXiv:1709.05584.

[7] P. Veličković, G. Cucurull, A. Casanova, A. Romero, P. Liò, and Y. Bengio, "Graph Attention Networks," (2018), arXiv:1710.10903.

[8] C. Morris, M. Ritzert, M. Fey, W. L. Hamilton, J. E. Lenssen, G. Rattan, and M. Grohe, "Weisfeiler and Leman Go Neural: Higher-order Graph Neural Networks," (2018), arXiv:1810.02244.

[9] W. L. Hamilton, R. Ying, and J. Leskovec, "Inductive representation learning on large graphs," in *Proceedings of the 31st International Conference on Neural Information Processing Systems*, NIPS'17 (2017) p. 1025–1035.

[10] A. Kapoor, X. Ben, L. Liu, B. Perozzi, M. Barnes, M. Blais, and S. O'Banion, "Examining covid-19 forecasting using spatio-temporal graph neural networks," (2020), arXiv:2007.03113.

[11] K. Xu, W. Hu, J. Leskovec, and S. Jegelka, "How Powerful are Graph Neural Networks?" (2018), arXiv:1810.00826.

[12] G. Corso, L. Cavalleri, D. Beaini, P. Liò, and P. Veličković, "Principal Neighbourhood Aggregation for Graph Nets," (2020), arXiv:2004.05718.

[13] J. Skarding, B. Gabrys, and Musial K., "Foundations and modelling of dynamic networks using dynamic graph neural networks: A survey," (2020), arXiv:2005.07496.

[14] Thilo Gross and Bernd Blasius, "Adaptive coevolutionary networks: a review," J. R. Soc. Interface **5**, 259–271 (2008).

[15] S. Vashishth, S. Upadhyay, G. S. Tomar, and M. Faruqui, "Attention interpretability across nlp tasks," (2019), arXiv:1909.11218.

[16] S. Serrano and N. A. Smith, "Is attention interpretable?" in *Proceedings of the 57th Annual Meeting of the Association for Computational Linguistics* (Association for Computational Linguistics, Florence, Italy, 2019) pp. 2931–2951.

[17] R. Tibshirani, "Regression shrinkage and selection via the lasso," J. R. Stat. Soc. B **58**, 267–288 (1996).



\end{document}